\documentclass[10pt]{article}
\pdfoutput=1

\usepackage{amsmath}
\usepackage{amssymb}
\usepackage{mathtools}
\usepackage{stackrel}
\usepackage{scalerel}
\usepackage{leftidx}
\usepackage{xfrac}

\usepackage{euscript}
\let\mathbb\undefined
\usepackage{bbold}
\usepackage{authblk}

\usepackage[shortlabels]{enumitem}

\usepackage{xcolor}
\usepackage{tcolorbox}
\usepackage{graphicx}
\usepackage{subfigure}

\usepackage[all,2cell,arrow,matrix]{xy} \UseAllTwocells \SilentMatrices
\usepackage{tikz-cd}
\usepackage{tikz}
\usetikzlibrary{arrows,arrows.meta,%
decorations.markings,decorations.pathreplacing,%
decorations.pathmorphing,calc}

\usepackage{verbatim}
\usepackage{comment}

\usepackage[nottoc]{tocbibind} % References in ToC
\usepackage{appendix}

\usepackage{geometry}
\tikzset{->-/.style={decoration={markings,mark=at position #1 with {\arrow{Stealth}}},postaction={decorate}},->-/.default=0.55}
\usepackage[amsmath,amsthm,thmmarks]{ntheorem}
\theoremstyle{definition}

\newtheorem{thm}{Theorem}[section]
\newtheorem{prop}[thm]{Proposition}

\newtheorem{cor}[thm]{Corollary}
\newtheorem{lem}[thm]{Lemma}

{
\theoremsymbol{\mbox{${\blacksquare}$}}
\newtheorem{defn}[thm]{Definition}
}
{
\theoremsymbol{\mbox{$\heartsuit$}}
\newtheorem{expl}[thm]{Example}
}
{
\theoremsymbol{\mbox{$\diamondsuit$}}
\newtheorem{rem}[thm]{Remark}
}
\qedsymbol{\mbox{$\square$}}

\numberwithin{equation}{section}
\numberwithin{thm}{section}

\usepackage{hyperref}
\usepackage{cleveref}
\hypersetup{colorlinks=true, linkcolor=red!50!black, citecolor=green!50!black, urlcolor=blue!80!black}

\newcommand\be            {\begin{equation}}
\newcommand\ee            {\end{equation}}
\newcommand\bea           {\begin{eqnarray}}
\newcommand\eea         {\end{eqnarray}}
\newcommand\bnu          {\begin{enumerate}}
\newcommand\enu          {\end{enumerate}}
\newcommand\bit          {\begin{itemize}}
\newcommand\eit          {\end{itemize}}

\newcommand{\pf}{\begin{proof}}
\newcommand{\epf}{\qed\end{proof}}

\makeatletter
\providecommand{\leftsquigarrow}{%
  \mathrel{\mathpalette\reflect@squig\relax}%
}
\newcommand{\reflect@squig}[2]{%
  \reflectbox{$\m@th#1\rightsquigarrow$}%
}
\makeatother

\newcommand\Cb			{\mathbb{C}}

\newcommand\Nb			{\mathbb{N}}

\newcommand\Rb			{\mathbb{R}}
\newcommand\Zb			{\mathbb{Z}}

\newcommand\bk			{\mathbb{k}}

\newcommand\CA			{\EuScript{A}}

\newcommand\CC			{\EuScript{C}}
\newcommand\CD			{\EuScript{D}}

\newcommand\CG			{\EuScript{G}}

\newcommand\CU			{\EuScript{U}}
\newcommand\CV			{\EuScript{V}}
\newcommand\CW			{\EuScript{W}}

\newcommand\CY			{\EuScript{Y}}
\newcommand\CZ			{\EuScript{Z}}

\newcommand{\FZ}			{\text{\usefont{U}{euf}{m}{n}Z}}

\newcommand\SC			{\mathsf{C}}

\newcommand{\bfe}		{\mathbf{e}}
\newcommand{\bfu}		{\mathbf{u}}

\DeclareMathOperator{\Hom}{Hom}
\DeclareMathOperator{\End}{End}
\DeclareMathOperator{\Aut}{Aut}

\DeclareMathOperator{\Id}{Id}
\DeclareMathOperator{\id}{id}

\DeclareMathOperator{\nat}{Nat}
\DeclareMathOperator{\fun}{Fun}

\DeclareMathOperator{\Map}{Map}
\DeclareMathOperator{\Sing}{Sing}

\DeclareMathOperator{\LMod}{LMod}

\newcommand{\op}			{\mathrm{op}}
\newcommand{\rev}			{\mathrm{rev}}

\newcommand{\one}			{\mathbb{1}}

\newcommand\forget  {\text{\usefont{U}{euf}{m}{n}f}}

\newcommand\cat			{\mathrm{Cat}}
\newcommand\Set			{\mathrm{Set}}
\newcommand\set			{\mathrm{Set}}
\newcommand\sSet			{\mathrm{sSet}}

\newcommand\vect			{\mathrm{Vec}}

\newcommand\rep			{\mathrm{Rep}}
\newcommand\Topo			{\mathrm{Top}}
\newcommand\Ab			{\mathrm{Ab}}

\newcommand{\bscale}	{0.7}
\makeatletter
\newcommand{\ec}[2][]	{{\@ec{#1 |}{#2}}}
\newcommand{\bc}[2][]	{{\@ec{#1}{#2}}}
\newcommand{\@ec}[2]	{\mathchoice
  {\displaystyle \raise.9ex\hbox{$\scaleobj{\bscale}{#1}$} {#2}}%
  {\textstyle \raise.9ex\hbox{$\scaleobj{\bscale}{#1}$} {#2}}%
  {\scriptstyle \raise.55ex\hbox{$\scriptstyle \scaleobj{\bscale}{#1}$} {#2}}%
  {\scriptscriptstyle \raise.38ex\hbox{$\scriptscriptstyle \scaleobj{\bscale}{#1}$} {#2}}%
}
\makeatother

\newcommand{\ot}			{\one_c}
\newcommand{\mt}			{m_c}

\begin{document}

\title{\huge Finite 2-group gauge theory and \\ its 3+1D lattice realization}
\author{Mo Huang \thanks{Email: \href{mailto:mohuang@hku.hk}{\tt mohuang@hku.hk} \\ \phantom{testemail: } \href{mailto:jasmine.huang.527@gmail.com}{\tt jasmine.huang.527@gmail.com}}}
\affil[a]{Department of Physics, The University of Hong Kong, Pokfulam Road, Hong Kong, China}
\date{\vspace{-5ex}}

\maketitle

\begin{abstract}
In this work, we employ the Tannaka-Krein reconstruction to compute the quantum double $\CD(\CG)$ of a finite 2-group $\CG$ as a Hopf monoidal category. We also construct a 3+1D lattice model from the Dijkgraaf-Witten TQFT functor for the 2-group $\CG$, generalizing Kitaev's 2+1D quantum double model. Notably, the string-like local operators in this lattice model are shown to form $\CD(\CG)$. Specializing to $\CG = \mathbb{Z}_2$, we demonstrate that the topological defects in the 3+1D toric code model are modules over $\CD(\mathbb{Z}_2)$.
\end{abstract}

\tableofcontents

\part{Preliminaries}

\section{Introduction}

The 2+1D quantum double model for a finite group $G$ is an exactly solvable model proposed by Kitaev \cite{Kit03}. It is the Hamiltonian version of the 2+1D $G$ gauge theory \cite{DW90}. Kitaev also showed \cite{Kit03} that the local operators in the quantum double model is isomorphic to the quantum double algebra $D(G)$, and the category of particle-like topological defects is equivalent to the representation category $\rep(D(G))$, which is also the Drinfeld center \cite{Maj91,JS91} of $\rep(G)$.

It is natural to study the higher-dimensional quantum double model. This generalization should also be a 3+1D exactly solvable model, and it should be the Hamiltonian version of the 3+1D gauge theory with the gauge group $G$ replaced by a 2-group, which is a natural categorification of group. Just as a group is a set equipped with a group structure, a 2-group is a category equipped with a group structure. So it consists of two levels of ordinary groups: one level for objects, and the other one for morphisms.

There are some different models of 2-group: weak 2-group, strict 2-group and crossed module. A weak 2-group is a monoidal (or tensor) category satisfying some invertibility conditions (see Section \ref{sec_2-group} for more details). A strict 2-group is similar to a weak 2-group, but the invertibility conditions hold in a more strict way. A crossed module is essentially the same as a strict 2-group, but it even does not look like a category: it consists of two ordinary groups and some morphisms between them. These three models are equivalent, in the sense that they constitute the equivalent 2-categories (see \cite{BL04} for example).

There are some works on 3+1D lattice models realizing 2-group gauge theory in the language of crossed modules. For example, see \cite{BCKMM17,BCKMM19,DT18,DT19,BD20,Del22a,HS23,HS23a,HS24}. They should be the Hamiltonian version of Yetter's TQFT \cite{Yet93}. However, most of these works did not discuss the topological defects. In \cite{DT19}, the authors showed that the loop-like topological defects are irreducible modules over the generalized tube algebra, but did not consider the 2-category of these topological defects. In \cite{Del22a}, the author studied a 3+1D lattice model with a specific gauge 2-group $\CG$ and the 2-category of topological defects on a 2+1D boundary, and showed that it is equivalent to the fusion 2-category $2\rep(\CG)$ of 2-representations of the gauge 2-group. In \cite{HS23,HS23a,HS24}, the authors found many topological defects in the lattice model, but did not discuss the global categorical structure of topological defects.

We believe that the main difficulty of the above study of topological defects is that, although the crossed modules look very simple, its representation theory is complicated.
\bit
\item The notion of a crossed module can be defined without using the categorical language, but its modules (or 2-representations) are still categories. It is more natural to consider the 2-representation theory of weak 2-groups, not crossed modules.
\item Although crossed modules are equivalent to strict 2-groups, it is still not convenient to study the representation theory of strict 2-groups, because the size of a strict 2-group or a crossed module is usually too large. For example, for every group $G$ the identity map $G \to G$ defines a crossed module which is equivalent to the trivial 2-group. Given a 2-group $\CG$, the smallest 2-group equivalent to $\CG$ (that is, the skeletal one) is usually not strict.
\eit
As a concrete example, in this work we explicitly compute the quantum double or Drinfeld double $\CD(\CG)$ (as a Hopf monoidal category) of a finite 2-group $\CG$. This is very hard for a crossed module without using the categorical language.

\begin{rem}
Here we always consider the 2-representation theory of 2-groups using finite semisimple categories, that is, Kapranov-Voevodsky 2-vector spaces \cite{KV94}. There is another type of 2-vector spaces: the Baez-Crans 2-vector space \cite{BC04}. There are also some works on the 2-representation theory of crossed modules in Baez-Crans 2-vector spaces \cite{CG23,Che23}, including the computation of the Drinfeld double of a crossed module. However, those Drinfeld doubles are not categories in the usual sense. It is interesting to study the relation between these two types of 2-representation theory of 2-groups.
\end{rem}

In this work we study the gauge theory with a finite gauge weak 2-group $\CG$. It was also studied by Kapustin and Thorngren \cite{KT17a} at the level of topological actions. We give an explicit construction of the TQFT functor (including the partition function) and show that the quantum double model can be directly constructed from the TQFT functor. Then we show that the string-like topological defects in the 3+1D quantum double model for $\CG$ is equivalent to $\FZ_1(2\rep(\CG))$. The main method is to show that:
\bnu
\item In a lattice model realization of a topological order, the string-like topological defects are modules over the (multi-)fusion category of string-like local operators.
\item In the 3+1D quantum double model, the multi-fusion category of string-like local operators is the quantum double $\CD(\CG)$ of $\CG$.
\item By the definition of $\CD(\CG)$ we have $2\rep(\CD(\CG)) \simeq \FZ_1(2\rep(\CG))$ as braided fusion 2-categories.
\enu
The quantum double $\CD(\CG)$ is a Hopf monoidal category \cite{CF94,Neu97}. A Hopf monoidal category is a (linear) category equipped with some algebraic structures similar to a Hopf algebra: a tensor product functor (multiplication), a cotensor product functor (comultiplication), and an antipode. In addition, the axioms of a Hopf algebra are replaced by some natural isomorphisms (such as associator and coassociator for a Hopf monoidal category, and these natural isomorphisms are also the defining data. Similar to Hopf algebra, the modules over a Hopf monoidal category $\CC$ form a monoidal 2-category $2\rep(\CC)$ \cite{Neu97}. Moreover, when $\CC$ is finite semisimple and rigid, $2\rep(\CC)$ is a fusion 2-category \cite{DR18}. Conversely, the Tannaka-Krein duality says that for every fusion 2-category $\SC$ and a fiber 2-functor $\forget \colon \SC \to 2\vect$, there is a canonical Hopf monoidal structure on $\End(\forget)$ and $2\rep(\End(\forget)) \simeq \SC$ as fusion 2-categories \cite{Gre23}. We compute the Hopf monoidal structure of $\CD(\CG)$ in Section \ref{sec_quantum_double_construction}. The simple objects, the (co)tensor product on simple objects, and the (co)associator, are summarized in Theorem \ref{thm_fusion_rule_quantum_double}.

Our approach is a generalization of Kitaev's method of finding the particle-like (0+1D) topological defects in the 2+1D quantum double model. For a 2+1D quantum double model for a finite group $G$, Kitaev showed that \cite{Kit03}:
\bnu
\item The local operators on a site form the quantum double algebra $D(G)$.
\item These local operators commute with the Hamiltonian. So every excited space (eigenspace of the Hamiltonian) is invariant under the action of local operators, or equivalently, a module over $D(G)$.
\item Therefore, the particle-like topological defects form the category $\rep(D(G)) \simeq \FZ_1(\rep(G))$.
\enu
We believe that this idea can be further generalized. In an $(n+1)$D lattice model, for every $0 \leq k < n$, the local operators living on a $k$-dimensional disk should form a (multi-)fusion $k$-category. This $k$-category should describe a dimension-$k$ (generalized) symmetry. The $k$-dimensional topological defects should be invariant under the action of these local operators, so they are representations of the (generalized) symmetry. Mathematically, they should be described by the modules over the (multi-)fusion $k$-category of $k$-dimensional local operators.

\medskip
The organization of this work is as follows. In Section \ref{sec_finite_group_connection_gauge_theory}, we briefly review the finite group gauge theory, including the notion of flat connections and gauge transformations, and the construction of the partition function of Dijkgraaf-Witten theory. We generalize the notion of flat connections and gauge transformations to finite 2-group in Section \ref{sec_2-group_connection}, and also find a natural equivalence between gauge transformations. The partition function and the full TQFT functor of the finite 2-group gauge theory is constructed in Section \ref{sec_2-group_gauge_theory}. As a mathematical preliminary of the quantum double model, in Section \ref{sec_quantum_double_construction} we explicitly compute the Hopf monoidal structure of the quantum double $\CD(\CG)$ of a finite 2-group $\CG$. Then in Section \ref{sec_lattice_model_definition} we construct the 3+1D quantum double model for $\CG$ directly from the construction of the TQFT functor. In Section \ref{sec_string_operator_topological_defect_quantum_double}, we first explain why the string-like topological defects are modules over the (multi-)fusion category of string-like local operators, then find the string-like local operators in the 3+1D quantum double model and show that they form a multi-fusion category $\CD(\CG)$. Finally, in Section \ref{sec_example}, we find the string-like local operators in the 3+1D toric code model, which is a quantum double model with $\CG = \Zb_2$, and explicitly identify the modules over $\CD(\Zb_2)$ with the string-like topological defects.

\medskip
\noindent
\textbf{Acknowledgement}: The author would like to thank Chenjie Wang for his support and guidance throughout the research assistantship in his group. This work was supported by Research Grants Council of Hong Kong under GRF 17311322.

%\medskip
%\noindent
%\textbf{Data Availability Statement}: No datasets were generated or analysed during the current study.
%
%\medskip
%\noindent
%\textbf{Competing Interests}: The authors declare no competing interests, financial or non-financial, that could influence the objectivity of this research.

\section{Review of finite group gauge theory} \label{sec_finite_group_connection_gauge_theory}

Let $G$ be a finite group. In this section we briefly recall some basic concepts and constructions in $G$ gauge theory.

\subsection{Basic notions} \label{sec_finite_group_connection}

Suppose $M$ is a manifold equipped with a triangulation. We also give a total order on the set of 0-simplices of $M$, which is also called a \emph{branching structure}. This induces a local orientation on each $k$-simplex of $M$ for $k>0$. For a $k$-simplex $x$, we denote the vertices of $x$ by $x_0 < x_1 < \cdots < x_k$. For $0 \leq j_0 < j_1 < \ldots < j_r \leq k$, we also use $[x_{j_0},\ldots,x_{j_r}]$ to denote the $r$-simplex spanned by the vertices $x_{j_0},\ldots,x_{j_r}$. In particular, $x = [x_0,\ldots,x_k]$. The set of $k$-simplices of $M$ is denoted by $M_k$.

A \emph{$G$-connection} (also called a \emph{$G$-gauge field}) on $M$ is a function $\tau \colon M_1 \to G$ which assigns an element $\tau(e) \in G$ to each (oriented) 1-simplex $e \in M_1$. A $G$-connection $\tau$ is \emph{flat} if
\[
\tau([p_0,p_2]) = \tau([p_1,p_2]) \tau([p_0,p_1])
\]
for all 2-simplices $p \in M_2$.

Given a $G$-connection $\tau$, one can construct another $G$-connection as follows. Suppose $\phi \colon M_0 \to G$ is a function. We define a $G$-connection $\tau'$ by
\[
\tau'(e) \coloneqq \phi(e_1) \tau(e) \phi(e_0)^{-1} .
\]
We say that $\phi$ is a \emph{gauge transformation} from $\tau$ to $\tau'$ and write $\tau' = T_\phi \tau$. Also, $\tau$ is flat if and only if $\tau'$ is flat. Thus the gauge transformations define an equivalence relation on the set of flat $G$-connections on $M$.

%\begin{rem}
%Let $C^k(M;G)$ be the set of function from $M_k$ to $G$. Then $C^1(M;G)$ is the set of $G$-connections on $M$. Taking the curvature defines a (pointed) map $\mathrm d \colon C^1(M;G) \to C^2(M;G)$, and its kernel $Z^1(M;G)$ is the set of flat $G$-connections. The gauge transformations define not only a (pointed) map $\mathrm d \colon C^0(M;G) \to C^1(M;G)$, but also an equivalence relation on $Z^1(M;G)$. The quotient set $H^1(M;G) \coloneqq Z^1(M;G) / \sim$ (the set $C_G(M)$ of $G$-structures on $M$) is the (non-abelian) cohomology of $M$ with coefficients in $G$.
%\end{rem}

The flat $G$-connections and gauge transformations also have a geometric interpretation. Recall that the classifying space of $G$, denoted by $\lvert \mathrm B G \rvert$, is a CW complex and can be obtained as follows:
\bit
\item At the beginning there is only one 0-cell $\ast$.
\item In the first step, we attach a 1-cell $x_g^1$ for each $g \in G$.
\item In the second step, for every $g,h \in G$ we attach a 2-cell $x_{g,h}^2$ whose boundary is the concatenation of three paths $x_h^1$, $x_g^1$ and $\overline{x_{gh}^1}$, where the overline denotes the reverse of the path.
\item \ldots
\item In the $k$-th step, we attach a $k$-cell for every $k$-tuple $(g_1,\ldots,g_k) \in G^k$ by identifying the $j$-th boundary $[0,\ldots,\hat j,\ldots,k] \subset \partial \Delta^k \simeq \partial D^k = S^k$ to the $(k-1)$-cell $x_{g_1,\ldots,g_{j-1},g_j g_{j+1},g_{j+2},\ldots,g_k}^{k-1}$ for all $0 \leq j \leq k$, where $\hat j$ means deleting the item $j$ in the sequence.
\item \ldots
\eit
By the cellular approximation theorem, every continuous map $M \to \lvert \mathrm B G \rvert$ is homotopic to a cellular map (which maps $k$-cells to $k$-cells). Since every $k$-cell of $\lvert \mathrm B G \rvert$ is determined by its boundaries for $k \geq 2$, a cellular map $\tau \colon M \to \lvert \mathrm B G \rvert$ is determined by $\tau(e) \in G$ for all $e \in M_1$. Therefore, a cellular map $M \to \lvert \mathrm B G \rvert$ is equivalent to a $G$-connection. Also, by the definition of 2-cells in $\lvert \mathrm B G \rvert$, this $G$-connection must be flat.

\begin{figure}[htbp]
\centering
\begin{tikzpicture}
\coordinate (a) at (0,0,0) ;
\coordinate (b) at (2,0,0) ;
\coordinate (c) at (0.7,0,-1.5) ;
\coordinate (d) at (0,2,0) ;
\coordinate (e) at (2,2,0) ;
\coordinate (f) at (0.7,2,-1.5) ;
\node[left] at (a) {$x_0$} ;
\node[right] at (b) {$x_1$} ;
\node[right] at (c) {$x_2$} ;
\node[left] at (d) {$x_0'$} ;
\node[right] at (e) {$x_1'$} ;
\node[right] at (f) {$x_2'$} ;
\draw (a)--(b)--(e)--(a)--(d)--(f)--(e)--(d) ;
\draw[dashed] (f)--(a)--(c)--(f)--(b)--(c) ;
\end{tikzpicture}
\caption{The standard triangulation of $x \times I$ for a 2-simplex $x$. Here we write $x_j$ for $(x_j,0)$ and $x_j'$ for $(x_j,1)$.}
\label{fig_standard_triangulation_group}
\end{figure}
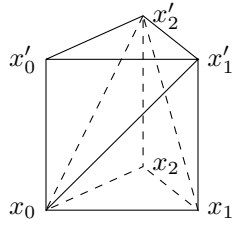

Given two cellular maps $\tau,\tau' \colon M \to \lvert \mathrm B G \rvert$, a homotopy between $\tau$ and $\tau'$ is a map $H \colon M \times I \to \lvert \mathrm B G \rvert$ such that $H(-,0) = \tau$ and $H(-,1) = \tau'$. Again we can assume that $H$ is a cellular map, where $M \times I$ is equipped with the following `standard' triangulation:
\bit
\item The subspaces $M \times \{0\}$ and $M \times \{1\}$ are equipped with the same triangulation with $M$.
\item If $x \in M_k$ is a $k$-simplex, $x \times I$ can be decompose as the union of $(k+1)$-simplex:
\[
x \times I = \bigcup_{j=0}^k [(x_0,0),\ldots,(x_j,0),(x_j,1),\ldots,(x_k,1)] .
\]
When $k = 2$, this decomposition is depicted in Figure \ref{fig_standard_triangulation_group}.
\eit

In particular, the set of 1-simplices of $M \times I$ is
\begin{multline*}
(M \times I)_1 = \{e \times \{0\} \mid e \in M_1\} \sqcup \{e \times \{1\} \mid e \in M_1\} \\
\sqcup \{[(v,0),(v,1)] \mid v \in M_0\} \sqcup \{[(e_0,0),(e_1,1)] \mid e \in M_1\} .
\end{multline*}
Thus the cellular map $H$ is determined by $H([(v,0),(v,1)]) \eqqcolon \phi(v) \in G$ for all $v \in M_0$. The flatness condition implies that
\begin{multline*}
\phi(e_1) \tau(e) = H([(e_1,0),(e_1,1)]) H(e \times \{0\}) = H([(e_0,0),(e_1,1)]) \\
= H(e \times \{1\}) H([(e_0,0),(e_0,1)]) = \tau'(e) \phi(e_0) .
\end{multline*}
Therefore, two cellular maps $\tau,\tau' \colon M \to \lvert \mathrm B G \rvert$ are homotopic if and only if there is a gauge transformation between two $G$-connections $\tau$ and $\tau'$.

We define a groupoid $\CC_G(M)$ as follows:
\bit
\item The objects are flat $G$-connections on $M$.
\item Given two $G$-connections $\tau,\tau'$ on $M$, a morphism $\phi \colon \tau \to \tau'$ is gauge transformation $\phi$ from $\tau$ to $\tau'$.
\item The composition of two gauge transformations $\phi \colon \tau \to \tau'$ and $\psi \colon \tau' \to \tau''$ is the pointwise multiplication $\psi \cdot \phi \colon \tau \to \tau''$.
\eit
On the other hand, we have the fundamental groupoid $\Pi_1(\Map(M,\lvert \mathrm B G \rvert))$ of the mapping space $\Map(M,\lvert \mathrm B G \rvert)$:
\bit
\item The objects are points in the mapping space $\Map(M,\lvert \mathrm B G \rvert)$, i.e., maps from $M$ to the classifying space $\lvert \mathrm B G \rvert$.
\item The morphisms are homotopy classes of paths in the mapping space $\Map(M,\lvert \mathrm B G \rvert)$, i.e., homotopy classes of homotopies between maps from $M$ to the classifying space $\lvert \mathrm B G \rvert$.
\item The composition of morphisms is given by the composition of homotopies.
\eit
Since flat $G$-connections and gauge transformations can be viewed as cellular maps and homotopies, there is a natural functor $\CC_G(M) \to \Pi_1(\Map(M,\lvert \mathrm B G \rvert))$.
The cellular approximation theorem implies that this functor is an equivalence.

%\begin{rem}
%We define a \emph{simple gauge transformation} as a gauge transformation $\phi$ which vanishes except at only one 0-simplex $v$. It only changes a $G$-connection $\tau$ on the edges adjacent to $v$:
%\[
%(T_\phi \tau)(e) = \begin{cases}
%\tau(e) \phi(v)^{-1} , & v = e_0 , \\
%\phi(v) \tau(e) , & v = e_1 , \\
%\tau(e) , & \text{otherwise} .
%\end{cases}
%\]
%Clearly every gauge transformation is the composition of simple gauge transformations.
%\end{rem}

\subsection{Dijkgraaf-Witten theory} \label{sec_finite_group_gauge_theory}

For every $n \geq 2$ and $\omega \in Z^n(G;U(1))$, we can define an $n$D oriented TQFT, called the Dijkgraaf-Witten theory \cite{DW90}. Here we give the construction of the partition functions.

Let $M$ be a $n$-dimensional triangulated oriented closed manifold. For every $G$-connection $\tau$ on $M$, define
\[
S(\tau) \coloneqq \prod_{x \in M_n} \omega(\tau([x_{n-1},x_n]),\ldots,\tau([x_0,x_1]))^{s_x} \in U(1) ,
\]
where $s_x = +1$ if the orientation of $x$ induced from $M$ is the same as that induced from the order of vertices, otherwise $s_x = -1$. Then we define the \emph{partition function}
\[
Z(M) \coloneqq \frac{1}{\lvert G \rvert^{\lvert M_0 \rvert}} \sum_{\tau \in \CC_G(M)} S(\tau) \in \Cb ,
\]
where the summation takes over all flat $G$-connections on $M$.

It is standard to show that $Z(M)$ is a topological invariant (i.e., independent of the triangulation on $M$) by proving that $Z(M)$ is invariant under Pachner moves. Let us briefly recall the notion of Pachner moves \cite{Pac87}. Let $\Delta_{n+1}$ be the $(n+1)$-simplex and $\partial_i \Delta_{n+1} \coloneqq [0,\ldots,\hat i,\ldots,n+1]$ be the $i$-th boundary $n$-simplex for $i = 0,\ldots,n+1$. For $1 \leq k \leq n+1$, the $k$-$(n+2-k)$-move is to replace a subcomplex of $M$ isomorphic to
\[
\partial_{<k} \Delta_{n+1} \coloneqq \bigcup_{0 \leq i \leq k-1} \partial_i \Delta_{n+1}
\]
by
\[
\partial_{\geq k} \Delta_{n+1} \coloneqq \bigcup_{k \leq i \leq n+1} \partial_i \Delta_{n+1} .
\]
In other words, a Pachner move on $M$ is to glue $M$ with an  $(n+1)$-disk $D^{n+1} \simeq \Delta_{n+1}$ along a subcomplex $K \subseteq M$ that is isomorphic to $\partial_{<k} \Delta_{n+1}$. The glued complex $M \cup_K \Delta_{n+1}$ has two boundaries: one is $M$, and the other one, denoted by $M' \coloneqq (M \cup_K \partial \Delta_{n+1}) \setminus K$, is the same as $M$ as a manifold but equipped with a different triangulation. Given a flat $G$-connection $\tau$ on $M$, it can be extended to a flat $G$-connection $\tilde \tau$ on $M \cup_K \Delta_{n+1}$ (not necessarily unique). Clearly $S(\tau)$ and $S(\tilde \tau \vert_{M'})$ are differed by $\omega(\tilde \tau \vert_{\partial \Delta_{n+1}}) = (\mathrm d \omega)(\tilde \tau \vert_{\Delta_{n+1}})$. Thus the cocycle condition of $\omega$ implies that $S$ is invariant under the Pachner moves. Furthermore, by counting the change of the number of flat $G$-connections and vertices after the Pachner moves, one can show that the partition function is a topological invariant.

This partition function $Z(M)$ can be viewed as a path integral (an weighted sum) of $S$ over the groupoid $\CC_G(M)$ of flat $G$-connections. Let us recall the definition of path integral over finite groupoids defined in \cite{FQ93}. Let $\CC$ be a finite groupoid (which has finitely many objects and morphisms) and $\beta$ be a function on $\CC$ which is constant on each isomorphism class of objects (called a \emph{locally constant function}), then we define the \emph{integral}
\[
\int_\CC \beta \coloneqq \sum_{[x]} \frac{\beta(x)}{\lvert \Aut(x) \rvert} ,
\]
where the summation takes over all isomorphism classes of objects $[x]$ in $\CC$. It is not hard to see that
\[
\int_\CC \beta = \sum_{x \in \CC} \frac{\beta(x)}{\lvert x \to \rvert} ,
\]
where $\lvert x \to \rvert$ is the number of morphisms whose sources are $x$. This integral is invariant under the equivalence of groupoids in the following sense. Suppose $F \colon \CC \to \CD$ is an equivalence of groupoids and $\beta$ is a locally constant function on $\CD$. Then $F^*(\beta)(x) \coloneqq \beta(F(x))$ defines a locally constant function $F^*(\beta)$ on $\CC$, and
\[
\int_\CC F^*(\beta) = \int_\CD \beta .
\]
Now we consider the finite groupoid $\CC_G(M)$ of flat $G$-connections on $M$. First we note that, $\omega$ can be viewed an $n$-cocycle on the classifying space $\lvert \mathrm B G \rvert$ (using the cellular cohomology). Then the $G$-connection $\tau$, as a cellular map from $M$ to $\lvert \mathrm B G \rvert$, pullbacks $\omega$ to an $n$-cocycle $\tau^*(\omega) \in Z^n(M;\bk^\times)$. Then clearly
\[
S(\tau) = \langle \tau^*(\omega),[M] \rangle ,
\]
where $[M] \in Z_n(M;\Zb)$ is the fundamental class of $M$. It follows immediately that $S(\tau)$ only depends on the cohomology class of $\omega$ and the homotopy class of $\tau$. Therefore, $S$ is a locally constant function on $\CC_G(M)$. We claim that
\[
Z(M) = \int_{\CC_G(M)} S .
\]
This is because a morphism in $\CC_G(M)$ with source $\tau$ is a gauge transformation from $\tau$ to another connection, and the number of such gauge transformations is clearly equal to $\lvert G \rvert^{\lvert M_0 \rvert}$.

\begin{rem}
Note that we have an equivalence of groupoids $\CC_G(M) \simeq \Pi_1(\Map(M,\lvert \mathrm B G \rvert))$. Then the partition function can be written as
\[
Z(M) = \int_{\Pi_1(\Map(M,\lvert \mathrm B G \rvert))} S .
\]
Since the right hand side does not involve the triangulation, the topological invariance of the partition function is obvious.
\end{rem}

\part{Finite 2-group gauge theory}

\section{Flat \texorpdfstring{$\CG$}{G}-connections on \texorpdfstring{$n$}{n}D manifold} \label{sec_2-group_connection}

\subsection{Finite 2-group and classifying space} \label{sec_2-group}

First we briefly review the basics of finite 2-groups. More details can be found in \cite{HZ23}.

A \emph{(weak) 2-group} $\CG$ is a monoidal category in which all objects and morphisms are invertible. Its \emph{first homotopy group} $\pi_1(\CG)$ is the group of isomorphism classes of objects in $\CG$, and its \emph{second homotopy group} $\pi_2(\CG)$ is the automorphism group $\CG(\one,\one)$ of the tensor unit $\one \in \CG$. We say $\CG$ is \emph{finite} if both $\pi_1(\CG)$ and $\pi_2(\CG)$ are finite. In the following, we denote $G \coloneqq \pi_1(\CG)$ and $A \coloneqq \pi_2(\CG)$ for simplicity.

Moreover, $G$ acts on $A$ by conjugation, and the associator of $\CG$ is a 3-cocycle $\alpha \in Z^3(G;A)$. The cohomology class $[\alpha] \in H^3(G;A)$ is called the \emph{Postnikov class} or the \emph{$k$-invariant}. The 2-group structure of $\CG$ is completely determined by $G$, $A$, the conjugation action of $G$ on $A$ and the Postnikov class $[\alpha]$. More precisely, the 2-group $\CG$ is equivalent to skeletal 2-group $\CG(G,A,\alpha)$ defined as follows:
\bit
\item The set of objects is the underlying set of $G$.
\item For $g,h \in G$, the hom space between them is defined by
\[
\Hom(g,h) \coloneqq \begin{cases} A , & g = h , \\ \emptyset , & g \neq h .\end{cases}
\]
\item The composition of morphisms is given by the multiplication of $A$.
\item The tensor product functor $\otimes \colon \CG(G,A,\alpha) \times \CG(G,A,\alpha) \to \CG(G,A,\alpha)$ is defined by
\[
\bigl( g \xrightarrow{a} g \bigr) \otimes \bigl( h \xrightarrow{b} h \bigr) \coloneqq \bigl( gh \xrightarrow{a (g \triangleright b)} gh \bigr) , \quad g,h \in G , \, a,b \in A ,
\]
where $\triangleright$ is the $G$-action on $A$.
\item The associator is defined by $\alpha_{g,h,k} = \alpha(g,h,k) \in A$.
\item The left and right unitors are defined by $\lambda_g = \alpha(e,e,g)^{-1}$ and $\rho_g = \alpha(g,e,e)$.
\eit
Note that the pentagon equation in $\CG(G,A,\alpha)$ is precisely the cocycle condition for $\alpha$:
\[
(g \triangleright \alpha(h,k,l)) \alpha(g,hk,l) \alpha(g,h,k) = \alpha(g,h,kl) \alpha(gh,k,l) .
\]
We can further assume that $\alpha$ is normalized, i.e., $\alpha(g,h,k) = 1$ if one of $g,h,k$ is the unit $e$. Then the left and right unitors are identities. In the rest of this section, we always assume that $\CG = \CG(G,A,\alpha)$ is a skeletal 2-group.

The \emph{classifying space} $\lvert \mathrm B \CG \rvert$ of $\CG$ can be defined as the geometric realization of the Duskin nerve of the one-point delooping $\mathrm B \CG$. So it is an infinite-dimensional CW complex.
\bit
\item There is only one 0-cell $\ast$.
\item For every $g \in G$, there is a 1-cell $x_g^1$.
\item For every $g,h \in G$ and $a \in A$, there is a 2-cell $x_{g,h;a}^2$ correspond to the morphism $a \colon g \otimes h \to gh$ in $\CG$.
\item There is a 3-cell for every collection $\{g_{ij} \in G\}_{0 \leq i < j \leq 3}$ and $\{a_{ijk} \in A\}_{0 \leq i < j < k \leq 3}$ such that the following diagram commutes in $\CG$:
\[
\xymatrix{
(g_{23} \otimes g_{12}) \otimes g_{01} \ar[rr]^-{\alpha(g_{23},g_{23},g_{01})} \ar[d]_{a_{123}} & & g_{23} \otimes (g_{12} \otimes g_{01}) \ar[d]^{g_{23} \triangleright a_{012}} \\
g_{13} \otimes g_{01} \ar[r]^-{a_{013}} & g_{03} & g_{23} \otimes g_{02} \ar[l]_-{a_{023}}
}
\]
The boundary 2-cells are the ones corresponding to $a_{ijk} \colon g_{jk} \otimes g_{ij} \to g_{ik}$ for $0 \leq i < j < k \leq 3$.
%This is equivalent to the equation
%\[
%\alpha(g_{23},g_{12},g_{01}) \cdot (g_{23} \triangleright a_{012}) a_{013}^{-1} a_{023} a_{123}^{-1} = 1 .
%\]
\item For $k \geq 3$, every $k$-cell is determined by its boundaries.
\eit

\subsection{Finite 2-group flat connections}

Suppose $M$ is an $n$-dimensional manifold and equipped with a triangulation and a total order on the vertices. We use the same notation as in Section \ref{sec_finite_group_connection}.

Let us consider continuous maps $\tau \colon M \to \lvert \mathrm B \CG \rvert$. By the cellular approximation theorem, we can assume that $\tau$ is a cellular map. Since the $k$-cells in $\lvert \mathrm B \CG \rvert$ are uniquely determined by their boundaries, $\tau$ is determined by $\tau(e)$ for all 1-simplices $e \in M_1$ and $\tau(p)$ for all 2-simplices $p \in M_2$. So we define a \emph{$\CG$-connection} on $M$ to be a pair $\tau = (\tau_1,\tau_2)$, where $\tau_1 \colon M_1 \to G$ and $\tau_2 \colon M_2 \to A$ are functions. The $\CG$-connections induced from cellular maps $M \to \lvert \mathrm B \CG \rvert$ are called \emph{flat}.

By the construction of $\lvert \mathrm B \CG \rvert$, we have a more precise definition of the flatness. We say that a $\CG$-connection $\tau$ is \emph{1-flat} if for every $p \in M_2$ we have
\be \label{eq_1_flat_connection}
\tau_1([p_0,p_2]) = \tau_1([p_1,p_2]) \tau_1([p_0,p_1]) ,
\ee
and it is \emph{2-flat} if for every $t \in M_3$ we have (the $\otimes$ symbols are omitted)
\be \label{eq_2_flat_connection}
\begin{array}{c}
\xymatrix@C=2.5em{
\bigl( \tau_1([t_2,t_3]) \tau_1([t_1,t_2]) \bigr) \tau_1([t_0,t_1]) \ar[rr]^-{\alpha(\tau_1([t_2,t_3]),\tau_1([t_1,t_2]),\tau_1([t_0,t_1]))} \ar[d]_{\tau_2([t_1,t_2,t_3])} & & \tau_1([t_2,t_3]) \bigl( \tau_1([t_1,t_2]) \tau_1([t_0,t_1]) \bigr) \ar[d]|{\tau_1([t_2,t_3]) \triangleright \tau_2([t_0,t_1,t_2])} \\
\tau_1([t_1,t_3]) \tau_1([t_0,t_1]) \ar[r]^-{\tau_2([t_0,t_1,t_3])} & \tau_1([t_0,t_3]) & \tau_1([t_2,t_3]) \tau_1([t_0,t_2]) \ar[l]_-{\tau_2([t_0,t_2,t_3])}
}
\end{array}
\ee
%\be \label{eq_2_flat_connection}
%\xymatrix{
%\bigl( \tau_1([t_2,t_3]) \otimes \tau_1([t_1,t_2]) \bigr) \otimes \tau_1([t_0,t_1]) \ar[rr]^-{\alpha(\tau_1([t_2,t_3]),\tau_1([t_1,t_2]),\tau_1([t_0,t_1]))} \ar[d]_{\tau_2([t_1,t_2,t_3])} & & \tau_1([t_2,t_3]) \otimes \bigl( \tau_1([t_1,t_2]) \otimes \tau_1([t_0,t_1]) \bigr) \ar[d]^{\tau_1([t_2,t_3]) \triangleright \tau_2([t_0,t_1,t_2])} \\
%\tau_1([t_1,t_3]) \otimes \tau_1([t_0,t_1]) \ar[r]^-{\tau_2([t_0,t_1,t_3])} & \tau_1([t_0,t_3]) & \tau_1([t_2,t_3]) \otimes \tau_1([t_0,t_2]) \ar[l]_-{\tau_2([t_0,t_2,t_3])}
%}
%\ee
Intuitively, the 1-flat condition means that every $2$-simplex is filled with a morphism in $\CG$. Then for some adjacent $2$-simplices, their corresponding morphisms can also be composed. The 2-flat condition simply says that, if we cut the boundary sphere of a $3$-simplex into two disks, then the morphisms on them are equal (up to an associator). A $\CG$-connection is \emph{flat} if it is both 1-flat and 2-flat. Then the flat $\CG$-connections on $M$ are equivalent to cellular maps from $M$ to $\lvert \mathrm B \CG \rvert$.

\begin{expl}
Suppose $A$ is trivial, i.e., $\CG = G$ is a finite 1-group. In this case the 1-flatness condition \eqref{eq_1_flat_connection} is the usual flatness condition for $G$ connections, and the 2-flatness condition \eqref{eq_2_flat_connection} is trivial.
\end{expl}

\begin{expl}
Suppose $G$ is trivial, i.e., $\CG = \mathrm B A$. In this case a $\mathrm B A$-connection is a 2-cochain $\tau \in C^2(M;A)$ in the simplicial cohomology. The 2-flatness condition \eqref{eq_2_flat_connection} is the 2-cocycle equation. In other words, a flat $\mathrm B A$-connection is a 2-cocycle $\tau \in Z^2(M;A)$ in the simplicial cohomology.
\end{expl}

\subsection{Gauge transformations of flat 2-group connections} \label{sec_gauge_transformation_2-group}

The \emph{gauge transformations} between flat $\CG$-connections are defined by the homotopies between cellular maps from $M$ to $\lvert \mathrm B \CG \rvert$. Given two cellular maps $\tau,\tau' \colon M \to \lvert \mathrm B \CG \rvert$, a homotopy between $\tau$ and $\tau'$ is a map $H \colon M \times I \coloneqq M \times [0,1] \to \lvert \mathrm B G \rvert$ such that $H(-,0) = \tau$ and $H(-,1) = \tau'$. Again we can equip $M \times I$ a triangulation and assume that $H$ is a cellular map. Then the homotopy $H$ can be expressed by some discrete data. The details of this approach can be found in Appendix \ref{appendix_gauge_transformation_triangulation}.

\begin{figure}[htbp]
\centering
\begin{tikzpicture}
\coordinate (a) at (0,0,0) ;
\coordinate (b) at (2,0,0) ;
\coordinate (c) at (0.7,0,-1.5) ;
\coordinate (d) at (0,2,0) ;
\coordinate (e) at (2,2,0) ;
\coordinate (f) at (0.7,2,-1.5) ;
\node[left] at (a) {$x_0$} ;
\node[right] at (b) {$x_1$} ;
\node[right] at (c) {$x_2$} ;
\node[left] at (d) {$x_0'$} ;
\node[right] at (e) {$x_1'$} ;
\node[right] at (f) {$x_2'$} ;
\draw (a)--(b)--(e) (a)--(d)--(f)--(e)--(d) ;
\draw[dashed] (a)--(c)--(f) (b)--(c) ;
\end{tikzpicture}
\caption{The product CW structure of $x \times I$ for a 2-simplex $x$. Here we write $x_j$ for $(x_j,0)$ and $x_j'$ for $(x_j,1)$.}
\label{fig_standard_product_CW_complex}
\end{figure}
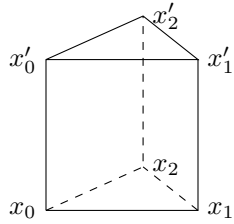

An equivalent but simpler approach is to equip $M \times I$ with the product CW complex structure, which is not necessarily a triangulation. More precisely, if $x$ is a $j$-cell ($j$-simplex) of $M$ and $y$ is a $k$-cell ($k$-simplex) of $I$, then $x \times y$ is a $(j+k)$-cell of $M \times I$, and $M \times I$ is the union of all the product cells. Here we take the simplest CW structure of $I$: it has one $1$-cell (the open interval) and two boundary $0$-cells. Then for every $k \geq 0$, the set of $(k+1)$-cells in $M \times I$ is
\[
\{y \times \{0\} \mid y \in M_{k+1}\} \sqcup \{y \times \{1\} \mid y \in M_{k+1}\} \sqcup \{z \times I \mid z \in M_k\} .
\]
Figure \ref{fig_standard_product_CW_complex} depicts the product CW structure of $x \times I$ for a 2-simplex $x$.

Then a cellular homotopy $H \colon M \times I \to \lvert \mathrm B \CG \rvert$ is determined by
\bit
\item $H(v \times I) \eqqcolon \phi_0(v) \in G$ for all $v \in M_0$;
\item $H(e \times I) \eqqcolon \phi_1(e) \in A$ for all $e \in M_1$.
\eit
Also the restriction of $H$ on $M \times \{0\}$ and $M \times \{1\}$ are $\tau$ and $\tau'$, respectively, by the definition of homotopy. The 1-flat condition for $H$ is
\begin{equation}  \label{eq_1-flat_gauge_transformation}
\phi_0(e_1) \tau_1(e) = \tau_1'(e) \phi_0(e_0)
\end{equation}
for every $e \in M_1$, because only in this case $\phi_1(e)$ can be viewed as a morphism $\phi_0(e_1) \tau_1(e) \to \tau_1'(e) \phi_0(e_0)$. Then the 2-flat condition means that the following diagram in $\CG$ commutes for every $p \in M_2$ (the $\otimes$ symbols are omitted):
\[
\xymatrix@C=5em{
\bigl( \phi_0(p_2) \tau_1([p_1,p_2]) \bigr) \tau_1([p_0,p_1]) \ar[r]^-{\phi_1([p_1,p_2])} \ar[d]_{\alpha} & \bigl( \tau_1'([p_1,p_2]) \phi_0(p_1) \bigr) \tau_1([p_0,p_1]) \ar[d]^{\alpha} \\
\phi_0(p_2) \bigl( \tau_1([p_1,p_2]) \tau_1([p_0,p_1]) \bigr) \ar[d]_{\phi_0(p_2) \triangleright \tau_2(p)} & \tau_1'([p_1,p_2]) \bigl( \phi_0(p_1) \tau_1([p_0,p_1]) \bigr) \ar[d]^{\tau_1' \triangleright \phi_1([p_0,p_1])} \\
\phi_0(p_2) \tau_1([p_0,p_2]) \ar[d]_{\phi_1([p_0,p_2])} & \tau_1'([p_1,p_2]) \bigl( \tau_1'([p_0,p_1]) \phi_0(p_1) \bigr) \ar[d]^{\alpha^{-1}} \\
\tau_1'([p_0,p_2]) \phi_0(p_0) & \bigl( \tau_1'([p_1,p_2]) \tau_1'([p_0,p_1]) \bigr) \phi_0(p_1) \ar[l]_-{\tau_2'(p)}
}
\]
%\[
%\xymatrix{
%\bigl( \phi_0(p_2) \tau_1([p_1,p_2]) \bigr) \tau_1([p_0,p_1]) \ar[r]^-{\phi_1([p_1,p_2])} \ar[d]_{\alpha} & \bigl( \tau_1'([p_1,p_2]) \phi_0(p_1) \bigr) \tau_1([p_0,p_1]) \ar[r]^-{\alpha} & \tau_1'([p_1,p_2]) \bigl( \phi_0(p_1) \tau_1([p_0,p_1]) \bigr) \ar[d]^{\tau_1' \triangleright \phi_1([p_0,p_1])} \\
%\phi_0(p_2) \bigl( \tau_1([p_1,p_2]) \tau_1([p_0,p_1]) \bigr) \ar[d]_{\phi_0(p_2) \triangleright \tau_2(p)} & & \tau_1'([p_1,p_2]) \bigl( \tau_1'([p_0,p_1]) \phi_0(p_1) \bigr) \ar[d]^{\alpha^{-1}} \\
%\phi_0(p_2) \tau_1([p_0,p_2]) \ar[r]^-{\phi_1([p_0,p_2])} & \tau_1'([p_0,p_2]) \phi_0(p_0) & \bigl( \tau_1'([p_1,p_2]) \tau_1'([p_0,p_1]) \bigr) \phi_0(p_1) \ar[l]_-{\tau_2'(p)}
%}
%\]
In other words, the 2-flat condition is the following equation for every $p \in M_2$:
\begin{multline} \label{eq_2-flat_gauge_transformation}
\tau_2'(p) (\tau_1'([p_1,p_2]) \triangleright \phi_1([p_0,p_1])) \phi_1([p_1,p_2]) = (\phi_0(p_2) \triangleright \tau_2(p)) \phi_1([p_0,p_2]) \\
\cdot \frac{\alpha(\tau_1'([p_1,p_2]),\tau_1'([p_0,p_1]),\phi_0(p_0)) \cdot \alpha(\phi_0(p_2),\tau_1([p_1,p_2]),\tau_1([p_0,p_1]))}{\alpha(\tau_1'([p_1,p_2]),\phi_0(p_1),\tau_1([p_0,p_1]))} .
\end{multline}
So we define a \emph{gauge transformation} between two $\CG$-connections $\tau$ and $\tau'$ as a pair $\phi = (\phi_0,\phi_1)$, where $\phi_0 \colon M_0 \to G$ and $\phi_1 \colon M_1 \to A$ are functions such that the equations \eqref{eq_1-flat_gauge_transformation} and \eqref{eq_2-flat_gauge_transformation} hold. Such a gauge transformation is denoted by $\phi \colon \tau \to \tau'$ or $T_\phi \tau = \tau'$.

\subsection{Equivalence of gauge transformations and the 2-groupoid \texorpdfstring{$\CC_\CG(M)$}{CG(M)}}

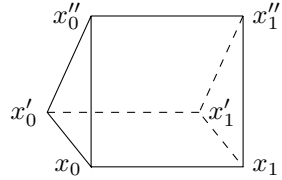
\begin{figure}[htbp]
\centering
\begin{tikzpicture}[rotate=90,xscale=-1]
\coordinate (a) at (0,0,0) ;
\coordinate (b) at (2,0,0) ;
\coordinate (c) at (0.7,0,-1.5) ;
\coordinate (d) at (0,2,0) ;
\coordinate (e) at (2,2,0) ;
\coordinate (f) at (0.7,2,-1.5) ;
\node[right] at (a) {$x_1''$} ;
\node[right] at (b) {$x_1$} ;
\node[right] at (c) {$x_1'$} ;
\node[left] at (d) {$x_0''$} ;
\node[left] at (e) {$x_0$} ;
\node[left] at (f) {$x_0'$} ;
\draw (a)--(b)--(e) (a)--(d)--(f)--(e)--(d) ;
\draw[dashed] (c)--(f) (a)--(c)--(b) ;
\end{tikzpicture}
\caption{The composition of two cylinders $[x_0,x_1] \times I$. The 2-simplices $[x_0,x_0',x_0'']$ and $[x_1,x_1',x_1'']$ are labeled by the unit $1 \in A$.}
\label{fig_composition_cylinder}
\end{figure}

Given two gauge transformation $\psi \colon \tau \to \tau'$ and $\phi \colon \tau' \to \tau''$, we define their \emph{composition} to be the composition of the corresponding homotopies. Geometrically, the composition of two gauge transformations is given by gluing two cylinders $M \times I$. Figure \ref{fig_composition_cylinder} depicts such a composition for a 1-simplex $x = [x_0,x_1]$ in $M$.

To compute the composition of $\phi$ and $\psi$, we label the two squares on the back by $\phi$ and $\psi$ respectively, and label the two triangles by the unit $1 \in A$. Then the labels on the front square can be computed by the 2-flat condition. It follows that the composite gauge transformation $\phi\psi \colon \tau \to \tau''$ is defined by
\be \label{eq_composition_gauge_transformation_0}
(\phi\psi)_0(v) \coloneqq \phi_0(v) \psi_0(v) , \quad \forall v \in M_0 ,
\ee
and
\be \label{eq_composition_gauge_transformation_1}
(\phi\psi)_1(e) \coloneqq \phi_1(e) (\phi_0(e_1) \triangleright \psi_1(e)) \cdot \frac{\alpha(\tau_1''(e),\phi_0(e_0),\psi_0(e_0)) \alpha(\phi_0(e_1),\psi_0(e_1),\tau_1(e))}{\alpha(\phi_0(e_1),\tau_1'(e),\psi_0(e_0))} .
\ee
Note that when $\alpha$ is nontrivial, the composition $\phi \psi$ depends on the domain $\tau$ of the gauge transformation. Moreover, one can verify that the composition of gauge transformations is not strictly associative. This suggests that there should be higher morphisms between gauge transformations.

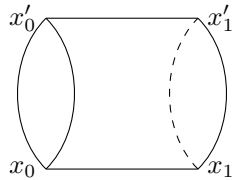
\begin{figure}[htbp]
\centering
\begin{tikzpicture}[rotate=90]
\draw (0,0) node[right] {$x_1$}--(0,2) node[left] {$x_0$} ;
\draw (2,0) node[right] {$x_1'$}--(2,2) node[left] {$x_0'$} ;
\draw (0,0) .. controls (0.5,-0.5) and (1.5,-0.5) .. (2,0) ;
\draw (0,2) .. controls (0.5,1.5) and (1.5,1.5) .. (2,2) ;
\draw[dashed] (0,0) .. controls (0.5,0.5) and (1.5,0.5) .. (2,0) ;
\draw (0,2) .. controls (0.5,2.5) and (1.5,2.5) .. (2,2) ;
%\draw (0,0) .. controls (0.5,0) and (1.5,1) .. (2,2) ;
%\draw[dashed] (0,0) .. controls (0.5,1) and (1.5,2) .. (2,2) ;
\end{tikzpicture}
\caption{The equivalence between two gauge transformations.}
\label{fig_2-morphism}
\end{figure}

These higher morphisms should be the homotopies between homotopies, i.e., maps from $M \times I \times I \to \lvert \mathrm B \CG \rvert$, and such a higher homotopy should be relative to $M \times \{0,1\}$. Thus it reduces to a map from the quotient $M \times I \times I / \sim$ to $\lvert \mathrm B \CG \rvert$, where the equivalence relation $\sim$ is defined by $(m,0,t) \sim (m,0,t')$ and $(m,1,t) \sim (m,1,t')$ for all $m \in M$ and $t,t' \in [0,1]$. Figure \ref{fig_2-morphism} depicts such a quotient space for a 1-simplex $x = [x_0,x_1]$ in $M$. Therefore, such a higher homotopy (again assumed to be a cellular map) is determined by the elements attached to the digons with vertices $\{x_0,x_0'\}$ and $\{x_1,x_1'\}$.

Hence we define an \emph{equivalence} between gauge transformations $\phi,\psi \colon \tau \to \tau'$ to be a map $\xi \colon M_0 \to A$ such that $\phi_0 = \psi_0$ and
\begin{equation} \label{eq_equivalence_gauge_transformation}
\phi_1(e) (\tau_1'(e) \triangleright \xi(e_0)) = \psi_1(e) \xi(e_1) , \quad \forall e \in M_1 .
\end{equation}
We denote such an equivalence by $\xi \colon \phi \Rightarrow \psi$.

\begin{expl} \label{expl_associator_gauge_transformation}
Let $\varphi \colon \tau \to \tau' , \psi \colon \tau' \to \tau'' , \phi \colon \tau'' \to \tau'''$ be gauge transformations. Then there is an equivalence
\[
\xi \colon (\phi \psi) \varphi \Rightarrow \phi (\psi \varphi)
\]
defined by
\[
\xi(v) \coloneqq \alpha(\phi_0(v),\psi_0(v),\varphi_0(v)) , \quad v \in M_0 .
\]
This equivalence plays the role of the associator for the composition of gauge transformations.
\end{expl}

Then we obtain a 2-groupoid, denoted by $\CC_\CG(M)$:
\bit
\item the objects are flat $\CG$-connections on $M$;
\item the 1-morphisms are gauge transformations;
\item the 2-morphisms are equivalences between gauge transformations.
\eit
By the cellular approximation theorem, this 2-groupoid is equivalent to the fundamental 2-groupoid $\Pi_2(\Map(M,\lvert \mathrm B \CG \rvert))$ of the mapping space $\Map(M,\lvert \mathrm B \CG \rvert)$. %In particular, the set $\pi_0(\Map(M,\lvert \mathrm B \CG \rvert))$ of homotopy classes of maps from $M$ to $\lvert \mathrm B \CG \rvert$ is isomorphic to the set of $\CG$-connections modulo the gauge transformations.

\begin{rem}
It is natural to consider the higher homotopies. However, they are essentially trivial because $\CG$ is only a 2-group. For example, if we want to define morphisms between equivalences between gauge transformations, it turns out that such a morphism must be unique (if exists). Equivalently, this suggests that the $k$-th homotopy groups of the mapping space $\Map(M,\lvert \mathrm B \CG \rvert)$ vanish for all $k \geq 3$.
\end{rem}

\begin{expl}
Suppose $A$ is trivial, i.e., $\CG = G$ is a finite 1-group. In this case the equivalences between gauge transformations are all trivial. Thus $\CC_G(M)$ is indeed a 1-groupoid, which is equivalent to the fundamental groupoid $\Pi_1(\Map(M,\lvert \mathrm B G \rvert))$.
\end{expl}

\begin{expl}
Suppose $G$ is trivial, i.e., $\CG = \mathrm B A$. In this case a gauge transformation $\phi \colon \tau \to \tau'$ is a 1-cochain $\phi \in C^1(M;A)$ in the simplicial cohomology such that $\tau = \tau' (\mathrm d \phi)$. An equivalence $\xi \colon \phi \Rightarrow \psi$ between two gauge transformations is a 0-cochain $\xi \in C^0(M;A)$ in the simplicial cohomology such that $\psi = \phi (\mathrm d \xi)$.
\end{expl}

\begin{expl} \label{expl_simple_gauge_transformation}
We define a \emph{simple 1-gauge transformation} as a gauge transformation $\phi$ such that $\phi_1$ vanishes and $\phi_0$ vanishes except at only one 0-simplex $v$. Then we have
\[
(T_\phi \tau)_1(e) = \begin{cases}
\tau_1(e) \phi_0(v)^{-1} , & v = e_0 , \\
\phi_0(v) \tau_1(e) , & v = e_1 , \\
\tau_1(e) , & \text{otherwise} .
\end{cases}
\]
and
\[
(T_\phi \tau)_2(p) = \begin{cases}
\tau_2(p) \alpha(\tau_1([p_1,p_2]),\tau_1([v,p_1]) \phi_0(v)^{-1},\phi_0(v)) , & v = p_0 , \\
\tau_2(p) \alpha(\tau_1([v,p_2]) \phi_0(v)^{-1},\phi_0(v),\tau_1([p_0,v])) , & v = p_1 , \\
(\phi_0(v) \triangleright \tau_2(p)) \alpha(\phi_0(v),\tau_1([p_1,v]),\tau_1([p_0,p_1])) , & v = p_2 , \\
\tau_2(p) , & \text{otherwise} .
\end{cases}
\]
Such a simple 1-gauge transformation is also denoted by $\phi_{g,v}$ where $g = \phi_0(v)$.

Similarly, we define a \emph{simple 2-gauge transformation} as a gauge transformation $\phi$ such that $\phi_0$ vanishes and $\phi_1$ except at only one 1-simple $e$. Then we have $(T_\phi \tau)_1 = \tau_1$ and
\[
(T_\phi \tau)_2(p) = \begin{cases}
\tau_2(p) (\tau_1([p_1,p_2]) \triangleright \phi_1(e))^{-1} , & e = [p_0,p_1] , \\
\tau_2(p) \phi_1(e)^{-1} , & e = [p_1,p_2] , \\
\tau_2(p) \phi_1(e) , & e = [p_0,p_2] , \\
\tau_2(p) , & \text{otherwise} .
\end{cases}
\]
Here we assume that the 3-cocycle $\alpha$ is normalized. If $\alpha$ is not normalized, the expression of $(T_\phi \tau)_2$ will be involved with $\alpha$ and more complicated. Such a simple 2-gauge transformation is also denoted by $\phi_{a,e}$ where $a = \phi_1(e)$.

It is clear that every gauge transformation is equivalent to the composition of simple 1-gauge transformations and simple 2-gauge transformations.
\end{expl}

%\begin{expl} \label{expl_associator_simple_gauge_transformation}
%Give $g,h,k \in G$ and a 0-simplex $v$, the associator
%\[
%\xi \colon (\phi_{g,v} \circ \phi_{h,v}) \circ \phi_{k,v} \Rightarrow  \phi_{g,v} \circ (\phi_{h,v} \circ \phi_{k,v})
%\]
%is given by $\xi_0(v) = \alpha(g,h,k)$ and $\xi_0(w) = 0$ for all $w \neq v$.
%\end{expl}

\section{Finite 2-group gauge theory} \label{sec_2-group_gauge_theory}

For every $n \geq 3$ and $\omega \in Z^n(\mathrm B \CG;U(1))$, we define the ($\omega$-twisted) $\CG$ gauge theory, which is an $n$D oriented TQFT and generalizes the Dijkgraaf-Witten theory for finite groups \cite{DW90}. When $\omega = 1$, this TQFT should be equivalent to Yetter's model \cite{Yet93}, which is a 2-group gauge theory defined in the language of crossed module. This TQFT was also studied by Kapustin and Thorngren \cite{KT17a} at the level of topological actions.

\subsection{Partition function} \label{sec_2-group_gauge_theory_partition_function}

In this subsection, we construct the partition function of the (twisted) $\CG$ gauge theory. The explicit construction of the TQFT functor is given in the next subsection.

Let $M$ be an $n$-dimensional triangulated oriented closed manifold. For every $\CG$-connection $\tau$, we define
\[
S(\tau) \coloneqq \prod_{x \in M_n} \omega(\tau(x))^{s_x} \in U(1) ,
\]
where $s_x = +1$ if the orientation of $x$ induced from $M$ is the same as that induced from the order of vertices, otherwise $s_x = -1$. By viewing $\omega$ as an $n$-cocycle on the classifying space $\lvert \mathrm B \CG \rvert$, we have
\[
S(\tau) = \langle \tau^* \omega , [M] \rangle \eqqcolon \int_M \tau^* \omega ,
\]
where $[M] \in Z^n(M;\Zb)$ is the fundamental class of $M$. It follows that $S(\tau)$ only depends on the homotopy class of $\tau$ (i.e., it is a gauge invariant) and the cohomology class of $\omega$. The partition function on $M$ is defined by
\[
Z(M) \coloneqq \frac{1}{\lvert G \rvert^{\lvert M_0 \rvert}} \frac{1}{\lvert A \rvert^{\lvert M_1 \rvert - \lvert M_0 \rvert}} \sum_{\tau \in \CC_\CG(M)} S(\tau) \in \Cb ,
\]
where the summation takes over all flat $\CG$-connections on $M$.

In the following we verify that $Z(M)$ is independent of the triangulation of $M$, by showing that it is invariant under the Pachner moves.

As in Section \ref{sec_finite_group_gauge_theory}, a Pachner move on $M$ is to glue $M$ with $\Delta_{n+1}$ along a subcomplex $K \subseteq M$ that is isomorphic to $\partial_{<k} \Delta_{n+1}$, and the new triangulation is denoted by $M'$. Any flat $\CG$-connection $\tau$ on $M$ can be extended to a flat $\CG$-connection $\tilde \tau$ on $M \cup_K \Delta_{n+1}$ (not necessarily unique). Again, $S(\tau)$ and $S(\tilde \tau \vert_{M'})$ are differed by $\omega(\tilde \tau \vert_{\partial \Delta_{n+1}}) = (\mathrm d \omega)(\tilde \tau \vert_{\Delta_{n+1}})$. Then the cocycle condition of $\omega$ implies that $S$ is invariant under the Pachner moves.

Now we count the change of the number of flat $\CG$-connections after the Pachner moves. Recall that the dimension of $M$ is $n \geq 3$.
\bit
\item After a $1$-$(n+1)$-move, there is one additional vertex $0$ and $(n+1)$ additional edges $[0,i]$ for $1 \leq i \leq n+1$. On the other hand, to extend $\tau$ to $\tilde \tau$, it suffices to determine $\tilde \tau_1([0,1]) \in G$ and $\tilde \tau_2([0,1,i]) \in A$ for $2 \leq i \leq n+1$, because the other elements $\tilde \tau_1([0,i])$ and $\tilde \tau_2([0,i,j])$ for $1<i<j \leq n+1$ can be determined by the flatness conditions on $[0,1,i,j]$. Therefore, every flat $\CG$-connection $\tau$ has $\lvert G \rvert \cdot \lvert A \rvert^n = \lvert G \rvert \cdot \lvert A \rvert^{(n+1)-1}$ different extensions $\tilde \tau$. Thus we have
\begin{multline*}
Z(M') = \frac{1}{\lvert G \rvert^{\lvert M_0 \rvert +1}} \frac{1}{\lvert A \rvert^{\lvert M_1 \rvert + (n+1) - \lvert M_0 \rvert - 1}} \sum_{\tilde \tau} S(\tilde \tau \vert_{M \cup_K \partial \Delta_{n+1}}) \\
= \frac{1}{\lvert G \rvert^{\lvert M_0 \rvert}} \frac{1}{\lvert A \rvert^{\lvert M_1 \rvert - \lvert M_0 \rvert}} \sum_{\tau \in \CC_\CG(M)} S(\tau) = Z(M) .
\end{multline*}
\item After a $2$-$n$-move, there is no additional vertex and one additional edge $[0,1]$. On the other hand, to extend $\tau$ to $\tilde \tau$, it suffices to determine $\tilde \tau_2([0,1,2])$, because the other elements $\tilde \tau_1([0,1])$ and $\tilde \tau_2([0,1,i])$ for $2<i \leq n+1$ can be determined by the flatness conditions on $[0,1,2,i]$. Therefore, every flat $\CG$-connection $\tau$ has $\lvert A \rvert$ different extensions $\tilde \tau$. Thus we have
\begin{multline*}
Z(M') = \frac{1}{\lvert G \rvert^{\lvert M_0 \rvert}} \frac{1}{\lvert A \rvert^{\lvert M_1 \rvert + 1 - \lvert M_0 \rvert}} \sum_{\tilde \tau} S(\tilde \tau \vert_{M \cup_K \partial \Delta_{n+1}}) \\
= \frac{1}{\lvert G \rvert^{\lvert M_0 \rvert}} \frac{1}{\lvert A \rvert^{\lvert M_1 \rvert - \lvert M_0 \rvert}} \sum_{\tau \in \CC_\CG(M)} S(\tau) = Z(M) .
\end{multline*}
\item For $3 \leq k \leq (n+2)/2$, a $k$-$(n+2-k)$-move does not add vertex nor edge. Also, the extension of $\tau$ to $\tilde \tau$ is unique. Then it is clear that $Z(M') = Z(M)$.
\item For $(n+2)/2 < k \leq n+1$, a $k$-$(n+2-k)$-move is the inverse of a $(n+2-k)$-$k$-move. Then the invariance of $Z$ reduces to the above discussions.
\eit
Hence, we have proved that the partition function $Z(M)$ is independent of the triangulation of $M$, thus a topological invariant.

\begin{expl}
Let us compute the partition function $Z(S^n)$ for $n \geq 3$. There is an obvious triangulation that $S^n \simeq \partial \Delta_{n+1}$. Since $\pi_n(\lvert \mathrm B \CG \rvert) = 0$ for $n \geq 3$, every flat $\CG$-connection $\tau$ on $S^n \simeq \partial \Delta_{n+1}$ can be extended to a flat $\CG$-connection $\tilde \tau$ on $D^{n+1} \simeq \Delta_{n+1}$. By the cocycle condition of $\omega$ we have
\[
S(\tau) = \langle \tau^* \omega,[S^n] \rangle = \langle \tilde \tau^* \omega,\partial [D^{n+1}] \rangle = \langle \mathrm d (\tilde \tau^* \omega),[D^{n+1}] \rangle = \langle \tilde \tau^*  (\mathrm d \omega),[D^{n+1}] \rangle = 1 .
\]
Then we only need to count the number of flat $\CG$-connections on $\partial \Delta_{n+1}$. There are $(n+2)$ vertices and $(n+2)(n+1)/2$ edges. To determine a flat $\CG$-connection $\tau$ on $\partial \Delta_{n+1}$, it suffices to determine $\tau_1([0,i]) \in G$ and $\tau_2([0,i,j]) \in A$ for $1 \leq i < j \leq n+1$. So the number of flat $\CG$-connections on $\partial \Delta_{n+1}$ is $\lvert G \rvert^{n+1} \cdot \lvert A \rvert^{(n+1)n/2}$. Hence
\be \label{eq_parition_function_sphere}
Z(S^n) = Z(\partial \Delta_{n+1}) = \frac{1}{\lvert G \rvert^{n+1}} \frac{1}{\lvert A \rvert^{(n+2)(n+1)/2 - (n+2)}} \lvert G \rvert^{n+1} \cdot \lvert A \rvert^{(n+1)n/2} = \frac{\lvert A \rvert}{\lvert G \rvert} .
\ee
Note that this is also independent of the choice of $\omega$.
\end{expl}

Similar to the construction in Section \ref{sec_finite_group_gauge_theory}, the partition function $Z(M)$ should be defined as a path integral (an weighted sum) of $S$ over the 2-groupoid $\CC_\CG(M)$ of flat $\CG$-connections. Suppose $\SC$ is a finite 2-groupoid. Its \emph{1-truncation} $\Pi_1(\SC)$ is a 1-groupoid whose objects are the same as $\SC$ and morphisms are the isomorphisms classes of 1-morphisms of $\SC$. The notion of \emph{locally constant functions} and \emph{integral} on $\SC$ are defined by those on the 1-truncation $\Pi_1(\SC)$. Let us compute the integral
\[
\int_{\CC_\CG(M)} S .
\]
By definition, this integral is equal to
\[
\sum_{\tau \in \CC_\CG(M)} \frac{S(\tau)}{\lvert \tau \rightarrow \rvert} ,
\]
where $\lvert \tau \rightarrow \rvert$ is the number of equivalence classes of gauge transformations with source $\tau$. A gauge transformation $\phi$ is determined by the maps $\phi_0 \colon M_0 \to G$ and $\phi_1 \colon M_1 \to A$, thus the set of all gauge transformations with source $\tau$ is isomorphic to $G^{\times M_0} \times A^{\times M_1}$. Similarly, we have a group $A^{\times M_0}$ of equivalences between gauge transformations, which acts on the set $G^{\times M_0} \times A^{\times M_1}$ by \eqref{eq_equivalence_gauge_transformation}, and the number of orbits is equal to $\lvert \tau \rightarrow \rvert$. To compute the number of orbits, we need to determine the stabilizers. By \eqref{eq_equivalence_gauge_transformation}, the stabilizer of a gauge transformation $\phi$ consists of the equivalences $\xi$ satisfying
\[
\tau'_1(e) \triangleright \xi(e_0) = \xi(e_1) , \quad \forall e \in M_1 .
\]
Once the value of $\xi$ on a 0-simplex is fixed, then the values on all 0-simplices in the same path-connected component of $M$ are determined by the above equation. Therefore, the stabilizer of  every gauge transformation is isomorphic to $A^{\times \pi_0(M)}$. It follows that every orbit has $\lvert A \rvert^{\lvert M_0 \rvert - \lvert \pi_0(M) \rvert}$ elements. Hence
\[
\lvert \tau \rightarrow \rvert = \lvert G \rvert^{\lvert M_0 \rvert} \cdot \lvert A \rvert^{\lvert M_1 \rvert - \lvert M_0 \rvert + \lvert \pi_0(M) \rvert} .
\]
So the partition function can be written as the integral of the action $S$ over the 2-groupoid $\CC_\CG(M)$:
\[
Z(M) = \lvert A \rvert^{\lvert \pi_0(M) \rvert} \cdot \int_{\CC_\CG(M)} S .
\]
Under the equivalence of 2-groupoids $\CC_\CG(M) \simeq \Pi_2(\Map(M,\lvert \mathrm B \CG \rvert))$, we have
\[
Z(M) = \lvert A \rvert^{\lvert \pi_0(M) \rvert} \cdot \int_{\Pi_2(\Map(M,\lvert \mathrm B \CG \rvert))} S .
\]
Obviously, this also implies the topological invariance of the partition function.

\subsection{\texorpdfstring{$n$}{n}D TQFT functor} \label{sec_TQFT_functor}

Now we construct the full TQFT functor of the (twisted) $\CG$ gauge theory. There is a standard way to define TQFTs from triangulations \cite{Yet93a} in two steps:
\bnu
\item Construct an $n$D \emph{lattice TQFT} $\tilde Z$. A lattice TQFT maps each triangulated $(n-1)$D closed manifold $\Sigma$ to a vector space $\tilde Z(\Sigma)$, and maps each triangulated $n$D cobordism $M \colon \Sigma_0 \to \Sigma_1$ to a linear map $\tilde Z(M) \colon \tilde Z(\Sigma_0) \to \tilde Z(\Sigma_1)$. Moreover, it has the following properties:
\bit
\item It is independent of the interior triangulation of $n$D cobordisms. However, it may depend on the triangulation of $(n-1)$D manifolds in general.
\item It preserves the composition: $\tilde Z(N \circ M) = \tilde Z(N) \circ \tilde Z(M)$ for composable $n$D cobordisms $M,N$. In particular, $\tilde Z(\Sigma \times I)$ is an idempotent on $\tilde Z(\Sigma)$ for each $(n-1)$D closed manifold $\Sigma$.
\item It preserves the tensor product: $\tilde Z(M \sqcup N) \simeq \tilde Z(M) \otimes \tilde Z(N)$ for $n$D cobordisms $M,N$.
\eit
\item Construct an $n$D TQFT $Z$ from $\tilde Z$ by taking the colimit over all triangulations on each manifold. There is a standard argument that every lattice TQFT $\tilde Z$ satisfying the above properties can produce a TQFT $Z$, without knowing the details of $\tilde Z$.
\enu
In the following we first give the lattice TQFT $\tilde Z$ of the 2-group gauge theory, then recall the standard construction from $\tilde Z$ to $Z$.

\medskip
For an $(n-1)$-dimensional triangulated oriented closed manifold $\Sigma$, we define $\tilde Z(\Sigma)$ to be the vector space spanned by all (not necessarily flat) $\CG$-connections on $\Sigma$. Thus
\[
\dim \tilde Z(\Sigma) = \lvert G \rvert^{\lvert \Sigma_1 \rvert} \lvert A \rvert^{\lvert \Sigma_2 \rvert} .
\]
For an $n$-dimensional triangulated oriented cobordism $M \colon \Sigma_0 \to \Sigma_1$, the linear map $\tilde Z(M) \colon \tilde Z(\Sigma_0) \to \tilde Z(\Sigma_1)$ is defined as follows. Given flat $\CG$-connections $\gamma_i$ on $\Sigma_i$ for $i = 0,1$, define
\[
\tilde Z(M;\gamma_0,\gamma_1) \coloneqq \frac{1}{\lvert G \rvert^{\lvert M_0 \rvert - \lvert (\partial M)_0 \rvert/2}} \frac{1}{\lvert A \rvert^{\lvert M_1 \rvert - \lvert M_0 \rvert - (\lvert (\partial M)_1 \rvert - \lvert (\partial M)_0 \rvert)/2}} \sum_{\substack{\tau \in \CC_\CG(M) \\ \tau \vert_{\Sigma_i} = \gamma_i}} S(\tau) ,
\]
where $\partial M = \Sigma_0 \sqcup \Sigma_1$ is the boundary of $M$. Then we define
\[
\tilde Z(M) \coloneqq \sum_{\gamma_0 \in \CC_\CG(\Sigma_0)} \sum_{\gamma_1 \in \CC_\CG(\Sigma_1)} \tilde Z(M;\gamma_0,\gamma_1) \cdot \lvert \gamma_1 \rangle \langle \gamma_0 \rvert .
\]
In other words,
\[
\tilde Z(M) \lvert \gamma_0 \rangle = \begin{cases}
\sum_{\gamma_1 \in \CC_\CG(M)} Z(M;\gamma_0,\gamma_1) \lvert \gamma_1 \rangle , & \text{$\gamma_0$ is flat,} \\
0 , & \text{otherwise.}
\end{cases}
\]
We claim that this defines a lattice TQFT $\tilde Z$.
\bit
\item By the same argument as in the discussion of the partition functions, the linear map $\tilde Z(M)$ is invariant under the $n$-dimensional Pachner moves, hence independent of the triangulation of $M \setminus \partial M$.
\item Given two triangulated cobordisms $M \colon \Sigma_0 \to \Sigma_1$ and $N \colon \Sigma_1 \to \Sigma_2$, we can glue them to obtain a cobordism $N \circ M = N \cup_{\Sigma_1} M \colon \Sigma_0 \to \Sigma_2$. For flat $\CG$-connections $\gamma_i$ on $\Sigma_i$ for $i = 0,1,2$, it is not hard to verify that
\[
\sum_{\gamma_1 \in \CC_G(\Sigma_1)} \tilde Z(N;\gamma_1,\gamma_2) \cdot \tilde Z(M;\gamma_0,\gamma_1) = \tilde Z(N \circ M;\gamma_0,\gamma_2) .
\]
It follows that $\tilde Z(N) \circ \tilde Z(M) = \tilde Z(N \circ M)$.
\item Clearly $\tilde Z$ preserves the tensor product.
\eit

\begin{expl} \label{expl_cylinder_projection}
Suppose $\omega = 1$ is trivial. Let us compute $\tilde Z(\Sigma \times I)$ for an $(n-1)$-dimensional triangulated oriented closed manifold $\Sigma$. Taking the product CW structure on $\Sigma \times I$, we have
\begin{gather*}
\lvert M_0 \rvert = 2 \lvert \Sigma_0 \rvert , \quad \lvert (\partial M)_0 \rvert = 2 \lvert \Sigma_0 \rvert , \\
\lvert M_1 \rvert = \lvert \Sigma_0 \rvert + 2 \lvert \Sigma_1 \rvert , \quad \lvert (\partial M)_1 \rvert = 2 \lvert \Sigma_1 \rvert .
\end{gather*}
Note that a flat $\CG$-connection on $\Sigma \times I$ is the same as a gauge transformation of flat $\CG$-connections on $\Sigma$. Thus
\[
\tilde Z(\Sigma \times I;\gamma_0,\gamma_1) = \frac{1}{\lvert G \rvert^{\lvert \Sigma_0 \rvert}} \frac{1}{\lvert A \rvert^{\lvert \Sigma_1 \rvert}} \cdot \#\{\phi \colon \gamma_0 \to \gamma_1\} .
\]
So the action of the projector $\tilde Z(\Sigma \times I)$ is
\[
\tilde Z(\Sigma \times I) \lvert \gamma_0 \rangle = \frac{1}{\lvert G \rvert^{\lvert \Sigma_0 \rvert}} \frac{1}{\lvert A \rvert^{\lvert \Sigma_1 \rvert}} \sum_{\phi} \lvert T_\phi \gamma_0 \rangle , \quad \forall \gamma_0 \in \CC_\CG(\Sigma) .
\]
Therefore, the states in $Z(\Sigma)$ are equal-weight superpositions of gauge-equivalent $\CG$-connections. As a corollary, $\dim Z(\Sigma)$ is equal to the number of gauge-equivalence classes of $\CG$-connections on $\Sigma$.
\end{expl}

The construction of the TQFT $Z$ from a lattice TQFT $\tilde Z$ in \cite{Yet93a} uses the colimit over all triangulations on a manifold. Here we give another different but equivalent construction, which helps us construct the lattice model of the 2-group gauge theory. Given an $(n-1)$-dimensional triangulated oriented closed manifold $\Sigma$, the linear map $\tilde Z(\Sigma \times I)$ is a projector on $\tilde Z(\Sigma)$ by the second property of $\tilde Z$. We define $Z(\Sigma) \subseteq \tilde Z(\Sigma)$ to be the image of $\tilde Z(\Sigma \times I)$.

First we show that $Z(\Sigma)$ is independent of the triangulation of $\Sigma$. Suppose there are two different triangulations $\mathcal T,\mathcal S$ on $\Sigma$. To be more precise, we use $(\Sigma,\mathcal T)$ and $(\Sigma,\mathcal S)$ to denote the manifold $\Sigma$ equipped with two different triangulations. Consider the cylinder $\Sigma \times I$ such that $\Sigma \times \{0\}$ is equipped with the triangulation $\mathcal T$ and $\Sigma \times \{1\}$ is equipped with the triangulation $\mathcal S$. Then we can extend the boundary triangulation to the whole manifold $\Sigma \times I$. Such an extension always exists because every $n$-simplex corresponds to a Pachner move on $\Sigma$. Such a triangulated cylinder, denoted by $(\Sigma \times I,\mathcal T,\mathcal S)$, may not be unique, but the linear map
\[
\tilde Z(\Sigma \times I,\mathcal T,\mathcal S) \colon \tilde Z(\Sigma,\mathcal T) \to \tilde Z(\Sigma,\mathcal S)
\]
is independent of the interior triangulation of $\Sigma \times I$. Similarly, we also have an well-defined linear map
\[
\tilde Z(\Sigma \times I,\mathcal S,\mathcal T) \colon \tilde Z(\Sigma,\mathcal S) \to \tilde Z(\Sigma,\mathcal T) .
\]
By the properties of $\tilde Z$ we have
\begin{gather*}
\tilde Z(\Sigma \times I,\mathcal S,\mathcal T) \circ \tilde Z(\Sigma \times I,\mathcal T,\mathcal S) = \tilde Z(\Sigma \times I,\mathcal T,\mathcal T) = \tilde Z((\Sigma,\mathcal T) \times I) , \\
\tilde Z(\Sigma \times I,\mathcal T,\mathcal S) \circ \tilde Z(\Sigma \times I,\mathcal S,\mathcal T) = \tilde Z(\Sigma \times I,\mathcal S,\mathcal S) = \tilde Z((\Sigma,\mathcal S) \times I) .
\end{gather*}
Therefore, after restricting to the images of two idempotents $\tilde Z(\Sigma \times I,\mathcal T,\mathcal S)$ and $\tilde Z(\Sigma \times I,\mathcal S,\mathcal T)$ induce canonical isomorphisms between the spaces $Z(\Sigma,\mathcal T)$ and $Z(\Sigma,\mathcal S)$.

Moreover, for a cobordism $M \colon \Sigma_0 \to \Sigma_1$ we have
\[
\tilde Z(M) = \tilde Z(M) \circ \tilde Z(\Sigma_0 \times I) = \tilde Z(\Sigma_1 \times I) \circ \tilde Z(M) .
\]
Therefore, $\tilde Z(M)$ induces a linear map from $Z(\Sigma_0)$ to $Z(\Sigma_1)$, denoted by $Z(M) \colon Z(\Sigma_0) \to Z(\Sigma_1)$. It is also independent of the triangulations. One can easily show that
\[
Z(N) \circ Z(M) = Z(N \circ M) , \quad Z(\Sigma \times I) = \Id_{Z(\Sigma)} , \quad Z(M \sqcup N) \simeq Z(M) \otimes Z(N) .
\]
Hence, the above construction defines a symmetric monoidal functor $Z \colon \mathrm{Cob}_n^{\mathrm{or}} \to \vect$.

\part{Quantum double model for finite 2-groups}

\section{Quantum double of finite 2-groups} \label{sec_quantum_double_construction}

\subsection{Hopf monoidal category} \label{sec_Hopf_monoidal_category}

Let us briefly review the notion of Hopf monoidal category \cite{CF94,Neu97}. For simplicity, we only consider finite semisimple Hopf monoidal categories, which are Hopf algebras in the 2-category $2\vect$ of finite semisimple categories. In general, one can consider the Hopf algebra in any symmetric (or even braided) monoidal 2-category.

A \emph{finite semisimple comonoidal category} $\CC$ is a coalgebra in $2\vect$, or equivalently, an algebra in $2\vect^{1\op}$ (see \cite{DS97,MC00} for the definition of algebra in monoidal 2-categories). This means that $\CC$ is a finite semisimple category equipped with a cotensor product linear functor $\CC \to \CC \boxtimes \CC$, a counit linear functor $\CC \to \vect$, coassociator and counitor natural isomorphisms satisfying some coherence relations dual to those of a monoidal category. A \emph{finite semisimple bimonoidal category} is a finite semisimple category equipped with both a monoidal structure and a comonoidal structure and some compatibility data. In particular, the tensor product and the unit functors are comonoidal functors, and the associator and the unitors are comonoidal natural transformations. Equivalently, the cotensor product and the counit functors are monoidal functors, and the coassociator and counitors are monoidal natural transformations. A \emph{finite semisimple Hopf monoidal category} is a finite semisimple bimonoidal category equipped with an antipode. We refer the readers to \cite{Neu97} for more details.

For a multi-fusion category $\CC$,  the 2-category $2\rep(\CC) \coloneqq \LMod_\CC(2\vect)$ of finite semisimple left $\CC$-modules, left $\CC$-module functors and left $\CC$-module natural transformations is a finite semisimple 2-category \cite{DR18}. Moreover, if $\CC$ is a finite semisimple rigid Hopf monoidal category, $2\rep(\CC)$ is a monoidal 2-category \cite{Neu97} and hence a fusion 2-category in the sense of \cite{DR18}, and the forgetful functor $2\rep(\CC) \to 2\vect$ is a fiber 2-functor (locally faithful monoidal 2-functor). The converse of this statement is the Tannaka-Krein reconstruction for fusion 2-categories \cite{Gre23}.

\begin{thm}[\cite{Gre23}]
Let $\SC$ be a fusion 2-category $\SC$ and $\forget \colon \SC \to 2\vect$ be a fiber 2-functor. Then the category $\End(\forget)$ of endo-natural transformations and modifications admits a canonical finite semisimple Hopf monoidal category structure, and $2\rep(\End(\forget))$ is equivalent to $\SC$ as fusion 2-categories.
\end{thm}

\subsection{Hopf monoidal categories for finite 2-groups}

Now we briefly recall some construction of Hopf monoidal categories for a finite 2-group $\CG$. For more details, see \cite{HZ23}.

A \emph{finite semisimple 2-representation} of $\CG$ is a finite semisimple category equipped with a $\CG$-action. The 2-category of finite semisimple 2-representations of $\CG$ is denoted by $2\rep(\CG)$. This is symmetric fusion 2-category whose tensor product is given by the Deligne tensor product $\boxtimes$.

The \emph{2-group 2-algebra} $\vect_\CG$ is obtained by first linearizing of the hom spaces of $\CG$ and then taking the Karoubi completion. It is a multi-fusion category and $2\rep(\CG)$ is equivalent to the 2-category of finite semisimple $\vect_\CG$-modules. When $\CG = \CG(G,A,\alpha)$ is skeletal, the simple objects of $\vect_\CG$ are of the form $(g,\rho)$, where $g \in G$ and $\rho \in \hat A$. The fusion rules are given by
\[
(g,\rho) \otimes (h,\sigma) = \delta_{\rho,g \triangleright \sigma} (gh,\rho) ,
\]
and the associator is
\[
\alpha_{(g,\rho),(h,g \triangleright \rho),(k,(gh) \triangleright \rho)} = (\rho \circ \alpha)(g,h,k) .
\]
As the linearization of $\CG$, the finite semisimple category $\vect_\CG$ admits a Hopf monoidal structure inherited from $\CG$, because $\CG$ is a Hopf algebra in the 2-category $\cat$ of categories. In particular, the cotensor product is given by
\[
\Delta(g) = g \boxtimes g , \quad g \in \CG \subset \vect_\CG .
\]

The \emph{regular algebra} $\mathcal F(\CG) \coloneqq \fun(\CG,\vect)$ is the functor category from $\CG$ to $\vect$ with the pointwise tensor product. It is dual to $\vect_\CG$:
\[
\fun(\CG,\vect) \simeq \fun_\bk(\vect_\CG,\vect) \simeq \vect_\CG^\op ,
\]
where the first equivalence is the universal property of Karoubi completion, and the second one is the Yoneda embedding. Therefore, it also admits a finite semisimple Hopf monoidal structure. The cotensor product can be written as
\[
\Delta(\Phi(g)) \simeq \int^{x,y \in \CG} \vect_\CG(g,x \otimes y) \odot \Phi(x) \boxtimes \Phi(y) \simeq \int^{x \in \CG} \Phi(x) \boxtimes \Phi(x^* \otimes g) ,
\]
where $\Phi(g) = \vect_\CG(g,-) \in \mathcal F(\CG)$. When $\CG = \CG(G,A,\alpha)$ is skeletal, the simple objects of $\mathcal F(\CG)$ are of the form $\Phi(x,\phi) \coloneqq \vect_\CG((x,\phi),-)$, where $x \in G$ and $\phi \in \hat A$. The fusion rules are given by
\[
\Phi(x,\phi) \otimes \Phi(y,\psi) = \delta_{x,y} \Phi(x,\phi \psi) ,
\]
and the associator is trivial.

For a finite 2-group $\CG$, we denote the composition of the forgetful 2-functors $\FZ_1(2\rep(\CG)) \to 2\rep(\CG) \to 2\vect$ by $\forget$, where $\FZ_1$ denotes the Drinfeld center of monoidal 2-categories \cite{KV94,BN96,Cra98}. We define the \emph{quantum double} $\CD(\CG)$ of $\CG$ (or the quantum double of $\vect_\CG$) to be the Hopf monoidal category $\End(\forget)$. By the Tannaka-Krein duality, we have $2\rep(\CD(\CG)) \simeq \FZ_1(2\rep(\CG))$. In the rest of this section, we give an explicit description of the Hopf monoidal structure of the quantum double $\CD(\CG)$.

\subsection{Half-braiding and \texorpdfstring{$\CG$}{G}-grading}

Suppose $\CZ = (\CZ,\zeta_{-,\CZ},\zeta_{-,-|\CZ}) \in \FZ_1(2\rep(\CG))$. Let $\varepsilon \in \fun_\bk(\vect_\CG,\vect) \simeq \fun(\CG,\vect)$ be the constant functor that maps every object in $\CG$ to $\bk \in \vect$ and every morphism to the identity map. It is also a $\CG$-module functor if $\vect_\CG$ is equipped with the left translation $\CG$-action and $\vect$ is equipped with the trivial $\CG$-action. Then $\zeta_{\varepsilon,\CZ}$ as depicted in the following diagram
\[
\xymatrix{
\vect_\CG \boxtimes \CZ \ar[r]^-{\zeta_{\vect_\CG,\CZ}} \ar[d]_{\varepsilon \boxtimes 1} \drtwocell<\omit>{\quad\zeta_{\varepsilon,\CZ}} & \CZ \boxtimes \vect_\CG \ar[d]^{1 \boxtimes \varepsilon} \\
\vect \boxtimes \CZ \ar[r]^-{\zeta_{\vect,\CZ}} & \CZ \boxtimes \vect
}
\]
gives an isomorphism
\begin{equation} \label{eq_G-grading_unitality}
(1 \boxtimes \varepsilon) \zeta_{\vect_\CG,\CZ}(\one \boxtimes z) \simeq z , \forall z \in \CZ .
\end{equation}
For every $x \in \CG$ and $z \in \CZ$, we define
\[
z_x \coloneqq (1 \boxtimes \Phi(x)) \zeta_{\vect_\CG,\CZ}(\one \boxtimes z) ,
\]
where $\Phi(x) \coloneqq \vect_\CG(x,-) \colon \vect_\CG \to \vect$. Then we have a linear functor
\begin{align}
\CZ \boxtimes \mathcal F(\CG) & \to \CZ \label{eq_G-grading_right_module_action} \\
z \boxtimes \Phi(x) & \mapsto z_x . \nonumber
\end{align}

\begin{lem}
The half-braiding on $\CZ$ is determined by the functor \eqref{eq_G-grading_right_module_action}.
\end{lem}

\pf
First we can check that
\[
(1 \boxtimes \Phi(y)) \biggl( \int^{x \in \CG} z_x \boxtimes x \biggr) \simeq \int^{x \in \CG} z_x \boxtimes \Phi(y)(x) \simeq \int^{x \in \CG} \vect_{\CG}(y,x) \odot z_x \simeq z_y , \quad \forall y \in \CG .
\]
This implies that
\[
\zeta_{\vect_\CG,\CZ}(\one \boxtimes z) \simeq \int^{x \in \CG} z_x \boxtimes x .
\]
For any $\CV \in 2\rep(\CG)$ and $v \in \CV$, by the following natural isomorphism
\[
\xymatrix{
\vect_\CG \boxtimes \CZ \ar[r]^-{\zeta_{\vect_\CG,\CZ}} \ar[d]_{(- \odot v) \boxtimes 1} \drtwocell<\omit>{\qquad\;\;\zeta_{(- \odot v),\CZ}} & \CZ \boxtimes \vect_\CG \ar[d]^{1 \boxtimes (- \otimes v)} \\
\CV \boxtimes \CZ \ar[r]^-{\zeta_{\CV,\CZ}} & \CZ \boxtimes \CV
}
\]
the half-braiding is given by
\begin{equation} \label{eq_half-braiding_G-grading}
\zeta_{\CV,\CZ}(v \boxtimes z) \simeq \int^{x \in \CG} z_x \boxtimes (x \odot v) .
\end{equation}
Hence we can always assume that the half-braiding has this form.
\epf

\begin{prop} \label{prop_G-grading_half-braiding}
The functor \eqref{eq_G-grading_right_module_action} defines a right $\mathcal F(\CG)$-module structure on $\CZ$. Equivalently,
\begin{align*}
\CZ & \to \CZ \boxtimes \vect_\CG \\
z & \mapsto \int^{x \in \CG} z_x \boxtimes x
\end{align*}
defines a right $\vect_\CG$-comodule structure on $\CZ$.
\end{prop}

\pf
The unitor is given by the isomorphism \eqref{eq_G-grading_unitality}. In particular, since
\[
\varepsilon \simeq \bigoplus_{x \in \pi_1(\CG)} \Phi(x,1) \simeq \lim_{x \in \CG^\op} \Phi \simeq \int_{x \in \CG} \Phi(x) ,
\]
we have
\[
z \simeq \int_{x \in \CG} z_x \simeq \lim_{x \in \CG^\op} z_x .
\]
Considering the half-braiding \eqref{eq_half-braiding_G-grading}, the natural isomorphism $\zeta_{\CV,\CW|\CZ}$ gives the isomorphism
\[
\int^{x,y \in \CG} (z_x)_y \boxtimes (y \odot v) \boxtimes (x \odot w) \simeq \int^{x \in \CG} z_x \boxtimes (x \odot v) \boxtimes (x \odot w) , \quad \forall x,y \in \CG , \, v \in \CV , \, w \in \CW , \, z \in \CZ .
\]
It follows that
\[
(z_x)_y \simeq \vect_\CG(y,x) \odot z_x , \quad \forall z \in \CZ , \, x,y \in \CG .
\]
This gives the associator of the right $\mathcal F(\CG)$-module structure on $\CZ$. This isomorphism can be written in a more functorial way: for every linear category $\CC$ and bilinear functor $Q \colon \CZ \times \vect_\CG \to \CC$ there is a canonical isomorphism
\begin{equation} \label{eq_grading_delta_function}
\int^{y \in \CG} Q((z_x)_y,y) \simeq Q(z_x,x) .
\end{equation}
The axioms of the right $\mathcal F(\CG)$-module structure follows from those of $\zeta_{-,-|\CZ}$.

The right $\vect_\CG$-comodule structure on $\CZ$ is induced by the right $\mathcal F(\CG)$-module structure. Also note that the coaction functor $\CZ \to \CZ \boxtimes \vect_\CG$ is equivalent to $\zeta_{\vect_\CG,\CZ}(\one \boxtimes -)$.
\epf

For a 1-group $G$, a $\fun(G)$-module structure or a $\bk[G]$-comodule structure on a vector space is the same as a $G$-grading. Therefore, we call the right $\mathcal F(\CG)$-module structure or the right $\vect_\CG$-comodule structure in Proposition \ref{prop_G-grading_half-braiding} the \emph{$\CG$-grading} on $\CZ$ induced by the half-braiding.

\begin{lem} \label{lem_2-group_rep_grading_compatible}
The $\CG$-grading on $\CZ$ is compatible with its left $\CG$-module structure in the sense that there is a natural isomorphism $g \odot z_x \simeq (g \odot z)_{(g \otimes x) \otimes g^*}$ for $g,x \in \CG$ and $z \in \CZ$, such that the following diagram commutes:
\[
\xymatrix{
(g \otimes h) \odot z_x \ar[r] \ar[d] & g \odot (h \odot z_x) \ar[r] & g \odot z_{(h \otimes x) h^*} \ar[d] \\
z_{((g \otimes h) \otimes x) \otimes (g \otimes h)^*} \ar[rr] & & z_{(g \otimes ((h \otimes x) \otimes h^*)) \otimes g^*} 
}
\]
\end{lem}

\pf
Since $\zeta_{\vect_\CG,\CZ}$ is a $\CG$-module functor, we have an isomorphism
\[
\int^{x \in \CG} (g \odot z)_x \boxtimes (x \otimes g) \simeq \zeta_{\vect_\CG,\CZ}(g \boxtimes (g \odot z)) \simeq g \odot \zeta_{\vect_\CG,\CZ}(\one \boxtimes z) \simeq \int^{x \in \CG} (g \odot z_x) \boxtimes (g \otimes x) .
\]
The commutative diagram follows from the axiom of $\CG$-module functors.
\epf

\begin{rem}
Therefore, an object in $\FZ_1(2\rep(\CG))$ is equivalent to a finite semisimple category equipped with a left $\vect_\CG$-module structure, a right $\vect_\CG$-comodule structure and a compatibility data as defined in Lemma \ref{lem_2-group_rep_grading_compatible}. This generalizes the classical notion of Yetter-Drinfeld module over a Hopf algebra.
\end{rem}

\begin{expl}
The category $\vect_\CG$ admits a canonical $\CG$-grading induced by the comultiplication $\vect_\CG \to \vect_\CG \boxtimes \vect_\CG$ which maps $g \in \CG$ to $g \boxtimes g$. In particular, $g_x = \vect_\CG(x,g) \odot g$ for all $g,x \in \CG$. The conjugation $\CG$-action on $\vect_\CG$ is compatible with this $\CG$-grading. This gives an object in $\FZ_1(2\rep(G))$:
\bit
\item The underlying category is $\vect_\CG$.
\item The left $\CG$-action is the conjugation $\CG$-action.
\item The half-braiding is
\begin{align*}
\gamma_{\CV,\vect_\CG} \colon \CV \boxtimes \vect_\CG & \to \vect_\CG \to \CV \\
v \boxtimes g & \mapsto g \boxtimes (g \odot v) .
\end{align*}
\item The natural isomorphism $\gamma_{\CV,\CW|\vect_\CG}$ is identity.
\eit
\end{expl}

Suppose $\CY = (\CY,\eta_{-,\CZ},\eta_{-,-|\CZ}) \in \FZ_1(2\rep(\CG))$ and $F \colon \CZ \to \CY$ is a 1-morphism in $\FZ_1(2\rep(\CG))$. Note that $F$ is equipped with a natural isomorphism
\[
\xymatrix{
\vect_\CG \boxtimes \CZ \ar[r]^-{\zeta_{\vect_\CG,\CZ}} \ar[d]_{1 \boxtimes F} \drtwocell<\omit>{} & \CZ \boxtimes \vect_\CG \ar[d]^{F \boxtimes 1} \\
\vect_\CG \boxtimes \CY \ar[r]^-{\eta_{\vect_\CG,\CZ}} & \CY \boxtimes \vect_\CG
}
\]
which induces an isomorphism $F(z_x) \simeq F(z)_x$ for every $z \in \CZ$ and $x \in \CG$. In other words, $F$ preserves the $\CG$-grading.

For any $y \in \CY$ and $z \in \CZ$, we also have
\[
(1 \boxtimes \zeta_{\vect_\CG,\CZ})(\eta_{\vect_\CG,\CY} \boxtimes 1)(\one \boxtimes y \boxtimes z) \simeq (1 \boxtimes \zeta_{\vect_\CG,\CZ}) \biggl(\int^{r \in \CG} y_r \boxtimes r \boxtimes z \biggr) \simeq \int^{r,s \in \CG} y_r \boxtimes z_s \boxtimes (s \otimes r) .
\]
On the other hand, by the definition of the half-braiding on the tensor product $\CY \boxtimes \CZ$, the left hand side is also isomorphic to
\[
\int^{t \in \CG} (y \boxtimes z)_t \boxtimes t .
\]
Thus
\[
(y \boxtimes z)_t \simeq (1 \boxtimes 1 \boxtimes \Phi(t)) \biggl( \int^{r,s \in \CG} y_r \boxtimes z_s \boxtimes (s \otimes r) \biggr) = \int^{r,s \in \CG} \vect_\CG(t,s \otimes r) \odot y_r \boxtimes z_s \simeq \int^{s \in \CG} y_{s^* \otimes t} \boxtimes z_s .
\]
This convolution encodes the (reversed) comonoidal structure of $\mathcal F(\CG)$.

\subsection{Quantum double \texorpdfstring{$\CD(\CG)$}{D(G)} as a representative object}

Let $\CD(\CG) \coloneqq \mathcal F(\CG) \boxtimes \vect_\CG \in 2\rep(\CG)$, where $\mathcal F(\CG)$ is equipped with the conjugation $\CG$-action and $\vect_\CG$ is equipped with the left translation $\CG$-action. For every $\CV \in 2\rep(\CG)$, we define a functor
\begin{align*}
\gamma_{\CV,\CD(\CG)} \colon \CV \boxtimes \CD(\CG) & \to \CD(\CG) \boxtimes \CV \\
v \boxtimes \Phi(x) \boxtimes g & \mapsto \Phi(x) \boxtimes g \boxtimes (x \odot v) .
\end{align*}
This functor is also a $\CG$-module functor:
\begin{multline*}
\gamma_{\CV,\CD(\CG)}(g \odot (v \boxtimes \Phi(x) \boxtimes h)) \simeq \Phi((g \otimes x) \otimes g^*) \boxtimes (g \otimes h) \boxtimes ((g \otimes x) \otimes g^*) \odot (g \odot v) \\
\simeq \Phi((g \otimes x) \otimes g^*) \boxtimes (g \otimes h) \boxtimes (g \odot (x \odot v)) \simeq g \odot \gamma_{\CV,\CD(\CG)}(v \boxtimes \Phi(x) \boxtimes h) , \quad \forall g,x,h \in \CG , \, v \in \CV .
\end{multline*}
Then it is easy to see that this defines a natural equivalence
\[
\gamma_{-,\CD(\CG)} \colon (- \boxtimes \CD(\CG)) \Rightarrow (\CD(\CG) \boxtimes -) \colon 2\rep(\CG) \to 2\rep(\CG) .
\]
For every $\CV,\CW \in 2\rep(\CG)$, note that $(\gamma_{\CV,\CD(\CG)} \boxtimes 1) \circ (1 \boxtimes \gamma_{\CW,\CD(\CG)})$ is canonically isomorphic to $\gamma_{\CV \boxtimes \CW,\CD(\CG)}$. Indeed, we have
\begin{multline*}
(\gamma_{\CV,\CD(\CG)} \boxtimes 1) (1 \boxtimes \gamma_{\CW,\CD(\CG)}) (v \boxtimes w \boxtimes \Phi(x) \boxtimes g) = (\gamma_{\CV,\CD(\CG)} \boxtimes 1) (v \boxtimes \Phi(x) \boxtimes g \boxtimes (x \odot w)) \\
= \Phi(x) \boxtimes g \boxtimes (x \odot v) \boxtimes (x \odot w) = \gamma_{\CV \boxtimes \CW,\CD(\CG)}(v \boxtimes w \boxtimes \Phi(x) \boxtimes g) .
\end{multline*}
We define $\gamma_{\CV,\CW|\CD(\CG)}$ to be this canonical natural isomorphism, as depicted in the following diagram:
\[
\xymatrix{
 & \CV \boxtimes \CD(\CG) \boxtimes \CW \ar[dr]^{\gamma_{\CV,\CD(\CG)} \boxtimes 1} \\
\CV \boxtimes \CW \boxtimes \CD(\CG) \ar[ur]^{1 \boxtimes \gamma_{\CW,\CD(\CG)}} \rrtwocell<\omit>{<-3.5>\quad\qquad\gamma_{\CV,\CW|\CD(\CG)}} \ar[rr]_-{\gamma_{\CV \boxtimes \CW,\CD(\CG)}} & & \CD(\CG) \boxtimes \CV \boxtimes \CW
}
\]
Then it is easy to see that $\CD(\CG) = (\CD(\CG),\gamma_{-,\CD(\CG)},\gamma_{-,-|\CD(\CG)})$ is an object in $\FZ_1(2\rep(\CG))$. In particular, the $\CG$-grading is given by
\[
(\Phi(x) \boxtimes g)_y = \vect_\CG(y,x) \odot \Phi(x) \boxtimes g , \quad \forall g,x,y \in \CG .
\]

\begin{lem}
For every $\CZ = (\CZ,\zeta_{-,\CZ},\zeta_{-,-|\CZ}) \in \FZ_1(2\rep(\CG))$, the hom category
\[
\FZ_1(2\rep(\CG))(\CD(\CG),\CZ)
\]
is equivalent to $\CZ$ as categories.
\end{lem}

\pf
For every $z \in \CZ$, we define a functor $F_z \colon \CD(\CG) \to \CZ$ by
\[
F_z(\Phi(x) \boxtimes g) \coloneqq (g \odot z)_x .
\]
The $\CG$-module functor structure is given by
\begin{multline*}
F_z(g \odot (\Phi(x) \boxtimes h)) \simeq ((g \otimes h) \odot z)_{(g \otimes x) \otimes g^*} \simeq (g \odot (h \odot z))_{(g \otimes x) \otimes g^*} \\ \simeq g \odot (h \odot z)_x = g \odot F_z(\Phi(x) \boxtimes h) , \quad \forall g,h,x \in \CG .
\end{multline*}
The compatibility data with the half-braiding is given by
\begin{multline*}
(F_z \boxtimes 1) \gamma_{\CV,\CD(\CG)} (v \boxtimes \Phi(x) \boxtimes g) = (F_z \boxtimes 1)(\Phi(x) \boxtimes g \boxtimes (x \odot v)) = (g \odot z)_x \boxtimes (x \odot v) \\
\simeq \int^{y \in \CG} \bigl((g \odot z)_x \bigr)_y \boxtimes (y \odot v) \simeq \zeta_{\CV,\CZ}(v \boxtimes (g \odot z)_x) \simeq \zeta_{\CV,\CZ} (1 \boxtimes F_z)(v \boxtimes \Phi(x) \boxtimes g)
\end{multline*}
It is tedious but not hard to check that $F_z$ is a 1-morphism in $\FZ_1(2\rep(\CG))$.

Then we show that given $z,z' \in \CZ$, the hom set $\Hom(F_z,F_{z'})$ in $\FZ_1(2\rep(\CG))$ is isomorphic to $\CZ(z,z')$. Suppose $\phi \colon F_z \Rightarrow F_{z'}$ is a 2-morphism in $\FZ_1(2\rep(\CG))$. Then $\phi_{\varepsilon \boxtimes \one}$ is a morphism from $z$ to $z'$. This gives a map $\Hom(F_z,F_{z'}) \to \CZ(z,z')$, which is clearly injective. To show that it is surjective, we need to construct a 2-morphism $\phi^f \colon F_z \Rightarrow F_{z'}$ such that $\phi^f_{\varepsilon \boxtimes \one} = f$. We define
\[
\phi^f_{\Phi(x) \boxtimes g} \coloneqq (g \odot f)_x \colon (g \odot z)_x \to (g \odot z')_x .
\]
It remains to show $\phi^f$ defined as above is a 2-morphism in $\FZ_1(2\rep(\CG))$. It is easy to see that $\phi^f$ is a $\CG$-module natural transformation, and its compatibility with the half-braiding follows from the naturality of \eqref{eq_grading_delta_function}.

Finally, we need to show that for any 1-morphism $F \colon \CD(\CG) \to \CZ$ in $\FZ_1(2\rep(\CG))$, there exists an object $z \in \CZ$ such that $F \simeq F_z$. Indeed we can take $z \coloneqq F(\varepsilon \boxtimes \one)$. The compatibility data of $F$ with the half-braiding gives the isomorphism $F(\Phi(x) \boxtimes \one) \simeq z_x$, and the $\CG$-module functor structure of $F$ gives the isomorphism $F(\Phi(x) \boxtimes g) \simeq (g \odot z)_x$. The compatibility of this isomorphism with half-braiding is tautological.
\epf

\begin{cor} \label{cor_quantum_double_representable}
The forgetful 2-functor $\forget \colon \FZ_1(2\rep(\CG)) \to 2\vect$ is represented by $\CD(\CG) \in \FZ_1(2\rep(\CG))$. In other words, $\forget$ is equivalent to the hom 2-functor $\FZ_1(2\rep(\CG))(\CD(\CG),-)$.
\end{cor}

\subsection{The Hopf monoidal structure on \texorpdfstring{$\CD(\CG)$}{D(G)}} \label{sec_quantum_double_Hopf_structure}

By Corollary \ref{cor_quantum_double_representable} and the Yoneda lemma, we have an equivalence
\[
\End(\forget) \simeq \nat(\FZ_1(2\rep(\CG))(\CD(\CG),-),\forget) \simeq \forget(\CD(\CG)) = \CD(\CG) ,
\]
which maps a natural 2-transformation $\phi \colon \forget \Rightarrow \forget$ to $\phi_{\CD(\CG)}(\varepsilon \boxtimes \one)$. Therefore, we can transfer the Hopf monoidal structure from $\End(\forget) \to \CD(\CG)$. In the following we explicitly compute the quantum double $\CD(\CG)$.

The monoidal structure of $\End(\forget)$ is the composition of natural 2-transformations. By the Yoneda lemma, we have the equivalence of monoidal categories
\[
\End(\forget) \simeq \End(\FZ_1(2\rep(\CG))(\CD(\CG),-)) \simeq \End_{\FZ_1(2\rep(\CG))}(\CD(\CG))^\rev .
\]
For $x,g \in \CG$, It is known that there is a 1-morphism (unique up to isomorphism) $F_{x,g} \colon \CD(\CG) \to \CD(\CG)$ such that $F_{x,g}(\varepsilon \boxtimes \one) = \Phi(x) \boxtimes g$. Indeed we can take $F_{x,g}(\Phi(y) \boxtimes h) \coloneqq (h \odot (\Phi(x) \boxtimes g))_y$. Then we have
\[
(F_{y,h} \circ F_{x,g})(\varepsilon \boxtimes \one) = F_{y,h}(\Phi(x) \boxtimes g) = (g \odot (\Phi(y) \boxtimes h))_x \simeq \vect_\CG(x,(g \otimes y) \otimes g^*) \odot \Phi(x) \boxtimes (g \otimes h) .
\]
Therefore, the tensor product of $\CD(\CG)$ is given by
\[
(\Phi(x) \boxtimes g) \otimes (\Phi(y) \boxtimes h) \coloneqq \vect_\CG(x,(g \otimes y) \otimes g^*) \odot \Phi(x) \boxtimes (g \otimes h) \simeq (\Phi(x) \otimes (g \odot \Phi(y))) \boxtimes (g \otimes h) .
\]
In particular, the tensor unit is $\varepsilon \boxtimes \one$. The associator is induced from that of $\mathcal F(\CG)$ and $\vect_\CG$. Hence, the monoidal category $\CD(\CG)$ can be viewed as a `crossed product' of $\mathcal F(\CG)$ and $\vect_\CG$, where $\vect_\CG$ acts on $\mathcal F(\CG)$ by conjugation.

We can compute the composition in $\End(\forget)$ directly. For $x,g \in \CG$, denote $\phi^{x,g} \colon \forget \Rightarrow \forget$ the natural 2-transformation such that
\[
\phi^{x,g}_{\CD(\CG)}(\varepsilon \boxtimes \one) = \Phi(x) \boxtimes g \in \forget(\CD(\CG)) .
\]
Then by the naturality we have
\begin{multline*}
(\phi^{x,g} \cdot \phi^{y,h})_{\CD(\CG)}(\varepsilon \boxtimes \one) = \phi^{x,g}_{\CD(\CG)}(\Phi(y) \boxtimes h) = \phi^{x,g}_{\CD(\CG)}(F_{y,h}(\varepsilon \boxtimes \one)) \\
\simeq F_{y,h}(\phi^{x,g}_{\CD(\CG)}(\varepsilon \boxtimes \one)) = F_{y,h}(\Phi(x) \boxtimes g) \simeq \vect_\CG(x,(g \otimes y) \otimes g^*) \odot \Phi(x) \boxtimes (g \otimes h) .
\end{multline*}
Thus we obtain the same result.

The comonoidal structure on $\CD(\CG)$ is given by
\begin{multline*}
\Delta(\Phi(x) \boxtimes g) = \phi^{x,g}_{\CD(\CG) \boxtimes \CD(\CG)}(\varepsilon \boxtimes \one \boxtimes \varepsilon \boxtimes \one) = \phi^{x,g}_{\CD(\CG) \boxtimes \CD(\CG)} F_{\varepsilon \one \varepsilon \one}(\varepsilon \boxtimes \one) \\
\simeq F_{\varepsilon \one \varepsilon \one} \phi^{x,g}_{\CD(\CG)}(\varepsilon \boxtimes \one) = F_{\varepsilon \one \varepsilon \one}(\Phi(x) \boxtimes g) = (\varepsilon \boxtimes g \boxtimes \varepsilon \boxtimes g)_x \\
\simeq \int^{r,s \in \CG} \vect_\CG(x,s \otimes r) \odot \Phi(r) \boxtimes g \boxtimes \Phi(s) \boxtimes g \simeq \int^{s \in \CG} \Phi(s^* \otimes x) \boxtimes g \boxtimes \Phi(s) \boxtimes g .
\end{multline*}
In other words, the comonoidal structure on $\CD(\CG)$ is induced by the reversed comonoidal structure on $\mathcal F(\CG)$ and the comonoidal structure on $\vect_\CG$. In particular, the counit is $\one \boxtimes \varepsilon \colon \mathcal F(\CG) \boxtimes \vect_\CG \to \vect$.

We summarize the result in the following theorem.

\begin{thm} \label{thm_fusion_rule_quantum_double}
Let $\CG = \CG(G,A,\alpha)$ be a finite skeletal 2-group.
\bnu[(1)]
\item The simple objects of $\CD(\CG) = \mathcal F(\CG) \boxtimes \vect_\CG$ are $\Phi(x,\phi) \boxtimes (g,\rho)$, where $g,x \in G$ and $\rho,\phi \in \hat A$.
\item The fusion rules are
\[
(\Phi(x,\phi) \boxtimes (g,\rho)) \otimes (\Phi(y,\psi) \boxtimes (h,\sigma)) = \delta_{x,gyg^{-1}} \delta_{\rho,g \triangleright \sigma} \Phi(x,\phi (g \triangleright \psi)) \boxtimes (gh,\rho) .
\]
\item The associator on $\mathcal F(\CG)$ is trivial, and the associator on $\vect_\CG$ is induced by $\alpha$.
\item The comonoidal structure is given by
\[
\Delta(\Phi(x) \boxtimes g) \simeq \int^{s \in \CG} \Phi(s^* \otimes x) \boxtimes g \boxtimes \Phi(s) \boxtimes g .
\]
%\begin{multline*}
%\Delta(\Phi(x) \boxtimes g) \simeq \int^{s \in \CG} \Phi(s^* \otimes x) \boxtimes g \boxtimes \Phi(s) \boxtimes g .
%\end{multline*}
\item The coassociator on $\mathcal F(\CG)$ is induced by $\alpha$, and the coassociator on $\vect_\CG$ is trivial.
\enu
\end{thm}

The functor
\begin{align*}
\vect_\CG & \to \CD(\CG) \\
g & \mapsto \varepsilon \boxtimes g
\end{align*}
is a Hopf monoidal functor. Clearly it induces the forgetful 2-functor $\FZ_1(2\rep(\CG)) \simeq 2\rep(\CD(\CG)) \to 2\rep(\vect_\CG) \simeq 2\rep(\CG)$. Similarly, the functor
\begin{align*}
\mathcal F(\CG) & \to \CD(\CG) \\
\Phi(x) & \mapsto \Phi(x) \boxtimes \one
\end{align*}
is also a Hopf monoidal functor, and it induces the forgetful 2-functor $\FZ_1(2\vect_\CG) \simeq \FZ_1(2\rep(\CG)) \simeq 2\rep(\CD(\CG)) \to 2\rep(\mathcal F(\CG)) \simeq 2\vect_\CG$.

\begin{expl}
Suppose $\CG = G$ is a finite 1-group. The quantum double $\CD(G) = \CD(\vect_G)$ was also studied by Crane and Yetter \cite{CY98} (see also \cite{Neu97}). Its simple objects are of the form $\Phi(x) \boxtimes g$ for all $g,x \in G$. The fusion rule is
\[
(\Phi(x) \boxtimes g) \otimes (\Phi(y) \boxtimes h) = \delta_{x,gyg^{-1}} \Phi(x) \boxtimes (gh) .
\]
The cotensor product is given by
\[
\Delta(\Phi(x) \boxtimes g) = \bigoplus_{\substack{a,b \in G \\ ab = x}} \Phi(b) \boxtimes g \boxtimes \Phi(a) \boxtimes g .
\]
The associator and coassociator are identity.
\end{expl}

\begin{expl}
Suppose $\CG = \mathrm B A$ where $A$ is a finite abelian group. Then we have the equivalences of finite semisimple Hopf monoidal categories
\[
\vect_{\mathrm B A} \simeq \mathcal F(\hat A) \simeq \vect^{\oplus \hat A} , \quad \mathcal F(\mathrm B A) \simeq \vect_{\hat A} \simeq \rep(A) ,
\]
and they induce the equivalences of fusion 2-categories
\[
2\rep(\mathrm B A) \simeq 2\vect_{\hat A} , \quad 2\vect_{\mathrm B A} \simeq 2\rep(\hat A) .
\]
The quantum double $\CD(\mathrm B A)$ has the simple objects $\Phi(\phi) \boxtimes \rho$ for all $\rho,\phi \in \hat A$. The fusion rule is
\[
(\Phi(\phi) \boxtimes \rho) \otimes (\Phi(\psi) \boxtimes \sigma) = \delta_{\rho,\sigma} \Phi(\phi \psi) \boxtimes \rho .
\]
The cotensor product is given by
\[
\Delta(\Phi(\phi) \boxtimes \rho) = \bigoplus_{\substack{\sigma,\tau \in \hat A \\ \sigma \tau = \rho}} \Phi(\phi) \boxtimes \sigma \boxtimes \Phi(\phi) \boxtimes \tau .
\]
The associator and coassociator are identity. It is not hard to see that $\CD(\mathrm B A) \simeq \CD(\hat A)$ as finite semisimple Hopf monoidal categories.
\end{expl}

\begin{rem}
In \cite[Section 4.5]{BBG23}, the authors computed the ``tube 2-algebra'' associated to a finite 2-group $\CG = \CG(G,A,\alpha)$ and a 4-cocycle $\pi \in Z^4(G;U(1))$ (viewed as a cocycle in $Z^4(\CG;U(1))$). This tube 2-algebra should be the underlying monoidal category of the twisted quantum double $\CD^\pi(\CG)$, which is a `quasi-Hopf monoidal category'. When $\pi$ is trivial, the results in \cite{BBG23} looks different from ours. This is because the categories in \cite{BBG23} are usually not Karoubi complete.
\end{rem}

%\subsection{Loop 2-groupoid and quantum double}
%
%Recall \cite{HXZ24} that the loop 2-groupoid $\mathrm L \CG$ of a finite 2-group $\CG$ is the 2-functor 2-category
%\[
%\mathrm L \CG \coloneqq \fun(\mathrm B \Zb,\mathrm B \CG) ,
%\]
%and it is equivalent to the homotopy quotient $\CG \dquotient \CG$ with $\CG$ acting on itself by conjugation. Then we have
%\[
%\fun(\mathrm L \CG,2\vect) \simeq \fun(\CG \dquotient \CG,2\vect) \simeq \fun(\CG,2\vect)^\CG \simeq (2\vect_\CG)^\CG \simeq \FZ_1(2\vect_\CG) \simeq \FZ_1(2\rep(\CG)) .
%\]
%
%\mynote{(quantum double is the 2-groupoid 2-algebra of the loop 2-groupoid)}

\subsection{Quasi-triangular structure}

For a finite semisimple Hopf monoidal category $\CC$, it is natural to expect that the braiding structure on $2\rep(\CC)$ can be equivalently induced by some structures on $\CC$. Such structures are called the quasi-triangular structure on $\CC$. By \cite{Neu97}, a \emph{quasi-triangular structure} on $\CC$ consists of\footnote{Some of the defining data of the quasi-triangular structure in \cite{Neu97} are not necessary. So here our definition is slightly different.}
\bit
\item an object $R \in \CC \boxtimes \CC$, usually denoted by $R = R^1 \boxtimes R^2$,
\item a natural isomorphism $\sigma_x \colon R \otimes \Delta(x) \to \Delta(x) \otimes R$ for $x \in \CC$,
\item an isomorphism $R_l \colon R^{13} \otimes R^{12} \to (1 \boxtimes \Delta)(R)$, where $R^{13}$ denotes $R^1 \boxtimes \one \boxtimes R^2 \in \CC \boxtimes \CC \boxtimes \CC$,
\item an isomorphism $R_r \colon R^{13} \otimes R^{23} \to (\Delta \boxtimes 1)(R)$,
\eit
and these data should satisfy some coherence conditions. Given such a quasi-triangular structure, the braiding on $2\rep(\CC)$ is defined as follows.
\bit
\item For $\CV,\CW \in 2\rep(\CC)$, the braiding $\beta_{\CV,\CW}$ is given by
\begin{align*}
\beta_{\CV,\CW} \colon \CV \boxtimes \CW & \to \CW \boxtimes \CV \\
v \boxtimes w & \mapsto (R^2 \odot w) \boxtimes (R^1 \odot v) = R^{21} \odot (w \boxtimes v) = T_{\CV,\CW}(R \odot (v \boxtimes w)) ,
\end{align*}
where $T_{\CV,\CW} \colon \CV \boxtimes \CW \to \CW \boxtimes \CV$ is the braiding of $2\vect$ by swapping two components.
\item For $\CU,\CV,\CW \in 2\rep(\CG)$, the modifications
\begin{gather*}
\beta_{\CU|\CV,\CW} \colon (1 \boxtimes \beta_{\CU,\CW}) \circ (\beta_{\CU,\CV} \boxtimes 1) \Rightarrow \beta_{\CU,\CV \boxtimes \CW} \\
\beta_{\CU,\CV|\CW} \colon (\beta_{\CU,\CW} \boxtimes 1) \circ (1 \boxtimes \beta_{\CV,\CW}) \Rightarrow \beta_{\CU \boxtimes \CV|\CW}
\end{gather*}
are induced by $R_l$ and $R_r$, respectively. More precisely, for $u \in \CU$, $v \in \CV$ and $w \in \CW$, the morphism
\begin{multline*}
(\beta_{\CU|\CV,\CW})_{u \boxtimes v \boxtimes w} \colon(R^2 \odot v) \boxtimes (R^2 \odot w) \boxtimes (R^1 \odot R^1 \odot u) \\
= T_{\CU,\CV \boxtimes \CW}((R^{13} \otimes R^{12}) \odot (u \boxtimes v \boxtimes w)) \to T_{\CU,\CV \boxtimes \CW}(((1 \boxtimes \Delta)(R)) \odot (u \boxtimes v \boxtimes w)) \\
= (R^2_{(1)} \odot v) \boxtimes (R^2_{(2)} \odot w) \boxtimes (R^1 \odot u)
\end{multline*}
is given by $T_{\CU,\CV \boxtimes \CW}(R_l \odot 1)$. Similarly, $(\beta_{\CU,\CV|\CW})_{u \boxtimes v \boxtimes w} \coloneqq T_{\CU \boxtimes \CV,\CW}(R_r \odot 1)$.
\eit

For a finite 2-group $\CG$, the Drinfeld center $\FZ_1(2\rep(\CG))$ is a braided fusion 2-category, so there is a canonical quasi-triangular structure on $\CD(\CG)$ such that $2\rep(\CD(\CG)) \simeq \FZ_1(2\rep(\CG))$ as braided fusion 2-categories. Here we explicitly give this quasi-triangular structure. The object $R \in \CD(\CG) \boxtimes \CD(\CG)$ can be recovered as
\[
R = (T_{\CD(\CG),\CD(\CG)} \circ \gamma_{\CD(\CG),\CD(\CG)})(\one \boxtimes \one) .
\]
Then one can compute that
\[
R = \int^{x \in \CG} \Phi(x) \boxtimes \one \boxtimes \varepsilon \boxtimes x \in \CD(\CG) \boxtimes \CD(\CG) .
\]
The isomorphisms $R_l$ and $R_r$ are identities.

\section{The 3+1D quantum double model for finite 2-groups} \label{sec_lattice_model_definition}

In this section, we define the 3+1D quantum double model for a skeletal finite 2-group $\CG = \CG(G,A,\alpha)$, which is the Hamiltonian version of the $\CG$ gauge theory defined in Section \ref{sec_2-group_gauge_theory}.

First we arbitrarily choose a triangulation of the 3d space manifold $\Sigma$. On each edge (1-simplex) $e$ there is a local Hilbert space $\mathcal H_e \coloneqq \Cb[G]$, and on each plaquette (2-simplex) $p$ there is a local Hilbert space $\mathcal H_p \coloneqq \Cb[A]$. The total Hilbert space $\mathcal H_{\mathrm{tot}}$ is defined to be the tensor product of all local Hilbert spaces:
\[
\mathcal H_{\mathrm{tot}} \coloneqq \bigl(\bigotimes_{e \in \Sigma_1} \mathcal H_e \bigr) \otimes \bigl(\bigotimes_{p \in \Sigma_2} \mathcal H_p \bigr) = \bigl(\bigotimes_{e \in \Sigma_1} \Cb[G] \bigr) \otimes \bigl(\bigotimes_{p \in \Sigma_2} \Cb[A] \bigr) .
\]
Therefore, a $\CG$-connection $\tau$ represents a tensor product state $\lvert \tau \rangle$ in the total Hilbert space:
\[
\lvert \tau \rangle \coloneqq \bigl( \bigotimes_{e \in \Sigma_1} \lvert \tau_1(e) \rangle_e \bigr) \otimes \bigl( \bigotimes_{p \in \Sigma_2} \lvert \tau_2(p) \rangle_p \bigr) .
\]
Note that the total Hilbert space is exactly the space $\tilde Z(\Sigma)$ defined in Section \ref{sec_TQFT_functor}.

We want to find the Hamiltonian with the following form:
\[
H = - \sum_i P_i ,
\]
where $P_i$'s are local commuting projectors. If this is the Hamiltonian version of the $\CG$ gauge theory, we expect that the ground state subspace is the space $Z(\Sigma)$ defined in Section \ref{sec_TQFT_functor}, that is, the image of the projector $\tilde Z(\Sigma \times I)$. Since $P_i$'s are local commuting projectors, the ground state subspace of $H$ is the image of the projector $\prod_i P_i$. Therefore, to define the Hamiltonian, we need to find local commuting projectors $P_i$'s such that $\prod_i P_i = \tilde Z(\Sigma \times I)$.

We first define these local commuting projectors and then verify that their product is $\tilde Z(\Sigma \times I)$.
\bit
\item For every tetrahedron $t \in \Sigma_3$ and an element $\rho \in \hat A$, there is a projector $D_t$ that examines the 2-flatness of the states on $t$. In other words, $D_t$ is defined by
\[
D_t \lvert \tau \rangle = \delta_{( \tau_1([t_2,t_3]) \triangleright \tau_2([t_0,t_1,t_2]) ) \tau_2([t_0,t_2,t_3]) \alpha(\tau_1([t_2,t_3]),\tau_1([t_1,t_2]),\tau_1([t_0,t_1])) , \tau_2([t_0,t_1,t_3]) \tau_2([t_1,t_2,t_3])} \lvert \tau \rangle .
\]
Hence, $\prod_t D_t$ is the projector to the space of 2-flat $\CG$-connections.
\item For every edge $e \in \Sigma_1$ and an element $a \in A$, there is an operator $C_a(e)$ that acts on the Hilbert spaces attached to the plaquettes adjacent to $e$ by
\[
C_a(e) \lvert \tau \rangle \coloneqq \lvert T_{\phi_{a,e}} \tau \rangle ,
\]
where $\phi_a(e)$ is the simple 2-gauge transformation with $(\phi_{a,e})_1(e) = a$ as defined in Example \ref{expl_simple_gauge_transformation}. Clearly we have $C_a(e) C_b(e) = C_{ab}(e)$ for $a,b \in A$. Then the operator $C_e$ defined by
\[
C_e \coloneqq \frac{1}{\lvert A \rvert} \sum_{a \in A} C_a(e)
\]
is a projector.
\item For every plaquette $p \in \Sigma_2$, there is an operator $B_p$ that examines the 1-flatness of the states on $p$. In other words, $B_p$ is defined by acts on the Hilbert spaces attached to $p$ and the boundary edges of $p$ by
\[
B_p \lvert \tau \rangle \coloneqq \delta_{\tau_1([t_0,t_2]) , \tau_1([t_1,t_2]) \tau_1([t_0,t_1])} \lvert \tau \rangle .
\]
Hence, $\prod_p B_p$ is the projector to the space of 1-flat $\CG$-connections.
%\[
%\Biggl\lvert
%\begin{tikzpicture}[scale=0.5,baseline={([yshift=-1ex]current bounding box.center)}]
%\useasboundingbox (-0.2,-0.2) rectangle (2.1,1.2) ;
%%
%\draw[->-] (0,0)--(1,1.5) node[midway,left,scale=0.8] {$h$} ;
%\draw[->-] (1,1.5)--(2,-0.5) node[midway,right,scale=0.8] {$g$} ;
%\draw[->-] (0,0)--(2,-0.5) node[midway,below,scale=0.8] {$k$} ;
%%
%\node[scale=0.8] at (1,0.5) {$a$} ;
%\end{tikzpicture}
%\Biggr\rangle
%\mapsto
%\delta_{gh,k}
%\Biggl\lvert
%\begin{tikzpicture}[scale=0.5,baseline={([yshift=-1ex]current bounding box.center)}]
%\useasboundingbox (-0.2,-0.2) rectangle (2.1,1.2) ;
%%
%\draw[->-] (0,0)--(1,1.5) node[midway,left,scale=0.8] {$h$} ;
%\draw[->-] (1,1.5)--(2,-0.5) node[midway,right,scale=0.8] {$g$} ;
%\draw[->-] (0,0)--(2,-0.5) node[midway,below,scale=0.8] {$k$} ;
%%
%\node[scale=0.8] at (1,0.5) {$a$} ;
%\end{tikzpicture}
%\Biggr\rangle .
%\]
\item For every vertex $v \in \Sigma_0$ and an element $g \in G$, there is an operator $\tilde A_g(v)$ that acts on the Hilbert spaces attached to the edges and plaquettes adjacent to $v$ by
\[
\tilde A_g(v) \lvert \tau \rangle \coloneqq \lvert T_{\phi_{g,v}} \tau \rangle ,
\]
where $\phi_{g,v}$ is the simple 1-gauge transformation with $(\phi_g(v))_0(v) = g$ as defined in Example \ref{expl_simple_gauge_transformation}. Then we define
\[
A_g(v) \coloneqq \tilde A_g(v) \cdot \prod_{v \in \partial e} C_e = \tilde A_g(v) \cdot \prod_{v \in \partial e} \frac{1}{\lvert A \rvert} \sum_{a \in A} C_a(e) .
\]
By \eqref{eq_composition_gauge_transformation_0} and \eqref{eq_composition_gauge_transformation_1} it is not hard to see that
\[
\tilde A_g(v) C_a(e) = \begin{cases}
C_{g \triangleright a}(e) \tilde A_g(v) , & v = e_1 , \\
C_a(e) \tilde A_g(v) , & \text{otherwise} .
\end{cases}
\]
Thus $\tilde A_g(v)$ commutes with $C_e$. Although $\tilde A_g(v) \tilde A_h(v)$ may not be equal to $\tilde A_{gh}(v)$, one can verify that $A_g(v) A_h(v) = A_{gh}(v)$ for $g,h \in G$. Thus the operator $A_v$ defined by
\[
A_v \coloneqq \frac{1}{\lvert G \rvert} \sum_{g \in G} A_g(v)
\]
is a projector, and $A_v$ commutes with $C_e$. Moreover, if $v \in \partial e$, we have $A_v C_e = A_v = C_e A_v$. It is not hard to verify that
\[
A_v \lvert \tau \rangle \propto \sum_\phi \lvert T_\phi \tau \rangle ,
\]
where the summation takes over all gauge transformations $\phi$ satisfying $\phi_0(w) = e$ for all 0-simplex $w \neq v$ and $\phi_1(e) = 1$ for all 1-simplex $e$ not adjacent to $v$.
\eit

It is not hard to see that the operators $A_v,B_p,C_e,D_t$ are local commuting projectors. Then the ground state subspace is the image of the projector
\[
P = \prod_{v \in \Sigma_0} A_v \prod_{p \in \Sigma_2} B_p \prod_{e \in \Sigma_1} C_e \prod_{t \in \Sigma_3} D_t = \prod_{v \in \Sigma_0} A_v \prod_{p \in \Sigma_2} B_p \prod_{t \in \Sigma_3} D_t .
\]
This projector $P$ is equal to the operator $\tilde Z(\Sigma \times I)$ defined in Section \ref{sec_TQFT_functor}
(see Example \ref{expl_cylinder_projection}). Indeed, the operator $\prod_p B_p \prod_t D_t$ is the projector to the space of flat $\CG$-connections, and given a flat $\CG$-connection $\tau$, the ground state $\prod_v A_v \lvert \tau \rangle$ is proportional to the equal-weight superposition (compare to Example \ref{expl_cylinder_projection})
\[
\sum_{\{\tau' \mid \exists \phi \colon \tau \to \tau'\}} \lvert \tau' \rangle .
\]
Also, every ground state is the linear combination of such states. Hence, the ground state subspace of the quantum double model on $\Sigma$ is the vector space $Z(\Sigma)$ assigned to $\Sigma$ in the $\CG$ gauge theory. As a corollary, the ground state degeneracy $\dim Z(\Sigma)$ is equal to the number of gauge-equivalence classes of $\CG$-connections on $\Sigma$.

So we explicitly write the Hamiltonian of the quantum double model:
\begin{equation} \label{eq_Hamiltonian_quantum_double}
H \coloneqq - \sum_{v \in \Sigma_0} A_v - \sum_{p \in \Sigma_2} B_p - \sum_{e \in \Sigma_1} C_e - \sum_{t \in \Sigma_3} D_t .
\end{equation}

\begin{rem}
For every $v \in \Sigma_0$, let $K_v$ be the closure of the star of $v \times I$ in $\Sigma \times I$. Then $K_v$ contains the $(n+1)$-simplices that have $v \times I$ as a face in the standard triangulation of $\Sigma \times I$ (see Figure \ref{fig_standard_triangulation}). Thus $\Sigma \times I = \bigcup_{v \in \Sigma_0} K_v$. We can think of this decomposition of $\Sigma \times I$ gives a decomposition of the operator
\[
\tilde Z(\Sigma \times I) = \prod_{v \in \Sigma_0} \tilde Z(K_v) ,
\]
in which $\tilde Z(K_v)$ is the operator $A_v$.
\end{rem}

\section{String-like local operators and string-like topological defects} \label{sec_string_operator_topological_defect_quantum_double}

\subsection{General discussion} \label{sec_string_topological_defect}

First we explain why the string-like topological defects are modules over the (multi-)fusion category of local operators.

Microscopically, in a concrete lattice model realization of a topological order, the topological defects can be defined as the excitations that can not be created nor annihilated by local operators. Equivalently, a topological defect can be defined as a subspace of the total Hilbert space that is invariant under the action of local operators, also called a topological sector of states. However, the notion of ``local operators'' is not very clear, and we only have some intuitions:
\bnu
\item For dimension-$k$ defects, the local operators defining them should be supported on dimension-$k$ submanifolds, or more precisely, a subregion of the lattice that looks like a dimension-$k$ submanifold macroscopically.
\item The local operators should be determined by the Hamiltonian. For example, the local operators on a smallest site should commute with the Hamiltonian.
\enu
Since a topological defect, as a subspace of the total Hilbert space, is invariant under the action of local operators, it is a module over the local operators algebra.

For particle-like (0+1D) topological defects, the 0d local operators are the usual notion of local operators. For example, in the 2+1D quantum double model for a finite group $G$, Kitaev showed that \cite{Kit03}:
\bnu
\item The local operators on a site form the quantum double algebra $D(G)$.
\item These local operators commute with the Hamiltonian. So every excited space (eigenspace of the Hamiltonian) is invariant under the action of local operators, or equivalently, a module over $D(G)$.
\item Therefore, the particle-like topological defects form the category $\rep(D(G)) \simeq \FZ_1(\rep(G))$.
\enu

In this work, we are mainly interested in the string-like (1+1D) topological defects. It turns out that a string-like topological defect is not only a subspace of the total Hilbert space, but also carries a structure of a category:
\bit
\item The objects are the states in the subspace.
\item The morphisms are the 0d local operators.
\eit
We call this category the \emph{state space} or the \emph{state category} of this string-like topological defect.

Let us illustrate the idea by the example of 1+1D Ising chain with $\Zb_2$-symmetry, which can be viewed as string-like topological defects in the 2+1D trivial phase. There is a spin on each site of the chain, so the total Hilbert space is
\[
\mathcal H_{\text{tot}} = \bigotimes_{i \in \Zb} \Cb^2 .
\]
The $\Zb_2$-symmetry is given by flipping the spin and realized by the operator
\[
U \coloneqq \prod_i X_i , \quad U^2 = 1 .
\]
This operator $U$ is a string-like local operator, and it generates a fusion category (the local operator ``algebra'') $\vect_{\Zb_2}$, which has two simple objects $\one,U$ with the $\Zb_2$ fusion rule $U \otimes U = \one$. The Hamiltonians of the Ising chain are
\begin{equation} \label{eq_Hamiltonian_Ising_1d_symmetric}
H_1 = -\sum_{i \in \Zb} X_i ,
\end{equation}
and
\begin{equation} \label{eq_Hamiltonian_Ising_1d_symmetry_broken}
H_2 = -\sum_{i \in \Zb} Z_i Z_{i+1} ,
\end{equation}
where $X_i$ and $Z_i$ are the Pauli operator on the $i$-th site.
\bit
\item For $H_1$, the ground state is unique:
\[
\lvert \Omega \rangle = \lvert \cdots +++ \cdots \rangle ,
\]
where $\lvert + \rangle$ is the eigenstate of $X$ with eigenvalue $+1$. It is the only simple object in te state category, so the state category is $\vect$. The operator $U$ acts on $\lvert \Omega \rangle$ invariantly, and this corresponds to the trivial action of $\vect_{\Zb_2}$ on $\vect$.
\item For $H_2$, the ground state subspace is two-dimensional:
\[
\lvert \Omega_\uparrow \rangle \coloneqq \lvert \cdots \uparrow \uparrow \uparrow \cdots \rangle , \quad
\lvert \Omega_\downarrow \rangle \coloneqq \lvert \cdots \downarrow \downarrow \downarrow \cdots \rangle ,
\]
where $\lvert \uparrow \rangle$ and $\lvert \downarrow \rangle$ are the eigenstates of $Z$ with eigenvalue $+1$ and $-1$, respectively. Therefore, the state category of this string-like topological defect is $\vect_{\Zb_2}$, which has two simple objects. The operator $U$ permutes the two states $\lvert \Omega_\uparrow \rangle$ and $\lvert \Omega_\downarrow \rangle$, and this corresponds to the regular action of $\vect_{\Zb_2}$ on itself. 
\eit
So we see that the actions of the string-like local operator $U$ on the ground states naturally induce the actions of the fusion category $\vect_{\Zb_2}$ on the state categories. In other words, we identify the string-like topological defects realized by the 1+1D Ising models \eqref{eq_Hamiltonian_Ising_1d_symmetric} and \eqref{eq_Hamiltonian_Ising_1d_symmetry_broken} with the modules over the local operator fusion category $\vect_{\Zb_2}$.

\begin{rem} \label{rem_fusion_category_different}
It is known that a 1+1D topological order can be described by the (multi-)fusion category of 0+1D topological defects (domain walls). This is not the same as the state category defined as above. The state category is a microscopic data and depends on the lattice model, while the (multi-)fusion category of 0+1D topological defects is a macroscopic data and independent of the lattice model. Also, the state category admits no tensor product in general. It is a module category over the local operator fusion category. In the following, we show that the (multi-)fusion category of 0+1D topological defects can be obtained from the state category as the module functor category.
\end{rem}

If we take the state categories $\CC,\CD$ of two string-like topological defects, a 0+1D topological defect between them should define a functor $F \colon \CC \to \CD$ as follows:
\bit
\item Given two states $\lvert \psi_x \rangle,\lvert \psi_y \rangle$ in two string-like topological defects, viewed as objects $x \in \CC$ and $y \in \CD$ respectively, the space of states on the 0+1D domain wall between them is the hom space $\Hom_\CD(F(x),y)$.
\eit
Usually $\CC,\CD$ are finite semisimple categories and $x,y$ are simple objects, so this hom space is also (non-canonically) isomorphic to $\Hom_\CD(y,F(x))$ and only counts the multiplicity of $y$ in $F(x)$. Note that this hom space is also isomorphic to $\Hom_\CC(F^{L/R}(y),x)$, where $F^L$ and $F^R$ are the left and right adjoint of $F$, respectively. Thus by reversing the orientation in the lattice model we can easily obtain the adjoint of $F$. Moreover, the action of string-like local operators should be transparent with respect to the 0+1D domain wall. This means that $F$ should be a module functor over the local operator fusion category.

For example, there are two simple 0+1D topological defects in the Hamiltonian $H_2$ \eqref{eq_Hamiltonian_Ising_1d_symmetry_broken}: one is the trivial defect, and the other one is the nontrivial defect realized by the domain wall states
\[
\lvert \Omega_{\uparrow\downarrow} \rangle \coloneqq \lvert \cdots \uparrow \uparrow \downarrow \downarrow \cdots \rangle , \quad
\lvert \Omega_{\downarrow\uparrow} \rangle \coloneqq \lvert \cdots \downarrow \downarrow \uparrow \uparrow \cdots \rangle .
\]
If we denote the simple objects of the state category $\vect_{\Zb_2}$ of the 1+1D Ising chain by $\Cb_\uparrow,\Cb_\downarrow$, then the domain wall states define a functor $F \colon \vect_{\Zb_2} \to \vect_{\Zb_2}$ by
\[
F(\Cb_\uparrow) = \Cb_\downarrow , \quad F(\Cb_\downarrow) = \Cb_\uparrow .
\]
These two equations correspond to the two states $\lvert \Omega_{\uparrow\downarrow} \rangle$ and $\lvert \Omega_{\downarrow\uparrow} \rangle$. This functor $F$ commutes with the action of the local operator fusion category $\vect_{\Zb_2}$ (i.e., permuting the simple object $\Cb_\uparrow$ and $\Cb_\downarrow$), so it is a $\vect_{\Zb_2}$-module functor. Clearly $F^2$ is identity, which corresponds to the trivial 0+1D topological defect. This shows that the fusion category of 0+1D topological defects in the 1+1D Ising chain \eqref{eq_Hamiltonian_Ising_1d_symmetry_broken} is also $\vect_{\Zb_2}$. As we discussed in Remark \ref{rem_fusion_category_different}, these three $\vect_{\Zb_2}$ (the fusion category of local operators, the state category, the fusion category of 0+1D topological defects) have different physical meaning.

%We claim that the 2-category of string-like topological defects in the 2+1D Ising model \eqref{eq_Hamiltonian_Ising_2d} is the 2-category of modules over the local operator fusion category $\vect_{\Zb_2}$, that is, $\LMod_{\vect_{\Zb_2}}(2\vect) \simeq 2\rep(\Zb_2)$. The following graph depicts the simple objects and hom categories of the 2-category $2\rep(\Zb_2)$:
%\[
%\xymatrix{
%\one \ar@(ul,ur)[]^{\rep(\Zb_2)}  \ar@/^/[rr]^{\vect} & & \ot \ar@(ul,ur)[]^{\vect_{\Zb_2}} \ar@/^/[ll]^{\vect}
%}
%\]
%The simple object $\one$ corresponds to the trivial 1+1D topological defect, and the simple object $\one_c$ corresponds to the 1+1D Ising chain \eqref{eq_Hamiltonian_Ising_1d}. In the above we have shown that the hom category $\Hom(\one_c,\one_c)$ is equivalent to $\vect_{\Zb_2}$. The other hom categories is left to the readers. One can also find similar hom categories in the 3+1D toric code model in Section \ref{sec_toric_code_operator_module}.

\subsection{String-like local operators in the 3+1D quantum double model}

First we define some local operators in the 3+1D quantum double model.
\bit
\item For every tetrahedron $t \in \Sigma_3$ and an element $\phi \in \hat A$, define
\begin{multline*}
D_\phi(t) \lvert \tau \rangle \coloneqq \phi \bigl(( \tau_1([t_2,t_3]) \triangleright \tau_2([t_0,t_1,t_2]) ) \cdot \tau_2([t_0,t_1,t_3])^{-1} \tau_2([t_0,t_2,t_3]) \\
\cdot \tau_2([t_1,t_2,t_3])^{-1} \alpha(\tau_1([t_2,t_3]),\tau_1([t_1,t_2]),\tau_1([t_0,t_1])) \bigr) \cdot \lvert \tau \rangle .
\end{multline*}
Then $D_\phi(t) D_\psi(t) = D_{\phi \psi}(t)$ for $\phi,\psi \in \hat A$. We also have $D_t = \lvert A \rvert^{-1} \sum_{\rho \in \hat A} D_\rho(t)$.
\item For every edge $e \in \Sigma_1$ and an element $\rho \in \hat A$, define
\[
C_\rho(e) \coloneqq \frac{1}{\lvert A \rvert} \sum_{a \in A} \rho(a)^{-1} C_a(e) .
\]
Then $C_\rho(e) C_\sigma(e) = \delta_{\rho,\sigma} C_\rho(e)$ for $\rho,\sigma \in \hat A$. We also have $C_e = C_1(e)$ where $1 \in \hat A$ is the trivial character.
\item For every pair $(v,p)$ of adjacent vertex $v \in \Sigma_0$ and plaquette $p \in \Sigma_2$, we define the \emph{holonomy} $\Phi^\tau(v,p)$ of a $\CG$-connection $\tau$ on $(v,p)$ by
\[
\Phi^\tau(v,p) \coloneqq
\begin{cases}
\tau_1([p_0,p_2])^{-1} \tau_1([p_1,p_2]) \tau_1([p_0,p_1]) , & v = p_0 , \\
\tau_1([p_0,p_1]) \tau_1([p_0,p_2])^{-1} \tau_1([p_1,p_2]) , & v = p_1 , \\
\tau_1([p_1,p_2]) \tau_1([p_0,p_1]) \tau_1([p_0,p_2])^{-1} , & v = p_2 .
\end{cases}
\]
Then $\tau$ is 1-flat if $\Phi^\tau(v,p) = e$ is trivial for every $(v,p)$. For $x \in G$, define
\[
B_x(v,p) \lvert \tau \rangle \coloneqq \delta_{\Phi^\tau(v,p) , x} \lvert \tau \rangle .
\]
Then $B_x(p) B_y(p) = \delta_{x,y} B_x(p)$ for $x,y \in G$. Note that $B_p = B_e(v,p)$ is independent of $v$. If $G$ is abelian, $B_x(v,p)$ is also independent of $v$.
\item For every vertex $v \in \Sigma_0$ and an element $g \in G$, there is an operator $\tilde A_g(v)$ defined by $\tilde A_g(v) \lvert \tau \rangle = \lvert T_{\phi_{g,v}} \tau \rangle$ (see Section \ref{sec_lattice_model_definition}).
\eit
%It is not hard to verify that all these operators commute with the Hamiltonian.

\begin{figure}[htbp]
\centering
\begin{tikzpicture}
\fill[blue!10,opacity=0.7] (0,0)--(-0.5,-0.5)--(3.5,-0.5)--(3,0)--cycle ;
\draw[thick] (0,0)--(3,0) ;
\fill (0,0) circle (0.07) node[above] {$v_0$} ;
\fill (1,0) circle (0.07) node[above] {$v_1$} ;
\fill (2,0) circle (0.07) node[above] {$v_2$} ;
\fill (3,0) circle (0.07) node[above] {$v_3$} ;
\node at (-0.7,0) {$s$} ;
\draw[thick] (0,0)--(0,-1) ;%node[midway,below right] {$e_0$} ;
\draw[thick] (1,0)--(1,-1) ;%node[midway,below right] {$e_1$} ;
\draw[thick] (2,0)--(2,-1) ;%node[midway,below right] {$e_2$} ;
\draw[thick] (3,0)--(3,-1) ;%node[midway,below right] {$e_3$} ;
\draw[dashed] (-0.5,-0.5)--(3.5,-0.5) ;
\draw[dotted] (0,0)--(-0.5,-0.5) ;
\draw[dotted] (0,0)--(0.5,-0.5) ;
\draw[dotted] (1,0)--(0.5,-0.5) ;
\draw[dotted] (1,0)--(1.5,-0.5) ;
\draw[dotted] (2,0)--(1.5,-0.5) ;
\draw[dotted] (2,0)--(2.5,-0.5) ;
\draw[dotted] (3,0)--(2.5,-0.5) ;
\draw[dotted] (3,0)--(3.5,-0.5) ;
\node[below] at (0.5,-0.5) {$p_1$} ;
\node[below] at (1.5,-0.5) {$p_2$} ;
\node[below] at (2.5,-0.5) {$p_3$} ;
\node at (-0.7,-0.5) {$\tilde s$} ;
\end{tikzpicture}
\caption{A ribbon consists of a string $s$ and a dual string $\tilde s$.}
\label{fig_ribbon}
\end{figure}
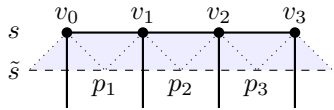

Let $s$ be a string in the lattice and $\tilde s$ be a string in the dual lattice such that they are adjacent. Their combination $(s,\tilde s)$ can be viewed as a ribbon, as depicted in Figure \ref{fig_ribbon}. %We also require that the strings $s,\tilde s$ and hence the ribbon $(s,\tilde s)$ are equipped with an orientation. 

We list the string-like local operators on this ribbon.
\bit
\item For every $g \in G$, there is a string-like local operators
\[
A_g(s) \lvert \tau \rangle \coloneqq \lvert T_{\phi_{g,s}} \tau \rangle ,
\]
where $\phi_{g,s}$ is the gauge transformation defined by $(\phi_{g,s})_1(e) = 1$ for all 1-simplex $e$ and
\[
(\phi_{g,s})_0(v) = \begin{cases}
g , & v \in s , \\
e , & v \notin s .
\end{cases}
\]
In general, $A_g(s)$ is not equal to $\prod_{v \in s} \tilde A_g(v)$. They are differed by some simple 2-gauge transformation on the edges on $s$.

%For every $g \in G$, there is a string-like local operators
%\[
%A_g(s) \coloneqq \prod_{v \in s} \tilde A_g(v) .
%\]
%Here the order of the product of operators is according to the orientation of the string $s$. As an example, if $s$ consists of 4 vertices as depicted in Figure \ref{fig_ribbon} and the orientation is from left to right, then $A_g(s) = \tilde A_g(v_3) \tilde A_g(v_2) \tilde A_g(v_1) \tilde A_g(v_0)$.
%
%By \eqref{eq_composition_gauge_transformation_0} and \eqref{eq_composition_gauge_transformation_1}, for two adjacent vertices $v,w$, the operators $\tilde A_g(v) \tilde A_g(w)$ and $\tilde A_g(w) \tilde A_g(v)$ are not equal in general, but are differed by a 0d local operator. So if we locally change the order of the product, we may obtain different operators, but they live in the same topological sector of string-like operators. 
%However, if we globally change the order of the product (for example, reverse the order), the difference

\item For every $x \in G$, there is a string-like local operators
\[
B_x(s,\tilde s) \coloneqq \prod_{(v,p) \in (s,\tilde s)} B_x(v,p) .
\]
\item For every $\rho \in \hat A$, there is a string-like local operators
\[
C_\rho(s) \coloneqq \prod_{e \in s} C_\rho(e) .
\]
\item For every $\phi \in \hat A$, there is a string-like local operators
\[
D_\phi(\tilde s) \coloneqq \prod_{t \in \tilde s} D_\phi(t) .
\]
\eit
Then we identify the operator $D_\phi(\tilde s) B_x(s,\tilde s) C_\rho(s) A_g(s)$ with the object $\Phi(x,\phi) \boxtimes (g,\rho) \in \mathcal F(\CG) \boxtimes \vect_\CG = \CD(\CG)$. Let us show that the fusion rules and the associators coincide with those of $\CD(\CG)$ (see Theorem \ref{thm_fusion_rule_quantum_double}).

First, it is easy to check the following fusion rules:
\begin{gather*}
%A_g(s) A_h(s) = A_{gh}(s) , \quad 
B_x(s,\tilde s) B_y(s,\tilde s) = \delta_{x,y} B_x(s,\tilde s) , A_g(s) B_x(s,\tilde s) = B_{gxg^{-1}}(s,\tilde s) A_g(s)  , \\
C_\rho(s) C_\sigma(s) = \delta_{\rho,\sigma} C_\rho(s) , \quad D_\phi(\tilde s) D_\psi(\tilde s) = D_{\phi \psi}(\tilde s) , \quad C_\rho(s) D_\phi(\tilde s) = D_\phi(\tilde s) C_\rho(s) , \\
A_g(s) C_\rho(s) = C_{g \triangleright \rho}(s) A_g(s) , \quad B_x(s,\tilde s) D_\phi(\tilde s) = D_\phi(\tilde s) B_x(s,\tilde s) , \\
A_g(s) D_\phi(\tilde s) = D_\phi(\tilde s) A_g(s) , \quad C_\rho(s) B_x(s,\tilde s) = B_x(s,\tilde s) C_\rho(s) .
\end{gather*}
It remains to check the fusion rule
\[
C_\rho(s) A_g(s) \otimes C_\sigma(s) A_h(s) = \delta_{\rho,g \triangleright \sigma} C_\rho(s) A_{gh}(s) .
\]
Indeed, the equation $C_\rho(s) A_g(s) \cdot C_\sigma(s) A_h(s) = \delta_{\rho,g \triangleright \sigma} C_\rho(s) A_{gh}(s)$ does not hold in general. The left hand side is equal to
\[
C_\rho(s) C_{g \triangleright \sigma}(s) A_g(s) A_h(s) = \delta_{\rho,g \triangleright \sigma} C_\rho(s) A_g(s) A_h(s) .
\]
However, $A_g(s) A_h(s)$ is not equal to $A_{gh}(s)$ in general because the gauge transformations $\phi_{g,s} \circ \phi_{h,s}$ and $\phi_{gh,s}$ may not equal. By \eqref{eq_composition_gauge_transformation_0} and \eqref{eq_composition_gauge_transformation_1}, $\phi_{g,s} \circ \phi_{h,s} = \psi \circ \phi_{gh,s}$ where $\psi$ is the composition of several simple 2-gauge transformations located at edges on $s$. So we have
\[
A_g(s) A_h(s) = \biggl(\prod_{e \in s} C_{a_e}(e) \biggr) \cdot A_{gh}(s)
\]
for some $a_e \in A$. Note that
\[
C_\rho(s) \biggl(\prod_{e \in s} C_{a_e}(e) \biggr) \cdot A_{gh}(s) = \prod_{e \in s} \rho(a_e) \cdot C_\rho(s) A_{gh}(s) \propto C_\rho(s) A_{gh}(s) .
\]
Therefore, although $C_\rho(s) A_g(s) \cdot C_\sigma(s) A_h(s)$ may not be equal to $\delta_{\rho,g \triangleright \sigma} C_\rho(s) A_{gh}(s)$, they are only differed by a scalar. So as objects in the multi-fusion category of string-like local operators, they are isomorphic. Hence we have shown that the fusion rules of these string-like operators coincide with those of $\CD(\CG)$. In particular:
\bit
\item The fusion rules of the operators $C_\rho(s) A_g(s)$ coincide with the fusion rules of $\vect_\CG$.
\item The fusion rules of the operators $D_\phi(\tilde s) B_x(s,\tilde s)$ coincide with the fusion rules of $\mathcal F(\CG)$.
\item The fusion rules between $C_\rho(s) A_g(s)$ and $D_\phi(\tilde s) B_x(s,\tilde s)$ coincide with the conjugation $\vect_\CG$-action on $\mathcal F(\CG)$.
\eit

The appearance of the associator may be mysterious because the product of operators is strictly associative. So before we compute the associator, let us explain how to `compute' the associator of a monoidal category $\CC$. Usually, the monoidal category $\CC$ is given a priori and its associator is also known. What we need to `compute' is the associator of a skeletal subcategory. More precisely, the computation includes the following steps.
\bnu
\item Choose a skeletal subcategory $\CC_0 \subseteq \CC$. This means that we need to choose an object in each isomorphism class of objects, and $\CC_0$ consists of these chosen objects. For $x \in \CC$, we denote the chosen object isomorphic to $x$ by $f(x)$.
\item Then we need to promote $f$ to an inverse of the inclusion functor $\CC_0 \hookrightarrow \CC$. This means that for every object $x \in \CC$, we need to choose an isomorphism $u_x \colon x \simeq f(x)$.
\item For $x,y \in \CC_0$, their tensor product $x \mathbin{\otimes^0} y$ in $\CC_0$ is defined to be $f(x \otimes y)$. In particular, there is an isomorphism $u_{x,y} \coloneqq u_{x \otimes y} \colon x \otimes y \to x \mathbin{\otimes^0} y$
\item For $x,y,z \in \CC_0$, the associator $\omega^0_{x,y,z}$ of $\CC_0$ is the unique isomorphism such that the following diagram commutes:
\begin{equation} \label{diag_associator_skeletal}
\begin{array}{c}
\xymatrix{
(x \otimes y) \otimes z \ar[r]^-{\omega_{x,y,z}} \ar[d]_{u_{x,y} \otimes 1} & x \otimes (y \otimes z) \ar[d]^{1 \otimes u_{y,z}} \\
(x \mathbin{\otimes^0} y) \otimes z \ar[d]_{u_{x \mathbin{\otimes^0} y,z}} & x \otimes (y \mathbin{\otimes^0} z) \ar[d]^{u_{x,y \mathbin{\otimes^0} z}} \\
(x \mathbin{\otimes^0} y) \mathbin{\otimes^0} z \ar[r]^-{\omega^0_{x,y,z}} & x \mathbin{\otimes^0} (y \mathbin{\otimes^0} z)
}
\end{array}
\end{equation}
Here $\omega$ is the associator of $\CC$. Note that $(x \mathbin{\otimes^0} y) \mathbin{\otimes^0} z$ and $x \mathbin{\otimes^0} (y \mathbin{\otimes^0} z)$ are the same object in $\CC_0$.
\enu

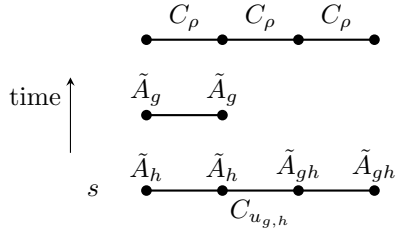
\begin{figure}[htbp]
\centering
\begin{tikzpicture}
\draw[thick] (0,0)--(3,0) ;
\fill (0,0) circle (0.07) node[above] {$\tilde A_h$} ;
\fill (1,0) circle (0.07) node[above] {$\tilde A_h$} ;
\fill (2,0) circle (0.07) node[above] {$\tilde A_{gh}$} ;
\fill (3,0) circle (0.07) node[above] {$\tilde A_{gh}$} ;
\node[below] at (1.5,0) {$C_{\tilde u_{g,h}}$} ;
\draw[thick] (0,1)--(1,1) ;
\fill (0,1) circle (0.07) node[above] {$\tilde A_g$} ;
\fill (1,1) circle (0.07) node[above] {$\tilde A_g$} ;
\draw[thick] (0,2)--(3,2) ;
\fill (0,2) circle (0.07) ;
\fill (1,2) circle (0.07) ;
\fill (2,2) circle (0.07) ;
\fill (3,2) circle (0.07) ;
\node[above] at (0.5,2) {$C_\rho$} ;
\node[above] at (1.5,2) {$C_\rho$} ;
\node[above] at (2.5,2) {$C_\rho$} ;
\draw[-stealth] (-1,0.5)--(-1,1.5) node[near end,left] {time} ;
\node at (-0.7,0) {$s$} ;
\end{tikzpicture}
\caption{The chosen isomorphim $\tilde u_{g,h} \colon A_g(s) \cdot A_h(s) \to A_{gh}(s)$. The time direction also denotes the order of the product of operators.}
\label{fig_associator_chosen_isomorphism}
\end{figure}

Now we consider the multi-fusion category of string-like local operators. The only nontrivial associator is that of $\vect_\CG$, then we need to compute the associator between the operators $C_\rho(s) A_g(s)$. The operators $C_\rho(s) A_g(s)$ are the chosen (simple) objects that form a skeletal subcategory. We have seen that the product of $C_\rho(s) A_g(s)$ and $C_\sigma(s) A_h(s)$ may not be equal to $\delta_{\rho,g \triangleright \sigma} C_\rho(s) A_{gh}(s)$, but they are isomorphic. So we need to fix an isomorphism $u_{(g,\rho),(h,g \triangleright \rho)} \colon C_\rho(s) A_g(s) \cdot C_{g \triangleright \rho}(s) A_h(s) \simeq C_\rho(s) A_{gh}(s)$. As depicted in Figure \ref{fig_associator_chosen_isomorphism}, we choose an element $\tilde u_{g,h} \in A$ and then define $u_{(g,\rho),(h,g \triangleright \rho)} \coloneqq \rho(\tilde u_{g,h})$.

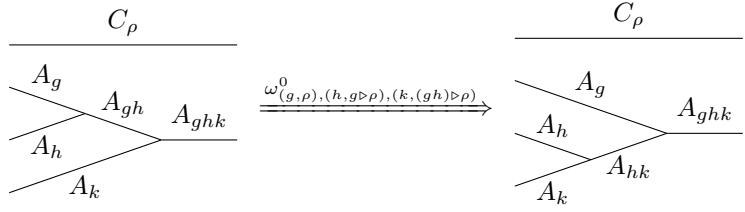
\begin{figure}[htbp]
\[
\begin{array}{c}
\begin{tikzpicture}[yscale=0.7]
\draw (0,2.8)--(3,2.8) node[midway,above] {$C_\rho$} ;
\draw (0,2)--(1,1.5) node[midway,above] {$A_g$} ;
\draw (0,1)--(1,1.5) node[midway,below] {$A_h$} ;
\draw (1,1.5)--(2,1) node[midway,above] {$A_{gh}$} ;
\draw (0,0)--(2,1) node[midway,below] {$A_k$} ;
\draw (2,1)--(3,1) node[midway,above] {$A_{ghk}$} ;
\end{tikzpicture}
\end{array}
\xRightarrow{\omega^0_{(g,\rho),(h,g \triangleright \rho),(k,(gh) \triangleright \rho)}}
\begin{array}{c}
\begin{tikzpicture}[yscale=0.7]
\draw (0,2.8)--(3,2.8) node[midway,above] {$C_\rho$} ;
\draw (0,2)--(2,1) node[midway,above] {$A_g$} ;
\draw (0,1)--(1,0.5) node[midway,above] {$A_h$} ;
\draw (0,0)--(1,0.5) node[midway,below] {$A_k$} ;
\draw (1,0.5)--(2,1) node[midway,below] {$A_{hk}$} ;
\draw (2,1)--(3,1) node[midway,above] {$A_{ghk}$} ;
\end{tikzpicture}
\end{array}
\]
\caption{The associator $\omega^0_{(g,\rho),(h,g \triangleright \rho),(k,(gh) \triangleright \rho)}$. The vertical direction is the time direction.}
\label{fig_associator}
\end{figure}

By the diagram \eqref{diag_associator_skeletal}, the associator $\omega^0_{(g,\rho),(h,g \triangleright \rho),(k,(gh) \triangleright \rho)}$ is the difference between $\rho(\tilde u_{g,h} \tilde u_{gh,k})$ and $\rho((g \triangleright \tilde u_{h,k}) \tilde u_{g,hk})$. Figure \ref{fig_associator} depicts this difference between. If we push the junctions in Figure \ref{fig_associator} to the same vertex $v$, this difference is given by the difference between the gauge transformations $\phi_{gh,v} \circ \phi_{k,v}$ and $\phi_{g,v} \circ \phi_{hk,v}$. By Example \ref{expl_associator_gauge_transformation}, this is exactly given by the associator $\alpha(g,h,k)$ of the 2-group $\CG$. Therefore, the associator $\omega^0_{(g,\rho),(h,g \triangleright \rho),(k,(gh) \triangleright \rho)}$ is equal to $(\rho \circ \alpha)(g,h,k)$, which coincides with the associator of $\vect_\CG$.

As a conclusion, we have shown that the multi-fusion category of string-like local operators is equivalent to $\CD(\CG)$. Since all of these string-like local operators commute with the Hamiltonian, as explained in Section \ref{sec_string_topological_defect}, the string-like topological defects are the modules over $\CD(\CG)$. Hence, the 2-category of string-like topological defects in the 3+1D quantum double model for $\CG$ is $2\rep(\CD(\CG)) \simeq \FZ_1(2\rep(\CG))$.

\section{Example: 3+1D toric code model} \label{sec_example}

%\subsection{3+1D toric code model}

\subsection{The 2-category of string-like topological defects}

The 3+1D toric code model \cite{HZW05} is the same as the 3+1D quantum double model for $\CG = \Zb_2$. We denote the elements of $\Zb_2$ by $\{\bfe,\bfu\}$ with $\bfu^2 = \bfe$.

In the 3+1D toric code model, there is a spin-$1/2$ on each edge of the lattice. In other words, the local Hilbert space is $\mathcal H_e = \Cb^2$ for each edge $e$. For every vertex $v$, there is an operator
\[
\mathbb A_v \coloneqq \prod_{v \in \partial e} X_e ,
\]
and for every plaquette $p$, there is an operator
\[
\mathbb B_p \coloneqq \prod_{e \in \partial p} Z_e ,
\]
where $X_e$ and $Z_e$ are Pauli $X$ and $Z$ operator acting on $\mathcal H_e$, respectively. The Hamiltonian is
\[
H \coloneqq - \sum_v \mathbb A_v - \sum_p \mathbb B_p .
\]

The local Hilbert space is isomorphic to the group algebra $\Cb[\Zb_2]$: the basis vector $\lvert \bfe \rangle, \lvert \bfu \rangle \in \Cb[\Zb_2]$ are identified with the eigenstates of Pauli $Z$ operator. So the Hilbert space of the 3+1D toric code model is isomorphic to that of the 3+1D quantum double model for $\CG = \Zb_2$. Then it is not hard to see that
\[
\mathbb A_v = A_\bfu(v) , \quad \mathbb B_p = B_\bfe(p) - B_\bfu(p) , \quad A_\bfe(v) = B_\bfe(p) + B_\bfu(p) = 1 .
\]
Hence the Hamiltonian of the 3+1D toric code model is also the same as that of the 3+1D quantum double model for $\CG = \Zb_2$ (up to a constant shift).

The topological defects in the 3+1D toric code model, including the strings and particles, form a braided fusion 2-category that is equivalent to $\FZ_1(2\rep(\Zb_2))$ \cite{KTZ20a}:
\bit
\item The objects are string-like (1+1D) topological defects. Microscopically, they are topological sectors of states, i.e., the subspaces of the total Hilbert space that are invariant under the local operator action. There are 4 simple ones:
\bit
\item The trivial defect $\one$ is generated by the ground state.
\item The $m$ string is generated by the excited state $\lvert \psi_m(\tilde s) \rangle$ supported on a string $\tilde s$ in the dual lattice, which is defined by
\[
\mathbb B_p \lvert \psi_m(\tilde s) \rangle = - \lvert \psi_m(\tilde s) \rangle , \quad \forall p \in \tilde s .
\]
\item The $\one_c$ string is generated by the ground states of the Hamiltonian
\be \label{eq_Hamiltonian_1_c}
H + \sum_{v \in s} \mathbb A_v - \sum_{e \in s} Z_e ,
\ee
where $s$ is a string in the lattice. In other words, the $\one_c$ defect is realized by remove the $\mathbb A_v$ operators and local Hilbert spaces (or equivalently, project the spins to the $Z = 1$ state) along the string $s$. The ground state subspace of \eqref{eq_Hamiltonian_1_c} is
\[
V_s \coloneqq \{\lvert \psi \rangle \mid \mathbb A_v \lvert \psi \rangle = \lvert \psi \rangle , \, \forall v \notin s , \, \mathbb B_p \lvert \psi \rangle = \lvert \psi \rangle , \, \forall p , \, Z_e \lvert \psi \rangle = \lvert \psi \rangle , \, \forall e \in s\} .
\]
We also define
\[
W_s \coloneqq \{\lvert \psi \rangle \mid \mathbb A_v \lvert \psi \rangle = \lvert \psi \rangle , \, \forall v \notin s , \, \mathbb B_p \lvert \psi \rangle = \lvert \psi \rangle , \, \forall p\} .
\]
Then $V_s$ is a subspace of $W_s$. The Hamiltonian \eqref{eq_Hamiltonian_1_c}, or simply $H_{\text{eff}} = - \sum_{e \in s} Z_e$, can be viewed as an effective Hamiltonian on $W_s$ with the ground state subspace $V_s$. Note that $\mathbb A_v$ anti-commutes with $Z_e$ if $v$ is adjacent to $e$. Thus this effective Hamiltonian $H_{\text{eff}}$ can be mapped to the symmetry-breaking Hamiltonian of the Ising chain $H_2 = -\sum_i Z_i Z_{i+1}$ (see \eqref{eq_Hamiltonian_Ising_1d_symmetry_broken}):
\[
\mathbb A_v \rightsquigarrow X_i , \quad Z_e \rightsquigarrow Z_i Z_{i+1} .
\]
Therefore, the ground state subspace $V_s$ is 2-dimensional, and the operator $\prod_{v \in s} \mathbb A_v$ exchanges two ground states.
\item The $m_c$ string is the fusion of $\one_c$ and $m$ string.
\eit
\item The 1-morphisms are particle-like (0+1D) topological defects. We list the simple 1-morphisms:
\bit
\item There are two simple 1-morphisms between $\one$ and itself. One is the trivial particle $1_\one$. Another one is the $e$ particle generated by the excited state $\lvert \psi_e(v) \rangle$ supported on a vertex $v$, which is defined by
\[
\mathbb A_v \lvert \psi_e(v) \rangle = - \lvert \psi_e(v) \rangle .
\]
\item There are two simple 1-morphisms between $\one_c$ and itself. One is the trivial particle $1_{\one_c}$. Another one is the $z$ particle generated by the subspace
\[
\{X_e \lvert \psi \rangle \mid \lvert \psi \rangle \text{ is a ground state of \eqref{eq_Hamiltonian_1_c}}\} ,
\]
where $e$ is an edge in the string $s$. In other words, the $z$ particle is given by reversing the spin at $e$ to the $Z = -1$ state. After mapped to the Ising chain, the $z$ particle is generated by the domain wall states (see Section \ref{sec_string_topological_defect}).
\item There is only one simple 1-morphism between $\one$ and $\one_c$. Depending on the orientation, we denote it by $x \colon \one \to \one_c$ and $y \colon \one_c \to \one$.
\item There is no non-zero 1-morphism between $\one,\one_c$ and $m,m_c$, because the $m$ string can not end due to the constraint $\prod_{p \in \partial t} \mathbb B_p = 1$ for every 3-cell $t$. The 1-morphisms on $m,m_c$ strings can be obtained from the above 1-morphisms by attaching an $m$ string.
\eit
\item The 2-morphisms are 0D topological defects (operators) in the spacetime. By Schur's lemma, the morphism space between two simple 1-morphisms $f,g$ is $\Cb$ if $f \simeq g$ and $0$ otherwise.
\item The fusion rule is given by
\[
m \otimes m = \one , \quad \one_c \otimes \one_c = \one_c \oplus \one_c , \quad m \otimes \one_c = \one_c \otimes m = m_c .
\]
\item It is not easy to describe the braiding structure. We refer readers to \cite{KTZ20a} for the details. The simplest nontrivial braiding is that the double braiding of $e$ particle and $m$ string is $-1$.
\eit
The simple objects and simple 1-morphisms in is 2-category $\FZ_1(2\rep(\Zb_2))$ are depicted in the following graph:
\[
\xymatrix{
\one \ar@(ul,ur)[]^{\rep(\Zb_2) = \langle 1_\one,e \rangle}  \ar@/^/[rr]^{\vect = \langle x \rangle} & & \ot \ar@(ul,ur)[]^{\vect_{\Zb_2} = \langle 1_{\one_c},z \rangle} \ar@/^/[ll]^{\vect = \langle y \rangle}
& & m \ar@(ul,ur)[]^{\rep(\Zb_2) = \langle 1_m,e_m \rangle}  \ar@/^/[rr]^{\vect = \langle x_m \rangle} & & \mt \ar@(ul,ur)[]^{\vect_{\Zb_2} = \langle 1_{m_c},z_m \rangle} \ar@/^/[ll]^{\vect = \langle y_m \rangle}
} .
\]
Note that both two connected components are equivalent to $2\rep(\Zb_2) \simeq \LMod_{\vect_{\Zb_2}}(2\vect)$ as 2-categories.

\subsection{String-like local operators and the modules} \label{sec_toric_code_operator_module}

We list the string-like local operators in the 3+1D toric code model:
\bit
\item For a string $s$ in the lattice, there are string operators
\[
A_\bfe(s) = 1 , \quad A_\bfu(s) = \prod_{v \in s} A_\bfu(s) = \prod_{v \in s} \mathbb A_v .
\]
\item For a string $\tilde s$ in the dual lattice, there are string operators
\[
B_\bfe(\tilde s) = \prod_{p \in \tilde s} B_\bfe(p) = \prod_{p \in \tilde s} \frac{1 + \mathbb B_p}{2} , \quad B_\bfu(\tilde s) = \prod_{p \in \tilde s} B_\bfu(p) = \prod_{p \in \tilde s} \frac{1 - \mathbb B_p}{2} .
\]
\eit
The fusion rule is given by
\[
A_\bfu(s)^2 = 1 , \quad B_\bfe(\tilde s)^2 = B_\bfe(\tilde s) , \quad B_\bfu(\tilde s)^2 = B_\bfu(\tilde s) , \quad B_\bfe(s) B_\bfu(\tilde s) = B_\bfu(\tilde s) B_\bfe(s) = 0 .
\]
Therefore, the operators $A_\bfe(s),A_\bfu(s)$ form a fusion category that is equivalent to $\vect_{\Zb_2}$, and the operators $B_\bfe(t)$ and $B_\bfu(t)$ form a multi-fusion category that is equivalent to $\fun(\Zb_2,\vect) \simeq \vect \oplus \vect$. Thus all the string-like local operators form a multi-fusion category that is equivalent to $\vect_{\Zb_2} \boxtimes \fun(\Zb_2,\vect) \simeq \vect_{\Zb_2} \oplus \vect_{\Zb_2}$, denoted by $\CD(\Zb_2)$ or $\CD(\vect_{\Zb_2})$.

We denote the simple objects of $\vect_{\Zb_2}$ by $\varepsilon_\bfe,\varepsilon_\bfu$, and the simple objects of $\fun(\Zb_2,\vect)$ by $\delta_\bfe,\delta_\bfu$. Then the simple objects of $\CD(\Zb_2)$ are $\varepsilon_x \delta_y$ for $x,y = \bfe,\bfu$. We list the simple modules over $\CD(\Zb_2)$:
\bit
\item $\vect^\bfe$: there is only one simple object $\varepsilon$, and the action is given by
\[
\varepsilon_\bfu \odot \varepsilon = \varepsilon , \quad \delta_\bfe \odot \varepsilon = \varepsilon , \quad \delta_\bfu \odot \varepsilon = 0 .
\]
\item $\vect_{\Zb_2}^\bfe$: there are two simple objects $\varepsilon_\bfe,\varepsilon_\bfu$, and the action is given by
\[
\varepsilon_\bfu \odot \varepsilon_x = \varepsilon_{\bfu \cdot x} , \quad \delta_e \odot \varepsilon_x = \varepsilon_x , \quad \delta_\bfu \odot \varepsilon_x = 0 , \quad \forall x = \bfe, \bfu .
\]
\item $\vect^\bfu$: there is only one simple object $\varepsilon$, and the action is given by
\[
\varepsilon_\bfu \odot \varepsilon = \varepsilon , \quad \delta_\bfe \odot \varepsilon = 0 , \quad \delta_\bfu \odot \varepsilon = \varepsilon .
\]
\item $\vect_{\Zb_2}^\bfu$: there are two simple objects $\varepsilon_\bfe,\varepsilon_\bfu$, and the action is given by
\[
\varepsilon_\bfu \odot \varepsilon_x = \varepsilon_{\bfu \cdot x} , \quad \delta_e \odot \varepsilon_x = 0 , \quad \delta_\bfu \odot \varepsilon_x = \varepsilon_x , \quad \forall x = \bfe, \bfu .
\]
\eit
These simple modules are the state categories of the simple string-like topological defects $\one,\one_c,m,m_c$ respectively:
\bit
\item The trivial string $\one$ is generated by the ground state, which corresponds to the simple object of $\vect^\bfe$. The operator $A_\bfu(s)$ acts on the ground state as identity, $B_\bfe(\tilde s)$ acts as identity, and $B_\bfu(\tilde s)$ acts as zero. This coincides with the module action of $\CD(\Zb_2)$ on $\vect^\bfe$.
\item The $\one_c$ string is generated by the two ground states of the Hamiltonian \eqref{eq_Hamiltonian_1_c}, which correspond to the two simple objects of $\vect_{\Zb_2}^\bfe$. The action of operator $A_\bfu(s)$ exchanges two ground states, and this coincides with the module action of $\varepsilon_\bfu \in \CD(\Zb_2)$ on $\vect_{\Zb_2}^\bfe$.
\item The $m$ string is generated by a single state, on which $A_\bfu(s)$ and $B_\bfu(\tilde s)$ act as identity and $B_\bfe(\tilde s)$ acts as zero.
\item The $m_c$ string is generated by two states, on which $A_\bfu(s)$ acts as exchanging, $B_\bfe(\tilde s)$ acts as zero, and $B_\bfu(\tilde s)$ acts as identity.
\eit
This explains that the string-like topological defects in the 3+1D toric code model are modules over the category $\CD(\Zb_2)$ of string-like local operators.

We also list the simple $\CD(\Zb_2)$-module functors between these simple modules and compare them to the 0+1D topological defects:
\bit
\item There are two simple $\CD(\Zb_2)$-module functors $F_\pm \colon \vect^\bfe \to \vect^\bfe$. Their underlying functors are both the identity functor, but their $\CD(\Zb_2)$-module structures are different:
\begin{equation} \label{eq_module_functor_e}
F_\pm(\varepsilon_\bfu \odot \varepsilon) \xrightarrow{\pm 1} \varepsilon_\bfu \odot F_\pm(\varepsilon) .
\end{equation}
Therefore, $F_+$ is the identity module functor and $(F_-)^2 = F_+$. Indeed, we have the equivalence of fusion categories $\fun_{\CD(\Zb_2)}(\vect^\bfe,\vect^\bfe) \simeq \rep(\Zb_2)$.

The fact that the underlying functors of $F_\pm$ are identity corresponds to that both $1_\one$ and $e$ particles are generated by a single state:
\[
\Hom_{\vect^\bfe}(F_\pm(\varepsilon),\varepsilon) = \Cb .
\]
The different module structures on $F_\pm$ corresponds to the different actions of the local operator $A_\bfu$ on the two states. For example, the minus sign in \eqref{eq_module_functor_e} corresponds to the minus sign in
\[
A_\bfu(s) \lvert \psi_e(v) \rangle = \prod_{v' \in s} \mathbb A_{v'} \lvert \psi_e(v) \rangle = - \lvert \psi_e(v) \rangle .
\]

\item There are two simple $\CD(\Zb_2)$-module functors $\varepsilon_\bfe \otimes - , \varepsilon_\bfu \otimes - \colon \vect_{\Zb_2}^\bfe \to \vect_{\Zb_2}^\bfe$. The functor $\varepsilon_\bfe \otimes -$ is the identity module functor, and $\varepsilon_\bfu \otimes -$ permutes two simple objects of $\vect_{\Zb_2}^\bfe$. This corresponds to the fact that the states on the two sides of a $z$ particle are different. In the 3+1D toric code model this is not very clear. After mapping the $\one_c$ string to the 1+1D Ising chain, and the $z$ particle is mapped to the domain wall states. Then this correspondence is obvious (see Section \ref{sec_string_topological_defect}). Thus we have the equivalence of fusion categories $\fun_{\CD(\Zb_2)}(\vect_{\Zb_2}^\bfe,\vect_{\Zb_2}^\bfe) \simeq \vect_{\Zb_2}$.

\item There is only one simple $\CD(\Zb_2)$-module functors between $\vect^\bfe$ and $\vect_{\Zb_2}^\bfe$. The functor $\vect^\bfe \to \vect_{\Zb_2}^\bfe$ maps $\varepsilon$ to $\varepsilon_\bfe \oplus \varepsilon_\bfu$, and the functor $\vect_{\Zb_2}^\bfe \to \vect^\bfe$ maps both two simple object $\varepsilon_\bfe,\varepsilon_\bfu$ to $\varepsilon$. They correspond to the fact that the $x$ or $y$ particle connects the single state in $\one$ (the ground state) to the two states in $\one_c$.

\item The $\CD(\Zb_2)$-module functors between $\vect^\bfu,\vect_{\Zb_2}^\bfu$ are similar.

\item There is no non-zero $\CD(\Zb_2)$-module functor between $\vect^\bfe,\vect_{\Zb_2}^\bfe$ and $\vect^\bfu,\vect_{\Zb_2}^\bfu$. For example, a $\CD(\Zb_2)$-module functor $F \colon \vect_{\Zb_2}^\bfe \to \vect^\bfu$ should be equipped with an isomorphism
\[
F(\delta_\bfu \odot x) \to \delta_\bfu \odot F(x)
\]
for every $x \in \vect_{\Zb_2}^\bfe$. However, the left hand side must be $0$, and the right hand side is $F(x)$. The incompatibility of the actions of $\delta_\bfe,\delta_\bfu$ on the two sides corresponds to the fact that the $m$ string can not end.
\eit

\subsection{3+1D dual toric code model}

Now we consider the 3+1D dual toric code model. It is obtained from the 3+1D toric code model by taking the dual lattice and rotating the local spins. It is the same as the 3+1D quantum double model for $\CG = \mathrm B \Zb_2$.

In the 3+1D dual toric code model, there is a spin-$1/2$ on each plaquette of the lattice. So the local Hilbert space is $\mathcal H_p = \Cb^2$ for each plaquette $p$. For every edge $e$, there is an operator
\[
\mathbb C_e \coloneqq \prod_{e \in \partial p} X_p ,
\]
and for every cube $t$, there is an operator
\[
\mathbb D_t \coloneqq \prod_{p \in \partial t} Z_p .
\]
The Hamiltonian is
\[
H \coloneqq - \sum_e \mathbb C_e - \sum_t \mathbb D_t .
\]

The local Hilbert space is isomorphic to the group algebra $\Cb[\Zb_2]$. Then we have
\[
\mathbb C_e = C_\bfu(e) , \quad \mathbb D_t = D_\bfe(t) - D_\bfu(t) , \quad C_\bfe(e) = D_\bfe(t) + D_\bfu(t) = 1 .
\]
Hence the Hamiltonian of the 3+1D dual toric code model is the same as that of the 3+1D quantum double model for $\CG = \mathrm B \Zb_2$ (up to a constant shift). We denote the elements in $\widehat{\Zb_2}$ by $\{1,\varphi\}$ with $\varphi^2 = 1$. Then
\[
C_1(e) = \frac12(1 + \mathbb C_e) , \quad C_\varphi(e) = \frac12(1 - \mathbb C_e) , \quad D_1(t) = 1 , \quad D_\varphi(t) = \mathbb D_t .
\]

The topological defects are the same as those in the 3+1D toric code model. There are still 4 simple string-like topological defects:
\bit
\item the trivial string $\one$;
\item the $\one_c$ string obtained by removing the $\mathbb D_t$ operators and spins along a string $\tilde s$ in the dual lattice;
\item the $m$ string generated by the excited state with $\mathbb C_e = -1$ for all $e$ in a string $s$ in the lattice;
\item the fusion of $m$ and $\one_c$.
\eit

We list the string-like operators in the 3+1D dual toric code model:
\bit
\item For a string $s$ in the lattice, there are string operators
\[
C_1(s) = \prod_{e \in s} C_1(e) = \prod_{e \in s} \frac{1 + \mathbb C_e}{2} , \quad C_\varphi(s) = \prod_{e \in s} C_\varphi(e) = \prod_{e \in s} \frac{1 - \mathbb C_e}{2} .
\]
\item For a string $\tilde s$ in the dual lattice, there are string operators
\[
D_1(\tilde s) = 1 , \quad D_\varphi(\tilde s) = \prod_{t \in \tilde s} D_\varphi(t) = \prod_{t \in \tilde s} \mathbb D_t .
\]
\eit
The fusion rule is given by
\[
C_1(s)^2 = C_1(s) , \quad C_\varphi(s)^2 = C_\varphi(s) , \quad C_1(s) C_\varphi(s) = C_\varphi(s) C_1(s) = 0 , \quad D_\varphi(\tilde s)^2 = 1 .
\]
Therefore, the operators $C_1(s) , C_\varphi(s)$ form a multi-fusion category that is equivalent to $\fun(\widehat{\Zb_2},\vect) \simeq \vect \oplus \vect$, and the operators $D_1(\tilde s) , D_\varphi(\tilde s)$ form a fusion category that is equivalent to $\vect_{\widehat{\Zb_2}} \simeq \rep(\Zb_2)$. Thus all the string-like local operators form a multi-fusion category that is equivalent to $\fun(\widehat{\Zb_2},\vect) \boxtimes  \vect_{\widehat{\Zb_2}} \simeq \vect_{\widehat{\Zb_2}} \oplus \vect_{\widehat{\Zb_2}}$, denoted by $\CD(\vect_{\mathrm B \Zb_2})$. It is equivalent to $\CD(\Zb_2)$, and their module 2-categories are also equivalent.

\begin{rem}
By \eqref{eq_parition_function_sphere}, the partition function of the dual toric code model on the 4-sphere $S^4$ is $2$, but that of the toric code model on $S^4$ is $1/2$. However, on the lattice model there is no essential difference between these two models.
\end{rem}

\appendix

\section{Simplicial sets and cohomology} \label{appendix_cohomology}

In this appendix we recall some basic notion of simplicial sets and their cohomology, including the cohomology of groups and 2-groups.

\medskip
For any $n \in \Nb$, define $[n]$ to be the set $\{0,\ldots,n\}$ with the usual total order. Let $\Delta$ be the category whose objects are $[n]$ for all $n \in \Nb$ and morphisms are order-preserving maps.

\begin{defn}
Let $\CC$ be a category. A \emph{simplicial object in $\CC$} is a functor $\Delta^\op \to \CC$. In particular, a \emph{simplicial set} $X$ is a functor $X \colon \Delta^\op \to \Set$. %An element $x \in X_n := X([n])$ is called an \emph{$n$-simplex}. 
A \emph{simplicial map} between two simplicial object $X,Y \colon \Delta^\op \to \CC$ is a natural transformation $F \colon X \to Y$. The category of simplicial objects in $\CC$ and simplicial maps between them is denoted by $\mathrm s \CC$. In particular, the category of simplicial sets is denoted by $\sSet$.
\end{defn}

%\begin{rem}
%We can also consider the category of finite totally ordered sets $\hat \Delta$. Clearly $\Delta \hookrightarrow \hat \Delta$ is an equivalence, thus we can also define a simplicial set to be a functor $\hat \Delta^\op \to \Set$. In some cases this definition will be more convenient.
%\end{rem}

The morphisms in $\Delta$ can be written as the composition of some `basic' morphisms (called face maps and degeneracy maps). So there is a more concrete equivalent definition of simplicial objects.

\begin{defn}
A simplicial object $X$ in a category $\CC$ consists of the following data:
\bnu
\item a sequence of objects $\{X_n \in \CC\}_{n \in \Nb}$,
\item morphisms $d_i \colon X_n \to X_{n-1} \, (i = 0 , \ldots , n)$ for any $n \geq 1$, called the face maps,
\item morphisms $s_i \colon X_n \to X_{n+1} \, (i = 0 , \ldots , n)$ for any $n \geq 0$, called the degeneracy maps, 
\enu
and they satisfy the following identities:
\bnu
\item $d_i d_j = d_{j-1} d_i$ if $i < j$,
\item $s_i s_j = s_{j+1} s_i$ if $i \leq j$,
\item $d_i s_j = s_{j-1} d_i$ if $i < j$,
\item $d_i s_j = \id$ if $i = j$ or $i = j + 1$,
\item $d_i s_j = s_j d_{i-1}$ if $i > j + 1$.
\enu
\end{defn}

%\begin{rem}
%Dually, a cosimplicial object in a category $\CC$ is a functor $\Delta \to \CC$, and it can also be written as a sequence of sets with coface maps and codegeneracy maps satisfying the dual identities. Also we can define semisimplicial objects by deleting degeneracy maps in the definition of simplicial sets.
%\end{rem}

%\begin{defn}
%Let $X$ be a simplicial set. We say an $n$-simplex $x \in X_n$ is nondegenerate if it is not the image of an $(n-1)$-simplex under degeneracy maps.
%\end{defn}

\begin{defn} \label{defn_homology_simplicial_abelian_object}
Let $\CA$ be an abelian category and $M \in \mathrm{s} \CA$ be a simplicial object in $\CA$. The (unnormalized) chain complex $C_{\bullet}(M)$ associated to $M$ is defined as follows:
\bit
\item $C_n(M) \coloneqq M_n$ for all $n \in \Nb$;
\item the differential map $\partial_n \colon C_n(M) \to C_{n-1}(M)$ is the alternative sum of face maps
\[
\partial_n \coloneqq \sum_{i = 0}^n (-1)^i d_i \colon M_n \to M_{n-1} .
\]
\eit 
The \emph{homology} $H_{\bullet}(M)$ of $M$ are the homology of the chain complex $C_{\bullet}(M)$. More precisely, $H_n(M) \coloneqq \ker(\partial_n)/\mathrm{im}(\partial_{n+1})$ for all $n \in \Nb$.
\end{defn}

Let us check that $\partial_n \circ \partial_{n+1} = 0$. By the identity $d_i d_j = d_{j-1} d_i$ for $i < j$ we have
\begin{align*}
\partial_n \circ \partial_{n+1} & = \sum_{i = 0}^n \sum_{j = 0}^{n+1} (-1)^{i+j} d_i d_j = \sum_{i = 0}^n \left(\sum_{j = 0}^i + \sum_{j = i+1}^{n+1} \right) (-1)^{i+j} d_i d_j \\
& = \sum_{i = 0}^n \sum_{j = 0}^i (-1)^{i+j} d_i d_j + \sum_{i = 0}^n \sum_{j = i+1}^{n+1} (-1)^{i+j} d_{j-1} d_i \\
& = \sum_{i = 0}^n \sum_{j = 0}^i (-1)^{i+j} d_i d_j + \sum_{i = 0}^n \sum_{k = i}^n (-1)^{i+k+1} d_k d_i \\
& = \sum_{0 \leq j \leq i \leq n} (-1)^{i+j} d_i d_j - \sum_{0 \leq i \leq k \leq n} (-1)^{k+i} d_k d_i = 0 .
\end{align*}

Let $A$ be an abelian group. Given a simplicial set $X$, by applying the functor
\[
\Hom_{\set}(-,A) \colon \set^\op \to \Ab ,
\]
we obtain a simplicial object $\Hom(X,A)$ in the category $\Ab^\op$. Then the construction in Definition \ref{defn_homology_simplicial_abelian_object} gives rise to a cochain complex of abelian groups
\[
\xymatrix{
\cdots \ar[r] & \Hom(X_{n-1},A) \ar[r] & \Hom(X_n,A) \ar[r] & \Hom(X_{n+1},A) \ar[r] & \cdots
}
\]
The cohomology groups of this cochain complex are called the cohomology groups of $X$ with coefficients in $A$. This recovers many classical cohomology theory.

\begin{expl}
A \emph{simplicial complex} $K$ is a pair $K = (V(K),S(K))$, where $V(K)$ is a set, whose elements are called the \emph{vertices} of $K$, and $S(K)$ is a set of non-empty finite subsets of $V(K)$ satisfying the following conditions:
\bnu
\item $\{v\} \in S(K)$ for any $v \in V(K)$;
\item If $s \in S(K)$ and $\emptyset \neq t \subset s$, then $t \in S(K)$.
\enu
The elements of $S(K)$ are called \emph{simplices} of $K$. If $s \in S(K)$, any non-empty subset $t \subset s$ is called a \emph{face} of $s$. A simplex $s \in S(K)$ containing $(n+1)$ vertices is called an \emph{$n$-simplex}. The set of $n$-simplices of $K$ is denoted by $S(K)_n$. An \emph{ordered simplicial complex} is a simplicial complex $K$ with a total order on $V(K)$.

Given an ordered simplicial complex $K$, we can construct a simplicial set $\hat K$ as follows:
\bit
\item For each $n \in \Nb$, the set $\hat K_n$ consists of ordered $(n+1)$-tuples $(v_0,\ldots,v_n)$ where $v_0 \leq \cdots \leq v_n \in V(K)$, such that the underlying set $\{v_0,\ldots,v_n\} \in S(K)$ is a simplex of $K$.
\item For each morphism $f \colon [m] \to [n]$ in $\Delta$, the map $\hat K_f \colon \hat K_n \to \hat K_m$ maps $(v_0,\ldots,v_n)$ to $(v_{f(0)},\ldots,v_{f(m)})$. In particular, the face maps and the degeneracy maps are
\begin{align*}
d_i(v_0,\ldots,v_n) & = (v_0,\ldots,\hat{v_i},\ldots,v_n) , \\
s_i(v_0,\ldots,v_n) & = (v_0,\ldots,v_i,v_i,\ldots,v_n) ,
\end{align*}
where $\hat{v_i}$ means deleting this term.
\eit

For any abelian group $A$, the cochain complex $C^\bullet(X,A) \coloneqq \Hom(X_\bullet,A)$ is the cochain complex of the simplicial cohomology of $X$ with coefficients in $A$.
\end{expl}

\begin{expl}
The topological $n$-simplex $\Delta^n$ is the subspace of $\Rb^{n+1}$ defined by
\[
\Delta^n \coloneqq \{(x_0,\ldots,x_n) \mid \sum_{i = 0}^n x_i = 1 , \, 0 \leq x_i \leq 1\} .
\]
Equivalently, it is the convex hull of $e_0 \coloneqq (1,0,\ldots,0),\ldots,e_n \coloneqq (0,\ldots,0,1)$. We can identify the elements $i$ of $[n]$ with the vertices $e_i$ of $\Delta^n$, and extend any map $f \colon [n] \to [m]$ in $\Delta$ by linearity to get a continuous map $\Delta^f \colon \Delta^n \to \Delta^m$. This defines a functor $\Delta \to \Topo$ where $\Topo$ is the category of topological spaces, that is, a simplicial object in $\Topo^\op$. More explicitly, let us write the coface maps and codegeneracy maps $\delta_i \colon \Delta^{n-1} \to \Delta^n$ and $\sigma_i \colon \Delta_{n+1} \to \Delta_n$ for all $0 \leq i \leq n$ as follows:
\begin{align*}
\delta_i(x_0,\ldots,x_{n-1}) & = (x_0,\ldots,x_{i-1},0,x_i,\ldots,x_{n-1}) , \\
\sigma_i(x_0,\ldots,x_{n+1}) & = (x_0,\ldots,x_i + x_{i+1},\ldots,x_{n+1}) .
\end{align*}

For any topological space $T$, the functor $\Hom_{\Topo}(-,T) \colon \Topo^\op \to \set$ maps the above simplicial object in $\Topo^\op$ to a simplicial set, denoted by $\Sing T$, called the \emph{singular set} of $T$. Thus an $n$-simplex of $\Sing T$ is a continuous map $\Delta^n \to T$. For any abelian group $A$, the cochain complex $C^\bullet(T,A) \coloneqq \Hom((\Sing T)_\bullet,A)$ is the cochain complex of the singular cohomology of $T$ with coefficients in $A$.
\end{expl}

\begin{expl}
Let $\CC$ be a category. The \emph{nerve} $N(\CC)$ of $\CC$ is the simplicial set $N(\CC) \coloneqq \hom_{\cat}(-,\CC) \colon \Delta^\op \to \Set$, where all totally ordered sets $[n]$ are viewed as categories in a natural way. In other words, $N(\CC)_0$ is the set of all objects in $\CC$, and $N(\CC)_n$ for $n \geq 1$ is the set of $n$-tuples of composable morphisms in $\CC$. %In particular, $\Delta[n]$ is defined to be the representable presheaf $\Delta[n] := \hom_\Delta(-,[n])$.

Let $G$ be a group. Its \emph{one-point delooping} $\mathrm B G$ is the category with only one object $\ast$ and the hom space $\Hom(\ast,\ast) \coloneqq G$. Then its nerve is a simplicial set $N(\mathrm B G)$ with $N(\mathrm B G)_n = G^{\times n}$, and the face maps and degeneracy maps are
\begin{align*}
d_0(g_1,\ldots,g_n) & = (g_2,\ldots,g_n) , \\
d_i(g_1,\ldots,g_n) & = (g_1,\ldots,g_i g_{i+1},\ldots,g_n) , \, 0 < i < n , \\
d_n(g_1,\ldots,g_n) & = (g_1,\ldots,g_{n-1}) , \\
s_i(g_1,\ldots,g_n) & = (g_1,\ldots,g_i,e,g_{i+1},\ldots,g_n) , \, 0 \leq i \leq n .
\end{align*}
For any abelian group $A$, the cochain complex $C^\bullet(G,A) \coloneqq \Hom(N(\mathrm B G)_\bullet,A)$ is the usual chain complex (bar resolution) of the group cohomology of $G$ with coefficients in $A$.

More generally, we can define the cohomology of a groupoid in a similar way.
\end{expl}

\begin{expl}
Let $\SC$ be a 2-category. The \emph{Duskin nerve} $N(\SC)$ of $\SC$ is the simplicial set $N(\SC) \coloneqq \hom_{2\cat}(-,\SC) \colon \Delta^\op \to \Set$, where all totally ordered sets $[n]$ are viewed as 2-categories in a natural way. In other words, $N(\SC)_0$ is the set of all objects in $\SC$, $N(\SC)_1$ is the set of all 1-morphisms in $\SC$, and $N(\SC)_n$ for $n \geq 2$ is the set of $n$-simplices filled by commutative diagrams of 2-morphisms in $\SC$.

Let $\CG$ be a 2-group. Its \emph{one-point delooping} $\mathrm B \CG$ is the 2-category with only one object $\ast$ and the hom category $\Hom(\ast,\ast) \coloneqq \CG$. Then its Duskin nerve is a simplicial set $N(\mathrm B \CG)$. For any abelian group $A$, the cochain complex $C^\bullet(\CG,A) \coloneqq \Hom(N(\mathrm B \CG)_\bullet,A)$ is the chain complex of the cohomology of $\CG$ with coefficients in $A$. When $\CG = G$ is a 1-group, this recovers the usual group cohomology.

More generally, we can define the cohomology of a 2-groupoid in a similar way.
\end{expl}

\section{Triangulation approach to gauge transformations} \label{appendix_gauge_transformation_triangulation}

In this appendix, we study the gauge transformation of flat 2-group connections on $M$ (see Section \ref{sec_gauge_transformation_2-group}) by using the triangulation of $M \times I$.

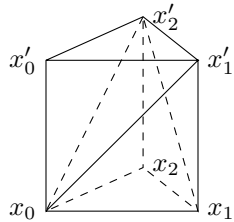
\begin{figure}[htbp]
\centering
\begin{tikzpicture}
\coordinate (a) at (0,0,0) ;
\coordinate (b) at (2,0,0) ;
\coordinate (c) at (0.7,0,-1.5) ;
\coordinate (d) at (0,2,0) ;
\coordinate (e) at (2,2,0) ;
\coordinate (f) at (0.7,2,-1.5) ;
\node[left] at (a) {$x_0$} ;
\node[right] at (b) {$x_1$} ;
\node[right] at (c) {$x_2$} ;
\node[left] at (d) {$x_0'$} ;
\node[right] at (e) {$x_1'$} ;
\node[right] at (f) {$x_2'$} ;
\draw (a)--(b)--(e)--(a)--(d)--(f)--(e)--(d) ;
\draw[dashed] (f)--(a)--(c)--(f)--(b)--(c) ;
\end{tikzpicture}
\caption{The standard triangulation of $x \times I$ for a 2-simplex $x$. Here we write $x_j$ for $(x_j,0)$ and $x_j'$ for $(x_j,1)$.}
\label{fig_standard_triangulation}
\end{figure}

\medskip
First, we take the following `standard' triangulation on $M$:
\bit
\item The subspaces $M \times \{0\}$ and $M \times \{1\}$ are equipped with the same triangulation with $M$.
\item If $x \in M_k$ is a $k$-simplex, $x \times I$ is decomposed as the union of $(k+1)$-simplices:
\[
x \times I = \bigcup_{j=0}^k [(x_0,0),\ldots,(x_j,0),(x_j,1),\ldots,(x_k,1)] .
\]
When $k = 2$, this decomposition is depicted in Figure \ref{fig_standard_triangulation}. %We also denote $K_x^j \coloneqq [(x_0,0),\ldots,(x_j,0),(x_j,1),\ldots,(x_k,1)]$. In particular, the set of $(n+1)$-simplices of $M \times I$ is
%\[
%(M \times I)_{n+1} = \{K_x^j \mid x \in M_n , \, j = 0,\ldots, n\} .
%\]
%For every $v \in M_0$, we also denote
%\be \label{eq_decomposition_cylinder_vertex}
%K_v \coloneqq \bigcup_{\substack{v = x_j \\ x \in M_n}} K_x^j .
%\ee
%Then $M \times I = \bigcup_{v \in M_0} K_v$.
\eit

In particular, the set of 1-simplices of $M \times I$ is
\begin{multline*}
(M \times I)_1 = \{e \times \{0\} \mid e \in M_1\} \sqcup \{e \times \{1\} \mid e \in M_1\} \\
\sqcup \{[(v,0),(v,1)] \mid v \in M_0\} \sqcup \{[(e_0,0),(e_1,1)] \mid e \in M_1\} ,
\end{multline*}
and the set of 2-simplices is
\begin{multline*}
(M \times I)_2 = \{p \times \{0\} \mid p \in M_2\} \sqcup \{p \times \{1\} \mid p \in M_2\} \\
\sqcup \{[(e_0,0),(e_0,1),(e_1,1)] \mid e \in M_1\} \sqcup \{[(e_0,0),(e_1,0),(e_1,1)] \mid e \in M_1\} \\
\sqcup \{[(p_0,0),(p_1,1),(p_2,1)] \mid p \in M_2\} \sqcup \{[(p_0,0),(p_1,0),(p_2,1)] \mid p \in M_2\} .
\end{multline*}
Thus the cellular map $H$ is determined by
\bit
\item $H([(v,0),(v,1)]) \eqqcolon \phi_0(v) \in G$ for all $v \in M_0$;
\item $H([(e_0,0),(e_0,0),(e_1,1)]) \eqqcolon \phi_1^+(e) \in A$ for all $e \in M_1$;
\item $H([(e_0,0),(e_1,1),(e_1,1)]) \eqqcolon \phi_1^-(e) \in A$ for all $e \in M_1$.
\eit
The 1-flatness condition implies that
\begin{multline}
\phi_0(e_1) \tau_1(e) = H([(e_1,0),(e_1,1)]) H(e \times \{0\}) = H([(e_0,0),(e_1,1)]) \\
= H(e \times \{1\}) H([(e_0,0),(e_0,1)]) = \tau_1'(e) \phi_0(e_0) .
\end{multline}
This is the same as the 1-flatness condition \eqref{eq_1-flat_gauge_transformation} obtained in Section \ref{sec_gauge_transformation_2-group}.

The 2-flatness conditions on the three 3-simplices of $p \times I$ give
\begin{align*}
\phi_1^-([p_0,p_2]) \tau_2'(p) & = H([(p_0,0),(p_1,1),(p_2,1)]) (\tau_1'([p_1,p_2]) \triangleright \phi_1^-([p_0,p_1])) \\
 & \hspace{2cm} \cdot \alpha(\tau_1'([p_1,p_2]),\tau_1'([p_0,p_1]),\phi_0(p_0)) , \\
H([(p_0,0),(p_1,0),(p_2,1)]) \phi_1^-([p_1,p_2]) & = H([(p_0,0),(p_1,1),(p_2,1)]) (\tau_1'([p_1,p_2]) \triangleright \phi_1^+([p_0,p_1])) \\
 & \hspace{2cm} \cdot \alpha(\tau_1'[p_1,p_2],\phi_0(p_1),\tau_1([p_0,p_1])) , \\
H([(p_0,0),(p_1,0),(p_2,1)]) \phi_1^+([p_1,p_2]) & = \phi_1^+([p_0,p_2]) (\phi_0(p_2) \triangleright \tau_2(p)) \\
 & \hspace{2cm} \cdot \alpha(\phi_0(p_2),\tau_1([p_1,p_2]),\tau_1([p_0,p_1])) .
\end{align*}
Multiplying the three equations together we obtain
\begin{multline*}
\tau_2'(p) (\tau_1'([p_1,p_2]) \triangleright (\phi_1^+([p_0,p_1]) / \phi_1^-([p_0,p_1]))) (\phi_1^+([p_1,p_2]) / \phi_1^-([p_1,p_2])) \\
\hspace{-5cm} = (\phi_0(p_2) \triangleright \tau_2(p)) (\phi_1^+([p_0,p_2]) / \phi_1^-([p_0,p_2])) \\
\cdot \frac{\alpha(\tau_1'([p_1,p_2]),\tau_1'([p_0,p_1]),\phi_0(p_0)) \cdot \alpha(\phi_0(p_2),\tau_1([p_1,p_2]),\tau_1([p_0,p_1]))}{\alpha(\tau_1'([p_1,p_2]),\phi_0(p_1),\tau_1([p_0,p_1]))} .
\end{multline*}
Define $\phi_1(e) \coloneqq \phi_1^+(e) / \phi_1^-(e)$ for all $e \in M_1$. Then the above equation can be written as
\begin{multline}
\tau_2'(p) (\tau_1'([p_1,p_2]) \triangleright \phi_1([p_0,p_1])) \phi_1([p_1,p_2]) = (\phi_0(p_2) \triangleright \tau_2(p)) \phi_1([p_0,p_2]) \\
\cdot \frac{\alpha(\tau_1'([p_1,p_2]),\tau_1'([p_0,p_1]),\phi_0(p_0)) \cdot \alpha(\phi_0(p_2),\tau_1([p_1,p_2]),\tau_1([p_0,p_1]))}{\alpha(\tau_1'([p_1,p_2]),\phi_0(p_1),\tau_1([p_0,p_1]))} .
\end{multline}
This is the same as the 2-flatness condition \eqref{eq_2-flat_gauge_transformation} obtained in Section \ref{sec_gauge_transformation_2-group}.

\bibliography{ref}

\end{document}